\documentclass[11pt]{article}
\usepackage[utf8]{inputenc}
\pdfoutput=1
\usepackage{setspace}
\usepackage{palatino}
\usepackage{graphicx}
\usepackage{float}
\usepackage{titling} 
\usepackage{multirow}
\usepackage{lscape}
\usepackage{amsmath}
\usepackage{amssymb}
\usepackage[a4paper, total={6in, 9.5in}]{geometry}
\usepackage{array}
\usepackage[normalem]{ulem}
\fontfamily{ppl}\selectfont 
\usepackage{eqparbox}
\usepackage{arydshln}

\usepackage[american]{babel}

\usepackage{bm}
\usepackage{caption}
\usepackage{subcaption}

\newcommand{\appendixqqsection}[1]{\addtocounter{section}{1}
   \setcounter{table}{0}
   \setcounter{figure}{0}
   \setcounter{equation}{0}
   \setcounter{subsection}{0}
  \section*{Supplementary \Alph{section}: #1}
}

\newcommand\appendixqq{
   \setcounter{section}{0}
   \renewcommand\thesection{\Alph{section}}
   \renewcommand{\thesubsection}{\thesection. \arabic{subsection}}
   \renewcommand{\thesubsubsection}{\thesubsection .\arabic{subsubsection}}
   \renewcommand{\thefigure}{\thesection.\arabic{figure}}
   \renewcommand{\thetable}{\thesection.\arabic{table}}
}

\usepackage[natbibapa]{apacite}
\PassOptionsToPackage{hyperindex,breaklinks}{hyperref}
\usepackage{hyperref}
\usepackage{xcolor}
\definecolor{darkblue}{rgb}{0, 0, 0.5}
\hypersetup{colorlinks=true,citecolor=darkblue, linkcolor=darkblue, urlcolor=darkblue}

\usepackage[textsize=small]{todonotes}

\title{Improving information retrieval through correspondence analysis instead of latent semantic analysis
}

\author{Qianqian Qi\\
   {\small\raggedright Utrecht University}\\
   \href{mailto:q.qi@uu.nl}{\texttt{q.qi@uu.nl}} 
\and David J. Hessen\\
   {\small\raggedright Utrecht University}\\
   \href{mailto:d.j.hessen@uu.nl}{\texttt{d.j.hessen@uu.nl}}
\and Peter G. M. van der Heijden\\
    {\small\raggedright Utrecht University and University of Southampton}\\
\href{mailto:p.g.m.vanderheijden@uu.nl}{\texttt{p.g.m.vanderheijden@uu.nl}}
    }
    
\date{}
\date{\vspace{-5ex}}

\renewenvironment{abstract}
 {\par\noindent\textbf{\abstractname: }\ \ignorespaces}
 {\par\medskip}
\usepackage{soul}
\begin{document}
{\setstretch{.8}
\maketitle

\begin{abstract}

\noindent Both latent semantic analysis (LSA) and correspondence analysis (CA) are dimensionality reduction techniques that use singular value decomposition (SVD) for information retrieval. Theoretically, the results of LSA display both the association between documents and terms, and marginal effects; in comparison, CA only focuses on the associations between documents and terms. Marginal effects are usually not relevant for information retrieval, and therefore, from a theoretical perspective CA is more suitable for information retrieval.

In this paper, we empirically compare LSA and CA. The elements of the raw document-term matrix are weighted, and the weighting exponent of singular values is adjusted to improve the performance of LSA. We explore whether these two weightings also improve the performance of CA. In addition, we compare the optimal singular value weighting exponents for LSA and CA to identify what the initial dimensions in LSA correspond to.

The results for four empirical datasets show that CA always performs better than LSA. Weighting the elements of the raw data matrix can improve CA; however, it is data dependent and the improvement is small. Adjusting the singular value weighting exponent usually improves the performance of CA; however, the extent of the improved performance depends on the dataset and number of dimensions. In general, CA needs a larger singular value weighting exponent than LSA to obtain the optimal performance. This indicates that CA emphasizes initial dimensions more than LSA, and thus, margins play an important role in the initial dimensions in LSA.
 
\end{abstract}
}

\section{Introduction}\label{Intro}

In information retrieval, a given user query is matched with relevant documents \citep{kolda1998semidiscrete, zhang2011comparative, 7314225, guo2022semantic} and a collection of documents is usually represented as a document-term matrix. The similarity between the given user query and each document in the document-term matrix is calculated, and documents with highest similarity are returned. Unfortunately, the returned documents may not be relevant, because similarity is calculated based on word matching, and different words may have the same meaning and the same word may have different meanings.  

Latent semantic analysis (LSA) is a variant of the vector space model for information retrieval \citep{dumais1988using, deerwester1990indexing, dumais1991improving}. LSA replaces the original document-term matrix with latent semantic vectors using singular value decomposition (SVD). It attempts to capture the association between documents and terms, which helps discover the hidden semantic associations between documents. In LSA, documents can be similar to a query even though they do not have any word in common.

LSA has been used as a common baseline for information retrieval \citep{parali2019information, duan2021hybrid, chang2021using}, text mining \citep{dzisevivc2019text, bellaouar2021topic}, and natural language processing \citep{mohan2022comprehensive, rezende2022combining}. Furthermore, results obtained using LSA are comparable to those of neural network models \citep{levy2015improving, altszyler2016comparative, phillips2021comparing}. For example, \citet{altszyler2016comparative} compared the capabilities of Word2Vec (Skip-Gram model) and LSA to extract relevant semantic patterns in dream reports and showed that LSA achieved a better performance. \citet{phillips2021comparing} showed that LSA outperformed neural network methods (such as ELMo word embeddings) in text classification tasks for educational data. Moreover, new methods that rely on LSA are being proposed \citep{azmi2019aaee, 9395976, hassani2021text, suleman2021extending, horasan2022latent, patil2022word}. For example, \citet{9395976} proposed an algorithm for text summarization that uses LSA, TF-IDF keyword extractor, and BERT encoder model. The algorithm performed better than latent Dirichlet allocation. \citet{horasan2022latent} proposed a collaborative filtering-based recommendation system using LSA and achieved good performance. \citet{patil2022word} developed a new procedure for information retrieval using LSA and TF-IDF. The procedure helped identify the most and the least significant words from sentences in documents with an accuracy of 0.943.

Capturing associations between documents and terms appears necessary for the success of LSA in computing science; however, the solution of LSA is a mix of the associations between documents and terms, and marginal effects arising from the lengths of documents and marginal frequencies of terms \citep{qi2021comparison}. \citet{hu2003lsa} and \citet{qi2021comparison} show that margins play an important role in the first dimension extracted by LSA.

Correspondence analysis (CA) is another technique that uses SVD \citep{greenacre1984theory, greenacre2017correspondence, beh2021introduction}. In computing science, CA has not been explored as much as LSA. CA is usually used to make two-dimensional graphical displays \citep{hou_huang_2020, arenas2021convolutional, van2021correspondence}. For example, \citet{arenas2021convolutional} depicted a biplot using CA to show that the document encoding of convolutional neural encoder can emphasize the dissimilarity between documents belonging to different classes.
Unlike LSA, CA ignores the information on marginal frequency differences between documents and between terms from the solution by preprocessing the data, and it only focuses on the relationships between documents and terms \citep{qi2021comparison}. Thus, CA seems more suitable for information retrieval. 

\citet{seguela2011comparison} and \citet{qi2021comparison} experimentally compared LSA and CA for text clustering and text categorization, respectively, and they found that CA performed better than LSA. Although LSA was originally proposed for information retrieval, to the best of our knowledge, an empirical comparison between LSA and CA continues to remain lacking in this field. In this paper, we use four datasets, three English datasets and a Dutch dataset, to compare the performances of LSA and CA in information retrieval. 

Weighting the elements of the raw document-term matrix is a common and effective method to improve the performance of LSA \citep{dumais1991improving, article, Bacciu2019BotAG}. LSA usually involves the SVD of a raw or pre-processed document-term matrix. Further, LSA owes its popularity to its applicability to different matrices. In CA, it is unusual to weight the elements of the raw document-term matrix, and processing the raw document-term matrix is an integral part of CA \citep{greenacre1984theory, greenacre2017correspondence, beh2021introduction}. CA is based on the SVD of the matrix of standardized residuals. Here, we study the CA of document-term matrices whose entries are weighted to see if this has an impact on the performance of CA.

In addition to weighting the elements of the document-term matrix to improve information retrieval, \citet{caron2001experiments} proposed changing the weighting exponent of the singular values in LSA. His results showed that adjusting the weighting exponent of singular values improves the performance of information retrieval. Since \citet{caron2001experiments}, singular value weighting exponents have been 
widely studied and applied in word embeddings generated from word-context matrices \citep{bullinaria2012extracting, osterlund2015factorization, drozd2016word, yin2018dimensionality}.  Further, other variants that change the singular value weighting exponent are studied in word embeddings created by Word2Vec and GloVe \citep{Mu2018, liu2019unsupervised}. To the best of our knowledge, there is no study of CA where the weighting exponent of the singular values is varied. Based on the success of adjusting the weighting exponent of singular values in LSA, we will explore whether this is also successful in CA.  

The larger the weighting exponent of the singular
values, the higher is the emphasis given to the initial dimensions. The standard weighting exponent corresponds to 1. According to the experimental results of \citet{caron2001experiments}, giving more emphasis to initial dimensions can often improve the performance of information retrieval on standard test datasets, whereas giving more emphasis to initial dimensions can decrease the performance on question/answer matching. Papers about word embeddings tend to reduce the contribution of initial dimensions to improve performance \citep{bullinaria2012extracting, osterlund2015factorization, drozd2016word, yin2018dimensionality, Mu2018, liu2019unsupervised}, although the optimal value of the singular value weighting exponent is task dependent \citep{osterlund2015factorization}. \citet{bullinaria2012extracting} reported that assigning less weight to initial dimensions leads to improved performance for TOEFL, distance comparison, semantic categorization, and clustering purity tasks on a word-context matrix created from the ukWaC corpus \citep{baroni2009wacky}. They argued that the general pattern appears to be that the initial dimensions tend not to contribute the most useful information about semantics and have a large “noise” component that is best removed or reduced. However, it remains unclear what the initial dimensions correspond to. If the initial dimensions corresponds to marginal differences between documents and between terms, then CA may be the optimal procedure because it ignores this irrelevant information.

In summary, this work makes three contributions. First, to compare LSA and CA in information retrieval. The results show that CA yields better performance than LSA. Thus, CA may be used as common baseline and be widely applied in information retrieval, text mining, and natural language processing. Second, to explore whether weightings, including the weighting of the elements of the raw document-term matrix and the adjusting of the singular value weighting exponent, can improve the performance of CA. Third, to study what the initial dimensions of LSA correspond to and whether CA is effective in ignoring the useless information in the raw or pre-processed document-term matrix that contributes a large part of the initial dimensions extracted by LSA. We extensively compare the performances of LSA and CA applied to four datasets using Euclidean distance, dot similarity, and cosine similarity.

The remainder of this paper is arranged as follows. In Section~\ref{S: lsaca}, LSA and CA are described in brief. Section~\ref{S: methods} presents the methodology used in this paper. The results for Euclidean distance are presented in Section~\ref{S: resultseuc}, and the results for dot similarity and cosine similarity are presented in Section~\ref{S: resultsdotcos}. Finally, Section~\ref{S: condis} concludes and discusses the results.

\section{LSA and CA}\label{S: lsaca}

In this section, we briefly describe LSA and CA. We refer the readers to \citet{qi2021comparison} for a more detailed presentation of the methods.

\subsection{LSA}\label{Sub: LSA}

Consider a raw document-term matrix $\bm{F} = [f_{ij}]$ with $m$ rows $(i = 1, ...,m)$ and $n$ columns
$(j = 1, ..., n)$, where the rows represent documents and the columns represent terms.
Weighting might be used to prevent the differential lengths of documents from considerably affecting the representation, or to impose certain preconceptions about which
terms are more important \citep{deerwester1990indexing}. The weighted element $a_{ij}$ for term $j$ in
document $i$ is
\begin{equation}\label{Eqweighfunctioncom}
a_{ij} = L(i,j)\times G(j) \times N(i),
\end{equation}
where the local weighting term $L(i,j)$ is the weight of term $j$ in document $i$, $G(j)$ is the global weight of term $j$ in the entire set of documents, and $N(i)$ is the weighting component for document $i$. The choices $L(i,j) = f_{ij}$, $G(j) = 1$, and $N(i) = 1$ result in the raw document-term matrix $\textbf{\emph{F}}$. The popular TF-IDF can be written in the form $L(i,j)=f_{ij},
G(j)=1+\log_2(n\text{docs}/df_j)
, N(i)=1$, where $n$docs is the number of documents in the set and $df_j$ is the number of documents where term $j$ appears \citep{dumais1991improving}. The SVD of $\bm{A}=[a_{ij}]$ is
\begin{equation}\label{EqweighfunctioncomM}
\bm{A}
=\bm{U}\bm{\Sigma}\bm{V}^T
\end{equation}
where $\bm{U}^T\bm{U} = \mathbf{I}$, $\bm{V}^T\bm{V} = \mathbf{I}$, and $\bm{\Sigma}$ is a diagonal matrix with singular values on the diagonal in the descending order. We denote matrices that contain the first $k$ columns of $\bm{U}$, first $k$ columns of $\bm{V}$, and $k$ largest singular values of $\bm{\Sigma}$ by $\bm{U}_k$, $\bm{V}_k$, and $\bm{\Sigma}_k$, respectively. Then, $\bm{U}_k\bm{\Sigma}_k(\bm{V}_k)^T$ provides the optimal rank-$k$ approximation of $\bm{A}$ in a least-squares sense, which shows that SVD can be used for data reduction. In LSA, the rows of $\bm{U}_k\bm{\Sigma}_k$ and $\bm{V}_k\bm{\Sigma}_k$ provide the coordinates of row and column points, respectively. Euclidean distances between the rows of $\bm{U}_k\bm{\Sigma}_k$ ($\bm{V}_k\bm{\Sigma}_k$) approximate those between the rows (columns) of $\bm{A}$ \citep{deerwester1990indexing, parhizkar2013euclidean}.

Representing out-of-sample documents or queries in the $k$-dimensional subspace of LSA is important for many applications including information retrieval. Suppose that the new weighted document is a row vector $\bm{d}$. Since $\bm{V}^T\bm{V} = \mathbf{I}$ and $\bm{U}^T\bm{U} = \mathbf{I}$, we have 
\begin{equation}
\label{Etransitionsvdrow} 
\begin{aligned}
\bm{A}\bm{V}_k=\bm{U}_k\bm{\Sigma}_k
\end{aligned}
\end{equation}
and
\begin{equation}
\label{Etransitionsvdcolumn} 
\begin{aligned}
\bm{A}^T\bm{U}_k=\bm{V}_k\bm{\Sigma}_k
\end{aligned}
\end{equation}
Therefore, using Equation~(\ref{Etransitionsvdrow}), the coordinates of the out-of-sample document $\bm{d}$ in the $k$-dimensional subspace of LSA is $\bm{d}\bm{V}_k$. Similarly, using Equation~(\ref{Etransitionsvdcolumn}), the coordinates of the out-of-sample term $\bm{t}$ (represented as row vector) in the $k$-dimensional subspace of LSA is $\bm{t}\bm{U}_k$.

As in \citet{qi2021comparison}, we first use a small dataset to illustrate LSA. This small dataset is introduced in \citet{aggarwal2018machine} (see Table~\ref{T6*6dt}), and it contains 6 documents. For each document, we are interested in the frequency of occurrence of six terms. The first three documents primarily refer to cats, the last two primarily to cars, and the fourth to both. The fourth term, jaguar, is polysemous because it can refer to either a cat or a car.  

\begin{table}[H]
\centering  
\caption{A document-term matrix $\bm{F}$: size 6$\times$6} 
\label{T6*6dt}
\begin{tabular}{ccccccc}    
\hline
 & lion & tiger & cheetah &jaguar & porsche & ferrari\\  
\hline  
doc1 & 2 & 2 & 1 & 2 & 0 & 0 \\ 
doc2 & 2 & 3 & 3 & 3 & 0 & 0 \\  
doc3 & 1 & 1 & 1 & 1 & 0 & 0\\ 
doc4 & 2 & 2 & 2 & 3 & 1 & 1\\ 
doc5 & 0 & 0 & 0 & 1 & 1 & 1\\ 
doc6 & 0 & 0 & 0 & 2 & 1 & 2\\ 
\hline 
\end{tabular}  
\end{table}

\begin{figure}[htbp]
    \centering
      \begin{subfigure}[b]{0.45\linewidth}
         \centering
         \includegraphics[width=1\textwidth]{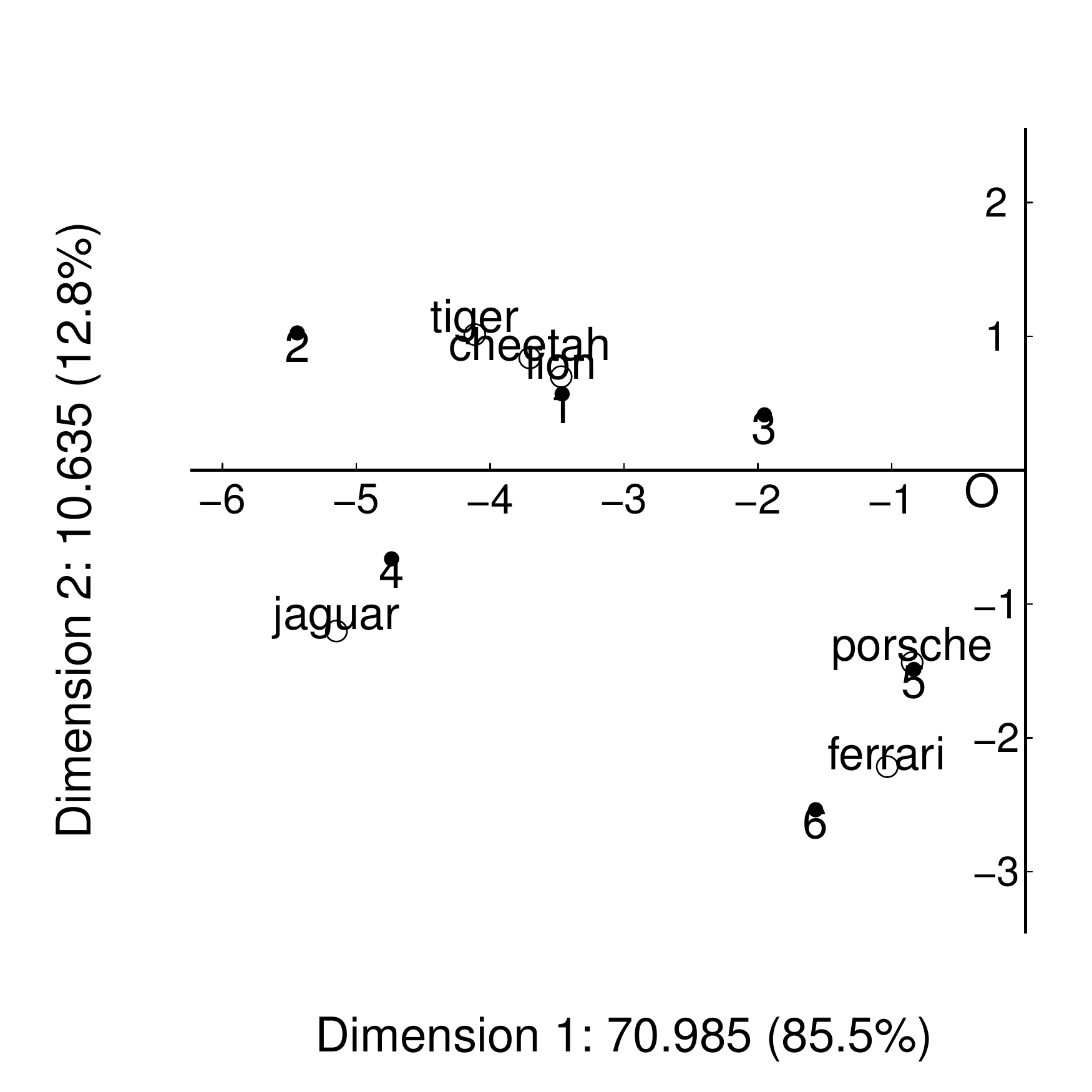}
       \caption{LSA-RAW}\label{F: lsaraw}
       \end{subfigure}
       \begin{subfigure}[b]{0.45\linewidth}
         \centering
         \includegraphics[width=1\textwidth]{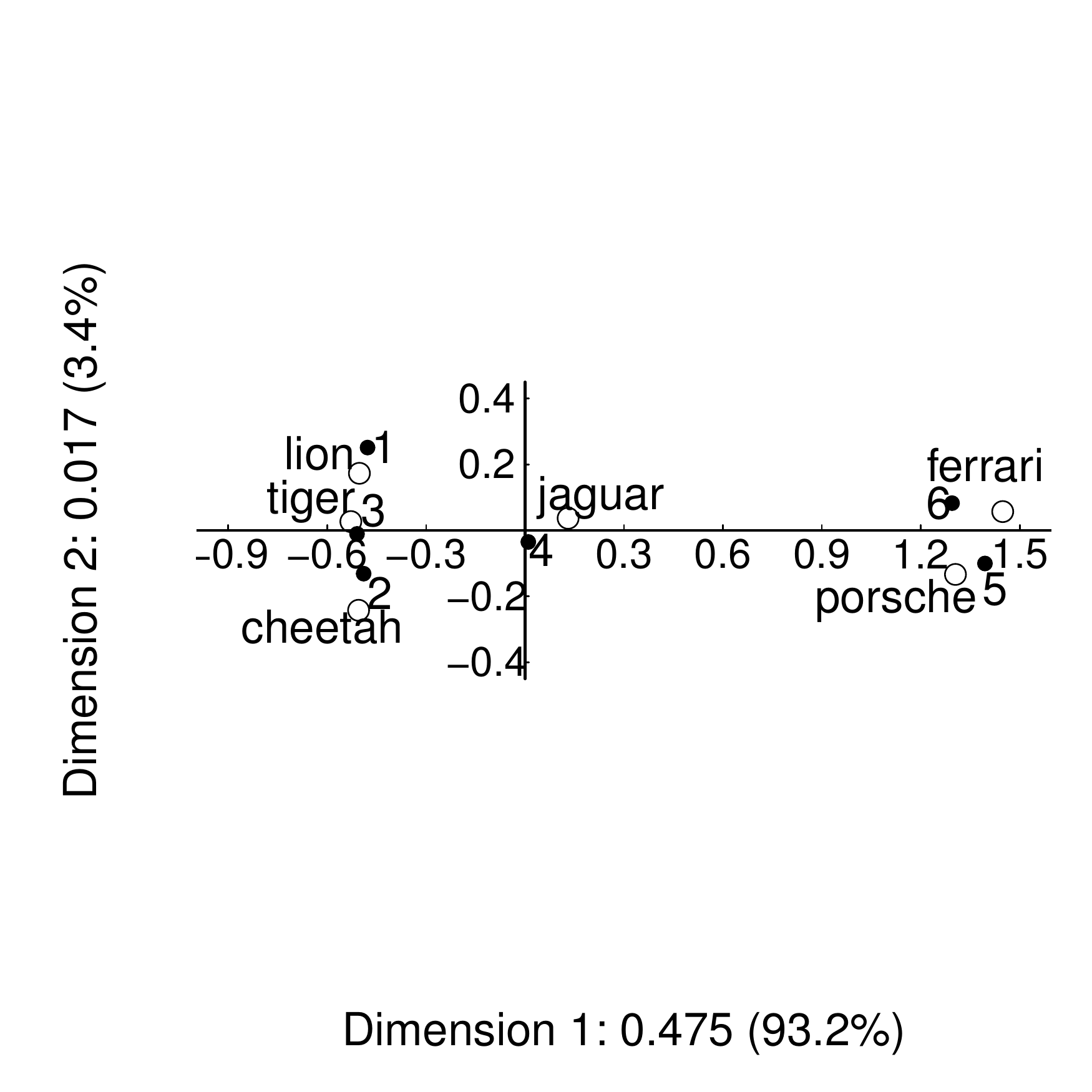}
         \caption{CA}\label{F: caraw}
         \end{subfigure}
    \caption{A two-dimensional plot of documents and terms for (a) LSA-RAW, (b) CA.}
    \label{F: lsaca}
\end{figure}

In the LSA of the raw document-term matrix (LSA-RAW), the rows and columns of $\bm{F}$ are not weighted, and therefore, we can replace $\bf{A}$ in Equation~(\ref{EqweighfunctioncomM}) by $\bm{F}$. The coordinates of the documents and of the terms for LSA-RAW in the first two dimensions are $\bm{U}_2\bm{\Sigma}_2$ and $\bm{V}_2\bm{\Sigma}_2$, respectively. Figure~\ref{F: lsaraw} shows the two-dimensional plot of the documents and terms. Cat terms (\emph{lion} and \emph{tiger}) are close together; car terms (\emph{jaguar}, \emph{porsche}, and \emph{ferrari}) are close together; car documents (5 and 6) are close together. However, the cat documents (1, 2, and 3) are not close together, neither is document 4 in between cat documents and car documents, and neither is \emph{jaguar} in between cat terms and car terms. This can be attributed to the fact that LSA displays both the relationships between documents and terms and the sizes of the documents and terms: for the latter,  \emph{jaguar},  for example, is used most often in the documents and is furthest away from the origin.

\subsection{CA}\label{Sub: CA}

In CA, an SVD is applied to the matrix of standarized residuals given by \citet{greenacre2017correspondence}
\begin{equation}
\label{CA}
\textbf{\emph{S}} = \textbf{\emph{D}}_r^{-\frac{1}{2}}(\textbf{\emph{P}}-\textbf{\emph{E}})\textbf{\emph{D}}_{c}^{-\frac{1}{2}}
\end{equation}
where $\textbf{\emph{P}} = [p_{ij}]$ is the matrix  of joint observed proportions with $p_{ij} = f_{ij}/\sum_{i}\sum_{j}f_{ij}$, $\textbf{\emph{D}}_r$ is a diagonal matrix with $r_i = \sum_jp_{ij}$ $(i = 1, 2, \cdots, m)$ on the diagonal, $\textbf{\emph{D}}_c$ is a diagonal matrix with $c_j = \sum_ip_{ij}$ $(j = 1, 2, \cdots, n)$ on the diagonal, and $\textbf{\emph{E}} = [r_ic_j]$ is the matrix of expected proportions under the statistical independence of the documents and the terms.  
The elements of $\textbf{\emph{D}}_r^{-\frac{1}{2}}(\textbf{\emph{P}}-\textbf{\emph{E}})\textbf{\emph{D}}_{c}^{-\frac{1}{2}}$ are standardized residuals under the statistical independence model. 
The sum of squares of these elements yields the total inertia, i.e., the Pearson $\chi^2$ statistic divided by sample size $N$. By taking the SVD of the matrix of standardized residuals, we get
\begin{equation}
\label{SP3}
\textbf{\emph{D}}_r^{-\frac{1}{2}}(\textbf{\emph{P}}-\textbf{\emph{E}})\textbf{\emph{D}}_{c}^{-\frac{1}{2}} = \textbf{\emph{U}} \boldsymbol{\Sigma} \textbf{\emph{V}}^T
\end{equation}
In CA, the rows of $\boldsymbol{\Phi}_k\boldsymbol{\Sigma}_k$ and $\boldsymbol{\Gamma}_k\boldsymbol{\Sigma}_k$ provide the coordinates of row and column points, respectively, where $\boldsymbol{\Phi}_k = \textbf{\emph{D}}_r^{-\frac{1}{2}}\textbf{\emph{U}}_k$ and $\boldsymbol{\Gamma}_k = \textbf{\emph{D}}_c^{-\frac{1}{2}}\textbf{\emph{V}}_k$. The weighted sum of the coordinates is 0: $\sum_ir_i\phi_{ik} = 0 = \sum_jc_j\gamma_{jk}$. Euclidean distances between the rows of $\boldsymbol{\Phi}_k\boldsymbol{\Sigma}_k$ ($\boldsymbol{\Gamma}_k\boldsymbol{\Sigma}_k$) approximate $\chi^2$-distances between the rows (columns) of $\textbf{\emph{F}}$, where the squared $\chi^2$-distance between rows $k$ and $l$ is
\begin{equation}
\label{chidistace}
\delta_{kl}^2 = \sum_j{\frac{\left(p_{kj}/r_k - p_{lj}/r_l\right)^2}{c_j}}
\end{equation}
In Equation~(\ref{chidistace}), the rows are transformed into vectors of conditional proportions adding up to 1 for each row, such as the $k$th row: $p_{kj}/r_k$, $j = 1, 2, \cdots, n$, and the differences between the column elements for column $j$ in the transformed rows are corrected for $c_j$, which represents the size of column $j$.

The transition formulas are
\begin{equation}
\label{transitiondecomposition1}
\textbf{\emph{D}}_r^{-1}\textbf{\emph{P}}\boldsymbol{\Gamma}_k = \boldsymbol{\Phi}_k\boldsymbol{\Sigma}_k
\end{equation}
and
\begin{equation}
\label{transitiondecomposition2}
\textbf{\emph{D}}_c^{-1}\textbf{\emph{P}}^T\boldsymbol{\Phi}_k = \boldsymbol{\Gamma}_k\boldsymbol{\Sigma}_k
\end{equation}
Equation~(\ref{transitiondecomposition1}) shows that the row points are in the weighted averages of the column points when rows of $\textbf{\emph{D}}_r^{-1}\textbf{\emph{P}}$ are used as weights when rows of $\textbf{\emph{D}}_r^{-1}\textbf{\emph{P}}$ are used as weights, and Equation~(\ref{transitiondecomposition2}) shows that the column points are in the weighted averages of the row points simultaneously.

According to Equation~(\ref{transitiondecomposition1}), a new document $\textbf{\emph{d}}$, represented by a row vector, can be projected onto the $k$-dimensional subspace by placing it in the weighted average of the column points using $(\textbf{\emph{d}}/\sum_{j=1}^nd_j)\boldsymbol{\Gamma}_k$. This can be similarly done for a new term $\textbf{\emph{t}}$. 

For the CA of Table \ref{T6*6dt}, the coordinates of the documents and terms for CA in the first two dimensions are $\boldsymbol{\Phi}_2\boldsymbol{\Sigma}_2$ and $\boldsymbol{\Gamma}_2\boldsymbol{\Sigma}_2$, respectively. Figure~\ref{F: caraw} shows a two-dimensional plot of the documents and terms. Cat terms (\emph{lion} and \emph{tiger}) are close together; car terms (\emph{jaguar}, \emph{porsche}, and \emph{ferrari}) are close together; \emph{jaguar} is in between cat and car terms; car documents (5 and 6) are close together, cat documents (1, 2, and 3) are close together; and document 4 is in between cat and car documents. All data properties are found in Figure~\ref{F: caraw}. A comparison of Figures~\ref{F: caraw} and \ref{F: lsaraw} suggests that CA provides a clearer visualization of the important aspects of the data than LSA. This is because the coordinates of each dimension are orthogonal to the margins due to $\sum_ir_i\phi_{ik} = 0 = \sum_jc_j\gamma_{jk}$, and CA focuses only on the relationship between the documents and the terms.

\section{Methodology}\label{S: methods}

In this section, we introduce the CA of a document-term matrix whose entries are weighted. We also discuss how the influence of the initial dimensions can be studied. Subsequently, we describe the study design, datasets, and evaluation methods used.

\subsection{CA of a document-term matrix of weighted frequencies}

Weighting the entries of the raw document-term matrix is an effective method for improving the performance of LSA, and this motivates us to study the weighting of the elements of the input matrix of CA.

The processing of the raw data matrix by $\textbf{\emph{D}}_r^{-\frac{1}{2}}(\textbf{\emph{P}}-\textbf{\emph{E}})\textbf{\emph{D}}_{c}^{-\frac{1}{2}}$ (see Equation~(\ref{CA})) is considered an integral part of CA. This processing step effectively eliminates the margins, which allows CA to focus on the relationships between documents and terms. The weighting of the entries of the raw document-term matrix in Equation~(\ref{Eqweighfunctioncom}), such as by TF-IDF, can be used to assign higher values to terms with
more indicative of the meaning of documents. Thus, the weighting of the entries of the raw document-term matrix may also be an effective method for improving the performance of CA.

To perform the CA of a document-term matrix of weighted frequencies, we first use Equation~(\ref{Eqweighfunctioncom}) to obtain a document-term matrix $\textbf{\emph{A}}$ of weighted frequencies, and then, we perform CA on this matrix $\textbf{\emph{A}}$ instead of $\textbf{\emph{F}}$. 

\subsection{Changing the contributions of the initial dimensions in SVD}\label{Sub: alpha}

\citet{caron2001experiments} proposed adjusting the relative strengths of vector components in LSA using $\bm{U}_k\bm{\Sigma}_k^\alpha$ or $\bm{V}_k\bm{\Sigma}_k^\alpha$ as coordinates instead of $\bm{U}_k\bm{\Sigma}_k$ or $\bm{V}_k\bm{\Sigma}_k$, where $\alpha$ is the singular value weighting exponent that adjusts the importance of the dimensions. The weighting exponent $\alpha$ determines how components are weighted relative to the standard $\alpha = 1$ case described in Section~\ref{Sub: LSA}. In comparison to $\alpha = 1$, $\alpha < 1$ gives less emphasis to initial dimensions, and $\alpha > 1$, more emphasis.

\citet{bullinaria2012extracting} used both weighting exponent $\alpha <1$ and the exclusion of initial dimensions, which led to performance improvements of a similar degree. They argued that the general pattern appears to be that the dimensions with the highest singular values tend not to contribute the most useful information about semantics and have a large “noise” component that is best removed or reduced. However, it is unclear what the initial dimensions actually correspond to. Given this context, we change the contributions of the initial dimensions extracted by both LSA and CA and compare their performances.

We use Table~\ref{T6*6dt} to illustrate the impact of $\alpha$ on singular values and coordinates. We use $\alpha = -0.5$, $\alpha = 0$, $\alpha = 0.5$, $\alpha = 1$, and $\alpha = 1.5$. In the literature, we regularly encounter $\alpha = 0.5$ because it relates to
\begin{equation}
\bm{F} = \bm{U}\bm{\Sigma}\bm{V}^T \\
= \left(\bm{U}\bm{\Sigma}^{1/2}\right)\left(\bm{\Sigma}^{1/2}\bm{V}^T\right) \\
\end{equation}
which can then be used for making biplots \citep{gabriel1971biplot} using coordinate pairs $\bm{U}_2\bm{\Sigma}^{1/2}_2$ and $\bm{V}_2\bm{\Sigma}^{1/2}_2$. Biplots represent the rows and columns of a matrix in the same graphic representation, and the other often-used coordinate pairs are $\bm{U}_2\bm{\Sigma}_2$ and $\bm{V}_2$, and pairs $\bm{U}_2$ and $\bm{V}_2\bm{\Sigma}_2$. In the pair $\bm{U}_2\bm{\Sigma}_2$ and $\bm{V}_2$, the Euclidean distances between the rows of $\bm{U}_2\bm{\Sigma}_2$ approximate the Euclidean distances between the rows of $\bm{F}$, and in the pair $\bm{U}_2$ and $\bm{V}_2\bm{\Sigma}_2$, the Euclidean distances between the rows of $\bm{V}_2\bm{\Sigma}_2$ approximate the Euclidean distances between the columns of $\bm{F}$. The coordinate pair $\bm{U}_2\bm{\Sigma}^{1/2}_2$ and $\bm{V}_2\bm{\Sigma}^{1/2}_2$ provides a compromise between these two solutions while still representing matrix $\bm{F}$ in a graphic representation. In practice, one often sees the use of the coordinate pair $\bm{U}_2\bm{\Sigma}_2$ and $\bm{V}_2\bm{\Sigma}_2$; however, this is not a biplot representation as $\bm{\Sigma}_2$ is used twice. In a biplot, if the row points are $\bm{U}_2\bm{\Sigma}_2^{a}$, then the column points are $\bm{V}_2\bm{\Sigma}_2^{1-a}$, i.e., any entry of the matrix is approximated by the inner product of the corresponding row and column vectors. Hereafter, we do not make a biplot; instead, we make a symmetric plot where documents and terms have the same value of $\alpha$ because symmetric coordinates are usually used in experiments \citep{dumais1988using, deerwester1990indexing, berry1995using, levy2015improving}.

Table~\ref{T6*6dtlsarawsingular} lists the singular values to the power $\alpha$: $\sigma^\alpha$, the squared singular values to the power $\alpha$: $\sigma^{2\alpha}$, and proportions $\sigma^{2\alpha}/\sum_\sigma \sigma^{2\alpha}$, where we refer to the total sum of squared singular values to the power of $\alpha$, $\sum_\sigma \sigma^{2\alpha}$, as $\alpha$--inertia. These proportions show how the sum of the Euclidean distances of all components to the origin is distributed over the components. The greater $\alpha$ is, the more emphasis is given to the initial components. When $\alpha$ is negative, the initial components have less emphasis than the later components. For example, for $\sigma^{-0.5}$, the first dimension accounts for $0.017$ of $\alpha$-inertia, whereas the fifth dimension accounts for $0.536$. When $\alpha$ is $0$, all components of the coordinates have the same emphasis, and because the rank of $\bm{F}$ is 5, each dimension accounts for $0.2$ of $\alpha$-inertia. As $\alpha$ increases, there is more emphasis on the initial components of the coordinates and less emphasis on the latter ones. The first dimension accounts for $0.623$, $0.855$, and $0.943$ of $\alpha$-inertia, while the fifth dimension accounts for $0.020$, $0.001$, and $0.000$, with $\alpha$ being $0.5$, $1$, and $1.5$, respectively. The standard LSA solution has $\alpha = 1$.

\begin{table}[htbp]
\centering  
\caption{The $\sigma^\alpha$, $\sigma^{2\alpha}$, and the proportion of explained $\alpha$-inertia $\sigma^{2\alpha}/\sum_\sigma \sigma^{2\alpha}$ for each dimension of LSA-RAW.} 
\label{T6*6dtlsarawsingular}
\begin{tabular}{lccccc}    
\hline
 & dim1 & dim2 & dim3 & dim4 & dim5 
 \\
\hline
$\sigma^{-0.5}$ & 0.345& 0.554& 1.006 &1.320& 1.917 \\
$\sigma^{-1}$ & 0.119& 0.307& 1.012& 1.741& 3.675 \\
$\sigma^{-1}/\sum_\sigma \sigma^{-1}$ & 0.017& 0.045& 0.148& 0.254& 0.536 \\
\hline
$\sigma^0$ & 1& 1& 1& 1 &1  \\
$\sigma^0$ & 1& 1& 1& 1 &1  \\
$\sigma^{0}/\sum_\sigma \sigma^{0}$ & 0.2& 0.2& 0.2& 0.2& 0.2  \\
\hline
$\sigma^{0.5}$ & 2.903& 1.806& 0.994& 0.758& 0.522 \\
$\sigma^{1}$ & 8.425& 3.261& 0.988& 0.574& 0.272  \\
$\sigma^{1}/\sum_\sigma \sigma^{1}$ &0.623 &0.241 &0.073& 0.042 &0.020\\
\hline
$\sigma^{1}$ & 8.425& 3.261& 0.988& 0.574 &0.272  \\
$\sigma^{2}$ & 70.985&  10.635 &  0.976&   0.330&   0.074  \\
$\sigma^{2}/\sum_\sigma \sigma^{2}$ & 0.855& 0.128& 0.012& 0.004& 0.001  \\
\hline
$\sigma^{1.5}$ & 24.455& 5.889&  0.982&  0.435&  0.142  \\
$\sigma^{3}$ & 598.063 & 34.684&   0.964&   0.189&   0.020 \\
$\sigma^{3}/\sum_\sigma \sigma^{3}$ & 0.943& 0.055& 0.002& 0.000& 0.000  \\
\hline 
\end{tabular}  
\end{table}

\begin{figure}[htbp]
    \centering
       \begin{subfigure}[b]{0.45\linewidth}
         \centering
         \includegraphics[width=1\textwidth]{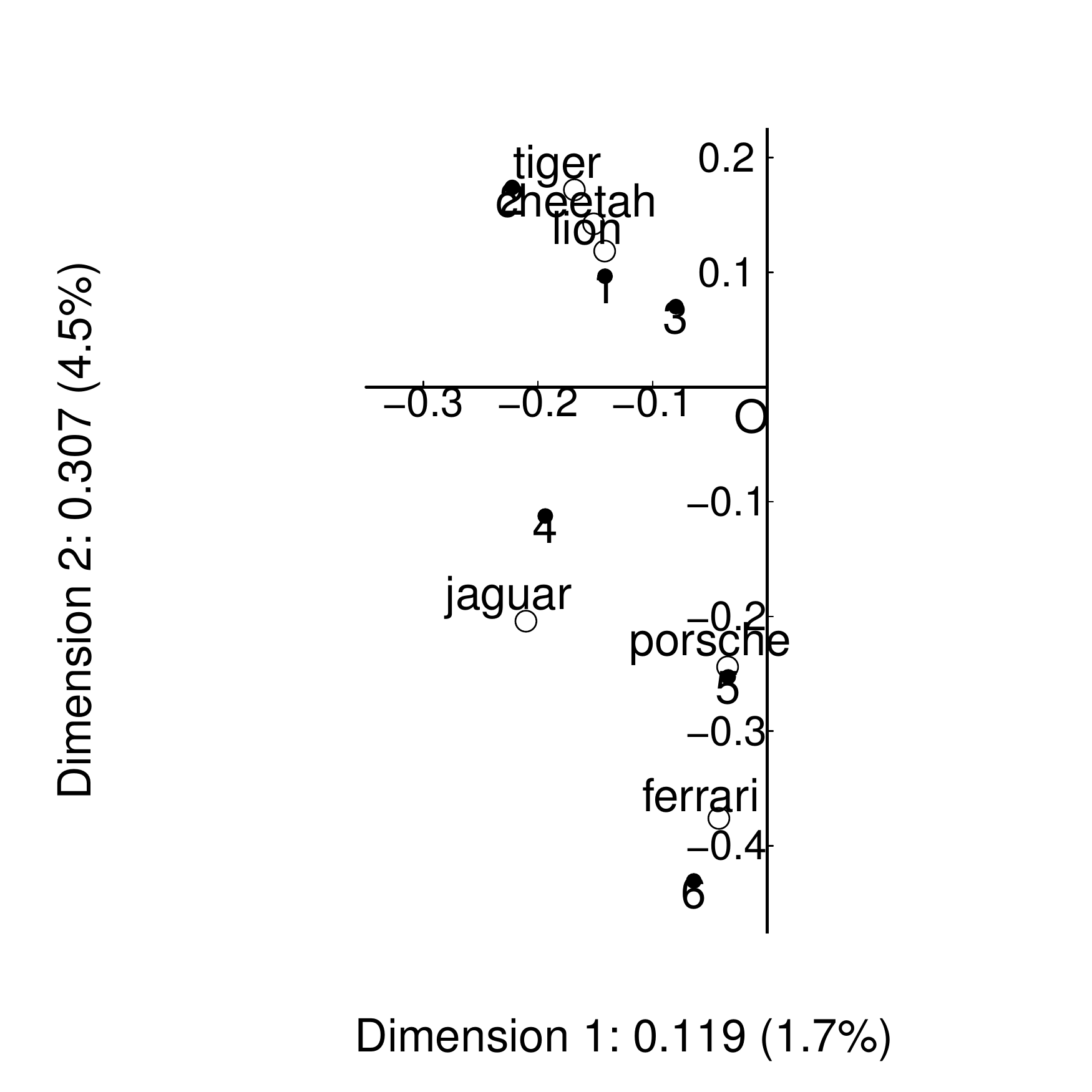}
         \caption{$\alpha = -0.5$}\label{F: lsarawalphaminus05}
         \end{subfigure}
       \hfill 
      \begin{subfigure}[b]{0.45\linewidth}
         \centering
         \includegraphics[width=1\textwidth]{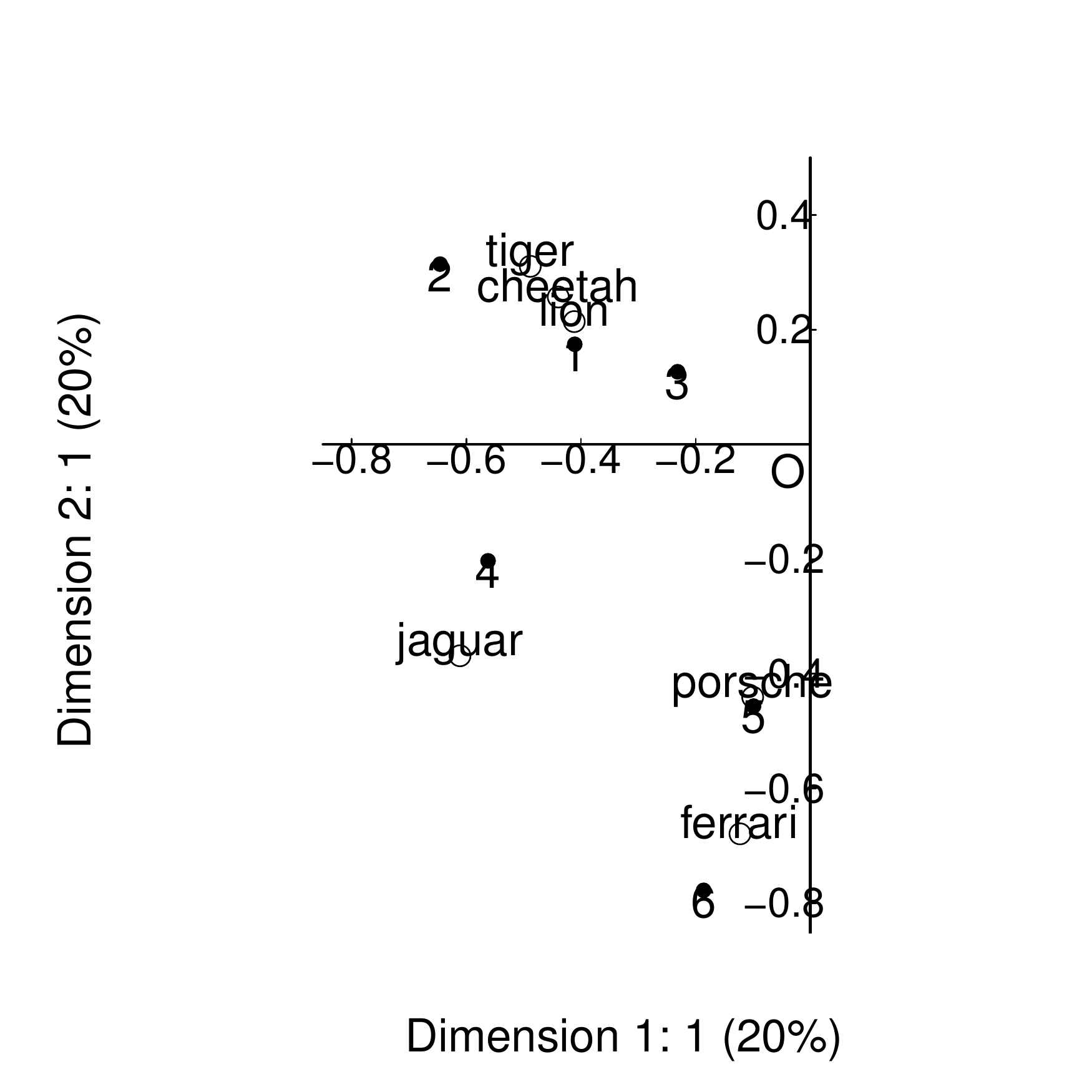}
         \caption{$\alpha = 0$}\label{F: lsarawalpha0}
         \end{subfigure}
         \hfill
       \begin{subfigure}[b]{0.45\linewidth}
         \centering
         \includegraphics[width=1\textwidth]{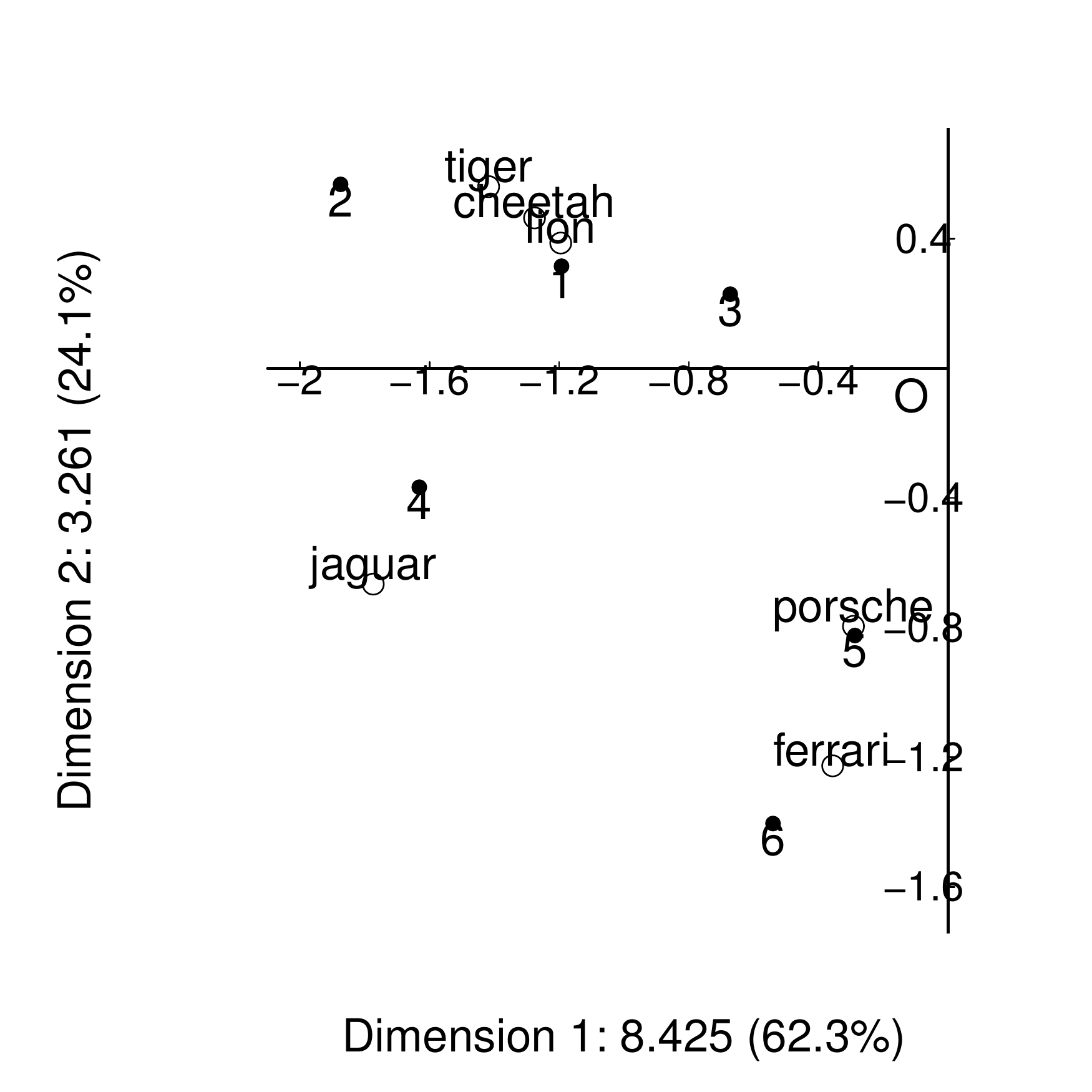}
         \caption{$\alpha = 0.5$}\label{F: lsarawalpha05}
         \end{subfigure}
         \hfill
      \begin{subfigure}[b]{0.45\linewidth}
         \centering
         \includegraphics[width=1\textwidth]{lsarawalpha1.pdf}
       \caption{$\alpha = 1$}\label{F: lsarawalpha1}
       \end{subfigure}
       \begin{subfigure}[b]{0.45\linewidth}
         \centering
         \includegraphics[width=1\textwidth]{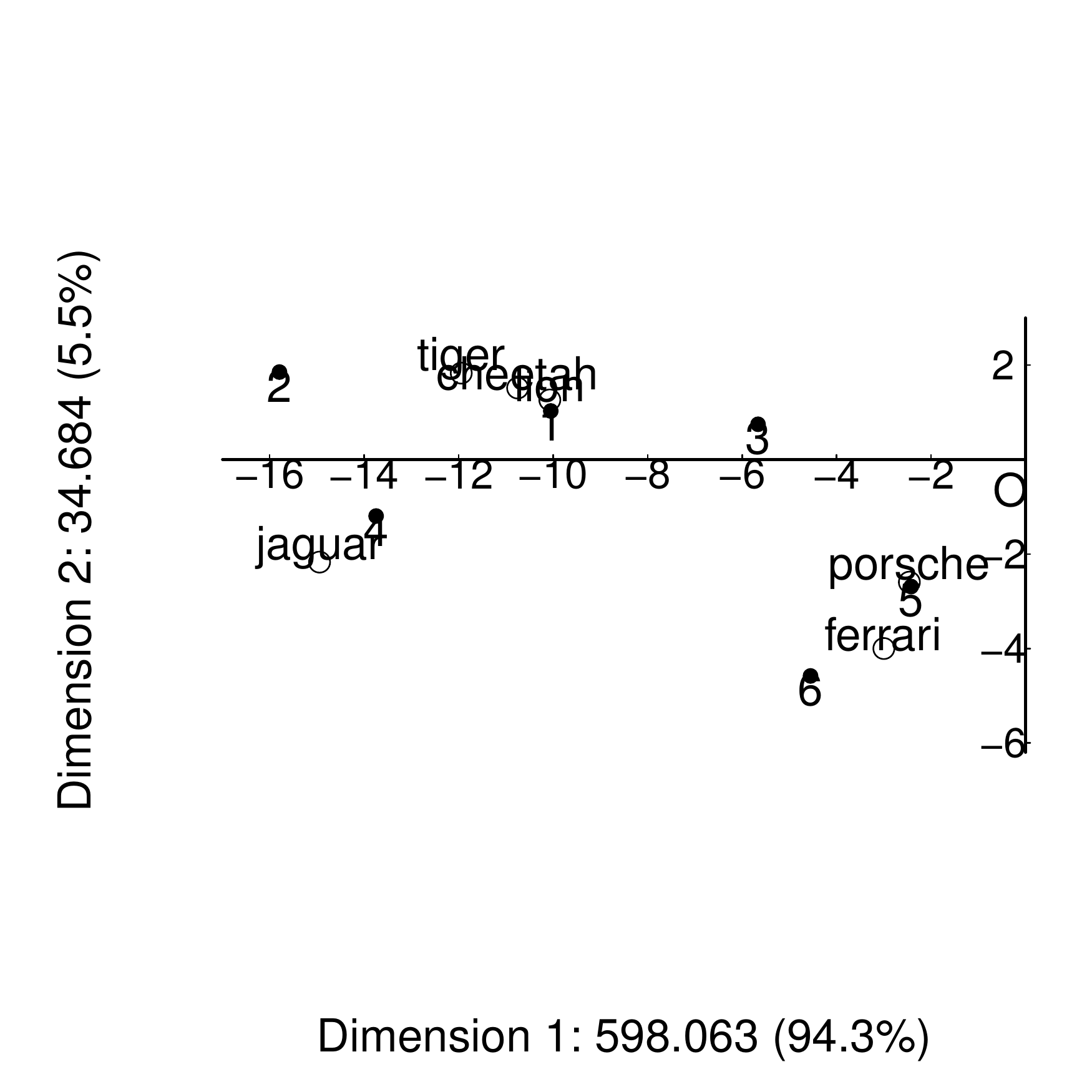}
         \caption{$\alpha = 1.5$}\label{F: lsarawalpha15}
         \end{subfigure}
    \caption{A two-dimensional plot of documents and terms for LSA-RAW with (a) $\alpha = -0.5$, (b) $\alpha = 0$, (c) $\alpha = 0.5$, (d) $\alpha = 1$, and (e) $\alpha = 1.5$.}
    \label{F: lsarawalpha}
\end{figure}

Figure~\ref{F: lsarawalpha} shows the two-dimensional plots of documents and terms for LSA-RAW with different choices of $\alpha$. Figure~\ref{F: lsarawalpha1} is the same as Figure~\ref{F: lsaraw}. As $\alpha$ increases, the Euclidean distances between row points (column points) on the first dimension increase relative to the second dimension.

\subsection{Design}\label{Sub: Experiment setup}

We compare the performances of LSA and CA for information retrieval, where two kinds of weightings are studied in LSA: the elements of the raw document-term matrix are weighted and the weighting exponent $\alpha$ is varied. We also explore the impact of these weightings in CA. Finally, we study mean average precision (MAP) as a function of $\alpha$ under an optimal number of dimensions for
 LSA and CA. We vary the number of dimension $k$ from 1, 2, $\cdots$, 20, 22, $\cdots$, 50, 60, $\cdots$ to 100 and the value of $\alpha$ from -6, -5.5, $\cdots$, -2, -1.8, $\cdots$, 4, 4.5, $\cdots$ to 8; we explore all $40\times 47 = 1,880$ combinations of parameter values. 

In the study of weighting the elements of the raw document-term matrix, we perform the LSA and CA of
\begin{itemize}
\item raw matrix $\textbf{\emph{F}}$, denoted by RAW, 
\item L1 row-normalized matrix  $\textbf{\emph{F}}^{L1}$ with $L(i,j)=f_{ij}$, $G(j)=1$, and $N(i)=1/\sum_{j=1}^n{f_{ij}}$, NROWL1, 
\item L2 row-normalized matrix $\textbf{\emph{F}}^{L2}$ with $L(i,j)=f_{ij}$, $G(j)=1$, and $N(i)=1/\sqrt{\sum_{j=1}^n{f_{ij}^2}}$, NROWL2, and
\item TF-IDF matrix $\textbf{\emph{F}}^{\text{TF-IDF}}$ described in Section~\ref{Sub: LSA}, TFIDF.
\end{itemize}
We refer to the combination of the CA and TF-IDF matrix as CA-TFIDF. Similarly, we obtain LSA-RAW, LSA-NROWL1, LSA-NROWL2, LSA-TFIDF, CA-RAW, CA-NROWL1, and CA-NROWL2. For performance comparison, RAW, NROWL1, NROWL2, and TFIDF are used for term matchings without dimensionality reduction.

\subsection{Datasets}\label{Sub:BBCsport}

LSA and CA are compared using three English datasets and one Dutch dataset. The three English datasets are the BBCSport \citep{greene2006practical}, BBCNews \citep{greene2006practical}, and 20 Newsgroups datasets (20-news-18846 bydata version) \citep{newsgroups20}. The Dutch dataset is the \emph{Wilhelmus} dataset \citep{mikekestemont2017}. 

Some statistics of the four datasets used are presented in Table~\ref{T: datasets}. The BBCNews dataset includes 2,225 documents that fall into one of five categories. The BBCSport dataset includes 731 documents that fall into one of five categories. The 20 Newsgroups dataset includes 18,846 documents that fall into one of 20 categories. This dataset is sorted into a training (60\%) and a test (40\%) set. We use a subset of this dataset to evaluate information retrieval. We randomly choose 600 documents from the training set of four categories (comp.graphics, rec.sport.hockey, sci.crypt, and talk.politics.guns) and 400 documents from the test set of these four categories. The \emph{Wilhelmus} dataset includes 186 documents divided into six categories.

\begin{table}[h]
\caption{Characteristics of datasets.}
    \begin{subtable}[h]{0.45\textwidth}
        \centering
        \begin{tabular}{l | l}
        \hline
        Categories& Data\\
        \hline  
        business& 510\\ 
        entertainment& 386 \\  
        politics& 417 \\  
        sport& 511\\  
        technology& 401\\  
       \hline
       \end{tabular}
       \caption{Characteristics of BBCNews.}
       \label{T: bbcnews}
    \end{subtable}
    \vspace{5mm}
    \begin{subtable}[h]{0.45\textwidth}
        \centering
        \begin{tabular}{l | l}
        \hline
        Categories& Data\\
        \hline  
        athletics& 101\\ 
        cricket& 124 \\  
        football”& 265 \\  
        rugby”& 147 \\  
        tennis& 100\\  
        \hline 
        \end{tabular}
        \caption{Characteristics of BBCSport.}
        \label{T: bbcsport}
     \end{subtable}
    \begin{subtable}[h]{0.45\textwidth}
        \centering
        \begin{tabular}{l | l | l}
        \hline
        Categories& Training data&Test data\\
        \hline  
        comp.graphics& 141&100\\ 
        rec.sport.hockey& 164 &99\\  
        sci.crypt& 161 &106\\  
        talk.politics.guns& 134&95\\ 
        \hline 
        \end{tabular}
        \caption{Characteristics of 20 Newsgroups.}
        \label{T: 20newsgroups}
     \end{subtable}
    \hfill
    \begin{subtable}[h]{0.45\textwidth}
        \centering
        \begin{tabular}{l | l}
        \hline
        Categories& Data\\
       \hline  
       datheen& 35\\ 
       marnix& 46 \\  
       heere& 23 \\  
       haecht& 35\\  
       fruytiers& 33\\  
       coornhert& 14 \\  
      \hline
    \end{tabular}
    \caption{Characteristics of Wilhelmus dataset.}
    \label{T: wilhelmus}
    \end{subtable}
\label{T: datasets}
\end{table}

To pre-process the three English datasets, we change all characters to lower case, remove punctuation marks, numbers, and stop words, and apply lemmatization. Subsequently, terms with frequencies lower than 10 are ignored. In addition, we remove unwanted parts of the 20 Newsgroups dataset, such as the header (including fields
like “From:” and “Reply-To:” followed by  email address), because these are almost irrelevant for information retrieval. The Dutch \emph{Wilhelmus} dataset is already pre-processed into tag-lemma pairs.

Since the \emph{Wilhelmus} and BBCSport datasets have a relatively low number of documents, we use leave-one-out cross-validation (LOOCV) for the \emph{Wilhelmus} dataset and five-fold cross-validation for the BBCSport dataset to evaluate LSA and CA \citep{gareth2021introduction}. The
BBCNews dataset is randomly divided into training (80\%) and validation (20\%) sets.

In the information retrieval part of the study, each document in the validation set is used as a query, where the category of the document is known. The documents in the training set that fall in the same category as the query are the relevant documents for this query.

\subsection{Evaluation}

We compare the MAP of each of the four versions of LSA and CA to explore the performance of these methods in information retrieval under changes in the contributions
of initial dimensions \citep{kolda1998semidiscrete}. The MAP is calculated as follows:
\begin{itemize}
\item The similarity is assessed between a query vector and each document vector of a document collection. We use three similarity metrics: Euclidean distance, dot similarity, and cosine similarity. As Euclidean distance is a key motivation for CA, we report results on Euclidean distance, and only report partial results for dot and cosine similarity in the main paper and the other results in the supplementary materials.
\item For Euclidean distance, the documents are ranked in an increasing order based on their similarity with the query vector (for dot and cosine similarity, the ranking is in the decreasing order); therefore, the first document has the highest similarity.
\item Precision-recall points are derived from the ordered list of documents. For a given query, Table~\ref{T: reandre} defines
four types of documents in the ordered list based on whether a document is relevant and retrieved:

$ \textbf{C} = \text{the set of relevant documents from the ordered list}$, i.e., documents that fall in the same category as the query

$ \textbf{D} = \text{the set of retrieved documents from the ordered list.}$, i.e., when 10 documents are returned, the set of retrieved documents consists of the first 10 documents in the ordered list.

\begin{table}[htbp]
\centering  
\caption{Retrieved and relevant documents.}
\label{T: reandre}
\begin{tabular}{lcr}    
& Relevant&Non-Relevant\\
\hline  
Retrieved & \textbf{C} $\cap$ \textbf{D}&$\overline{\rm  \textbf{C}}$  $\cap$ \textbf{D}	\\ 
Not Retrieved & \textbf{C} $\cap$ $\overline{\rm  \textbf{D}}$ &$\overline{\rm  \textbf{C}}$ $\cap$ $\overline{\rm  \textbf{D}}$\\  
\hline 
\end{tabular} 
\end{table}

Let $|.|$ denote the number of documents in a set. Then, precision and recall are defined as

\begin{equation}
\label{precision}
\text{precision} = \frac{|\textbf{C} \cap \textbf{D}|}{|\textbf{D}|}
\end{equation}
and
\begin{equation}
\label{recall}
\text{recall} =  \frac{|\textbf{C} \cap \textbf{D}|}{|\textbf{C}|}.
\end{equation}

Thus, precision is defined as the ratio of the number of relevant documents retrieved over the total
number of retrieved documents, and recall is defined as the ratio of the
number of relevant documents retrieved over the total number of relevant documents. For a given query, the set \textbf{C} is fixed. The set \textbf{D} is not fixed; if we return the first $i$ documents, then \textbf{D} consists of the first $i$ documents in the ordered list. Thus, for a given $i$, we can obtain a precision (see Equation~(\ref{precision})) and recall (see Equation~(\ref{recall})) pair. We run values of $i$ from 1 to $l$ (the number of documents in the ordered list), and obtain $l$ precision-recall pairs.
\item  Then, 11 pseudo-precisions are calculated under 11 recalls (0, 0.1, $\cdots$, 1.0), where a pseudo-precision at recall $x$ is the maximum precision from recall $x$ to recall 1. For example, pseudo-precision at recall 0.2 is the maximum precision from recall 0.2 to recall 1.
\item The average precision for the query is obtained by averaging the 11 pseudo-precisions. 
\item The MAP is the mean across all queries. 
\end{itemize}
Greater MAP values indicate a better performance.

\section{Results for Euclidean distance}\label{S: resultseuc}

\subsection{Comparing LSA and CA for information retrieval}\label{Sub: standardca}

\subsubsection{MAP as a function of the number of dimensions for the four versions of LSA with the standard weighting exponent $\alpha = 1$ and for CA}\label{Subsub: standardalpha}

We first investigate the performance of LSA and CA in terms of MAP, in their standard use, i.e., without varying the weighting exponent $\alpha$, i.e., $\alpha = 1$. Term matching without the preliminary use of LSA and CA, i.e., directly on the document-term matrix, is denoted by RAW. We expect that, in line with \citet{qi2021comparison}, the performance of LSA and CA will be better than that of RAW, and the performance of CA will be better than that of the four versions of LSA.

\begin{figure}[h]
    \centering
       \begin{subfigure}[b]{0.45\linewidth}
         \centering
         \includegraphics[width=1\textwidth]{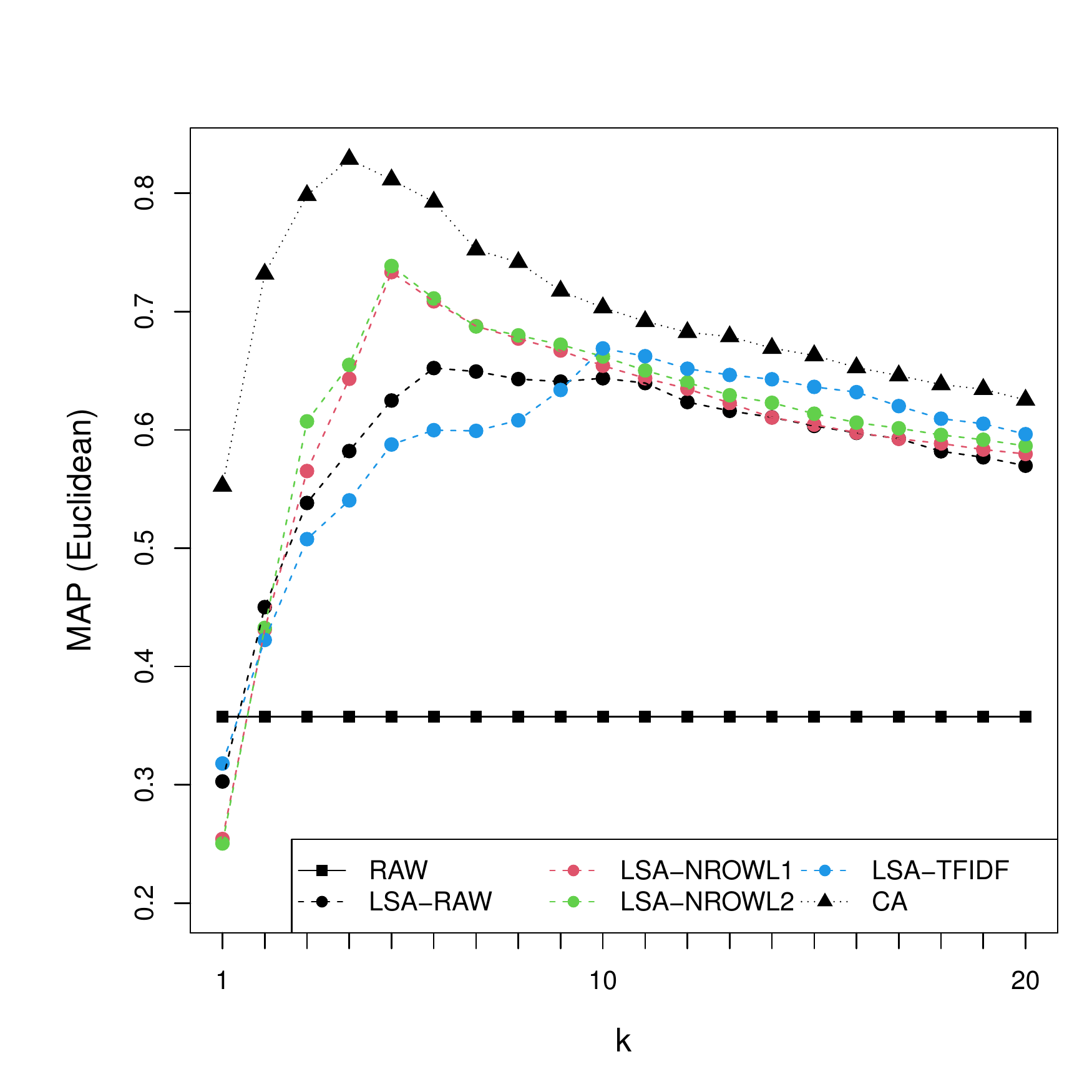}
         \caption{BBCNews}\label{F: eucstandardpBBCnewsdim1to20}
         \end{subfigure}
      \begin{subfigure}[b]{0.45\linewidth}
         \centering
         \includegraphics[width=1\textwidth]{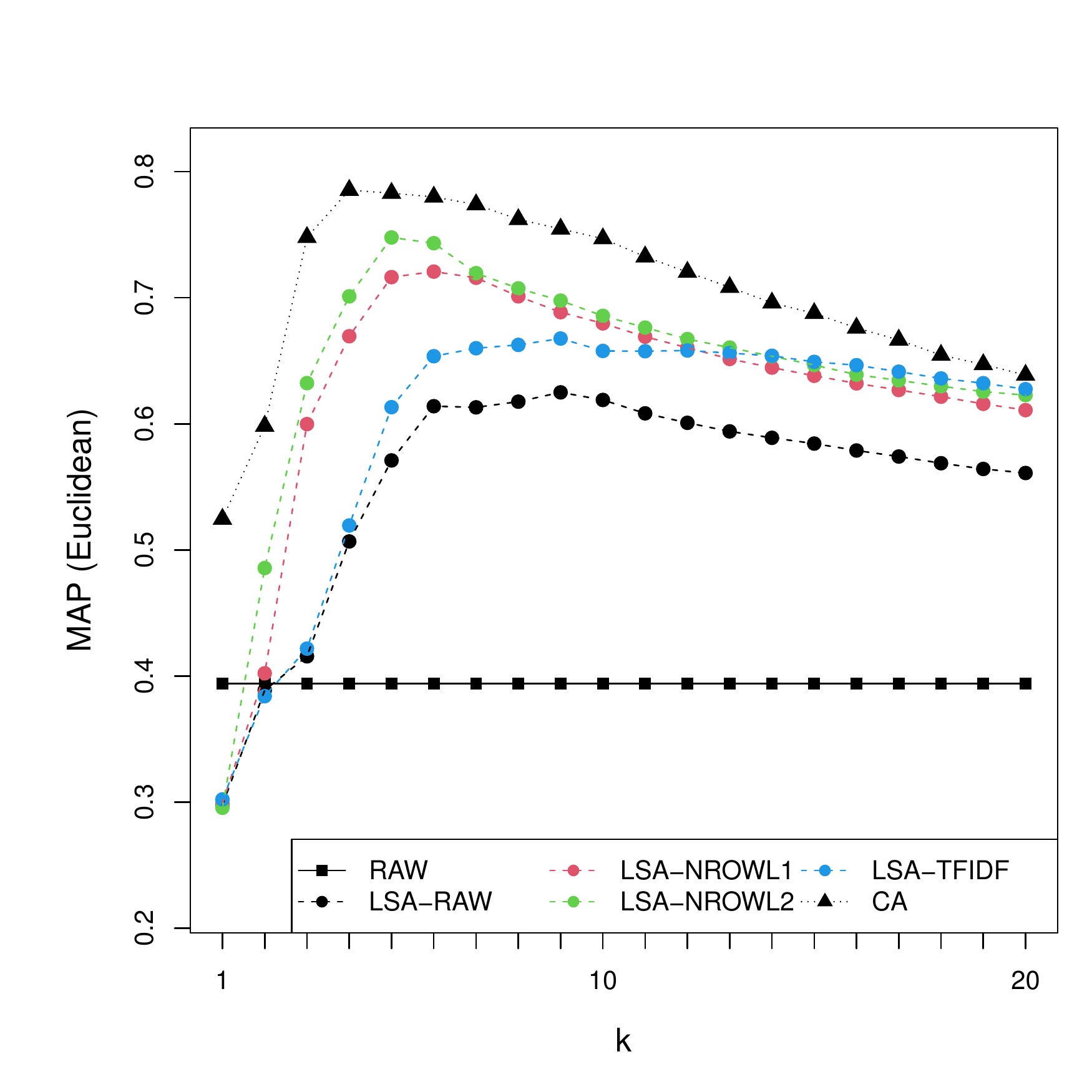}
         \caption{BBCSport}\label{F: eucstandardpBBCsportdim1to20}
         \end{subfigure}
     \begin{subfigure}[b]{0.45\linewidth}
         \centering
         \includegraphics[width=1\textwidth]{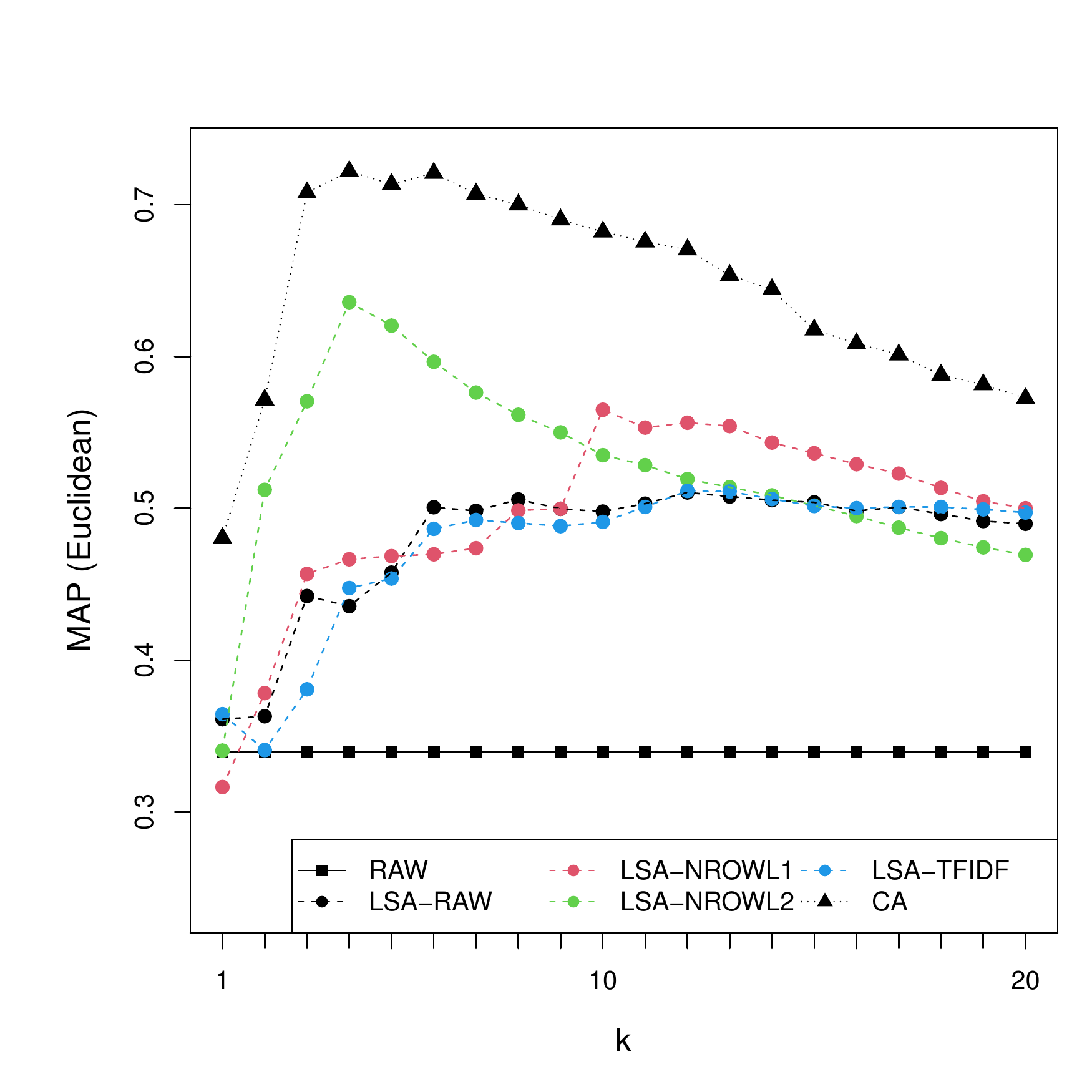}
         \caption{20 Newsgroups}\label{F: eucstandardp20newsgroupsdim1to20}
         \end{subfigure}
      \begin{subfigure}[b]{0.45\linewidth}
         \centering
         \includegraphics[width=1\textwidth]{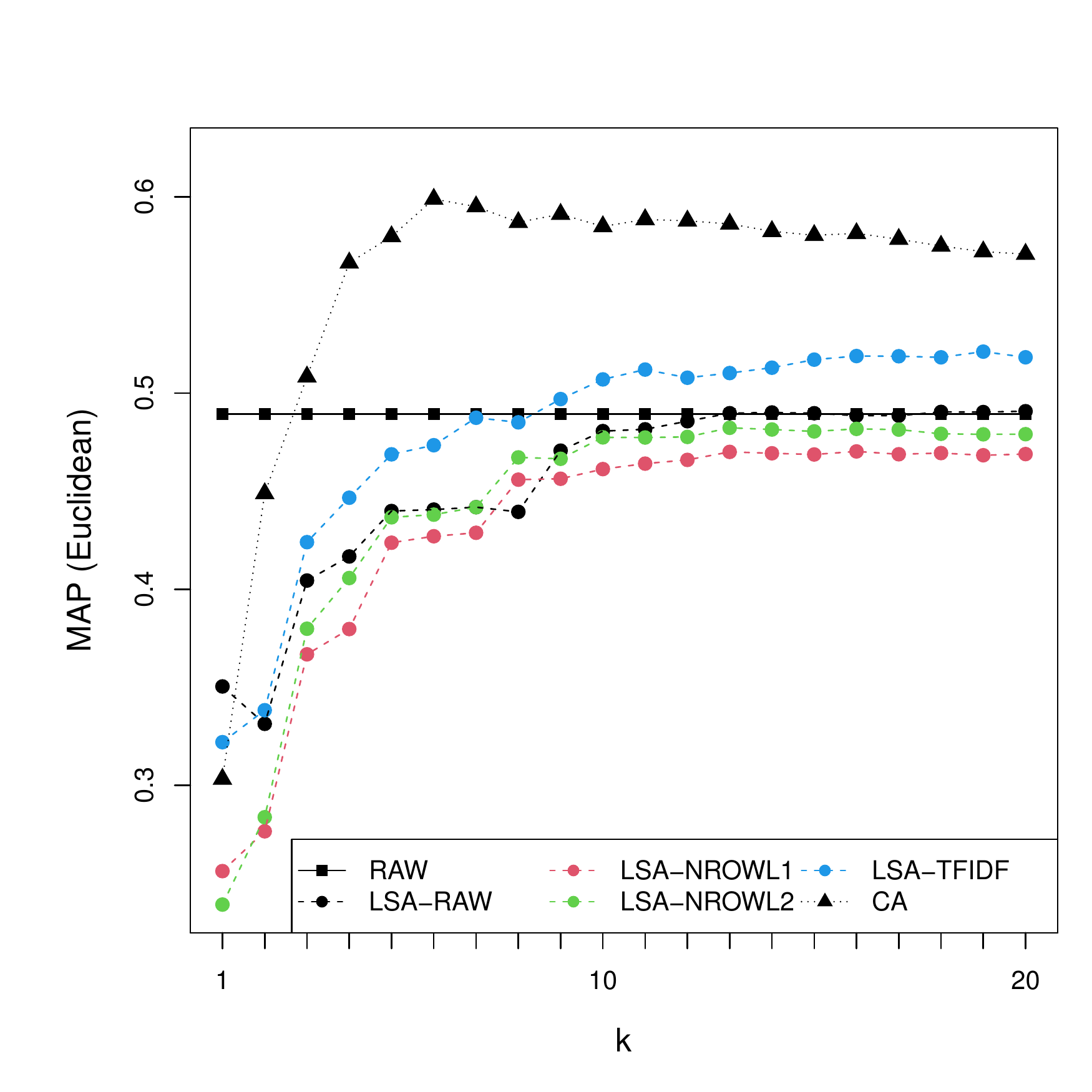}
         \caption{Wilhelmus}\label{F: eucstandardpwilhelmusdim1to20}
         \end{subfigure}
    \caption{MAP as a function of the number of dimensions $k$ under standard coordinates.}
    \label{F: eucstandardpdim1to20}
\end{figure}

Figure~\ref{F: eucstandardpdim1to20} shows MAP as a function of the number of dimensions $k$ for different weighting schemes of LSA, and for CA. We display only the first 20 dimensions, as all lines usually decrease after dimension 20. Figures with dimensionality up to 100 can be found in the supplementary materials. For the four versions of LSA, and  for CA, Table~\ref{TIReucalpha1} presents the dimension number for which the optimal MAP is reached, as well as the MAP values, in each of the four datasets. We conclude the following from Figure~\ref{F: eucstandardpdim1to20} and Table~\ref{TIReucalpha1}:
\begin{itemize}
\item Both LSA and CA result in better MAP than RAW, which results in a straight line when the full dimensional matrix is used.
\item For both LSA and CA, performance is a function of the number of dimensions $k$. Overall, MAP rises as a function of $k$ to reach a peak, and then, it goes down. For CA, the peak is reached at $k = 4$. In CA, the information used to calculate MAP increases in the first four dimensions in comparison to the noise. In the components of $k \geq 5$, the noise dominates the useful information, which results in the MAP going down from this point. 
\item CA results in a considerably better MAP than the four versions of LSA, which is in line with \citet{qi2021comparison}, who showed that the performance of CA is better than that of LSA for document-term matrices. This is because of the differential treatment of margins in LSA and CA. The margins provide irrelevant information for making queries. In CA, the margins are removed, and therefore, the relative amount of information in comparison to the noise, which we informally refer to as the information - noise ratio, is considerably larger in CA than in LSA. This explains the better MAP in CA.
\item The peaks for the four versions of LSA are usually found at higher dimensionality $k$ than the peaks for CA. This is because margins are noise for queries when we fix $\alpha = 1$; in LSA, this noise plays an important role in the first few dimensions. Hence, this earlier peak in CA is also explained by its better information - noise ratio.
\item The four LSA methods are not equally effective. In all four datasets, the performance of LSA can be significantly improved using weighting schemes. The improvements over LSA-RAW are data dependent. On average, across the four datasets, LSA-NROWL2 is the best, but for the \emph{Wilhelmus} dataset, LSA-NROWL1 and LSA-NROWL2 result in a somewhat worse MAP than that with LSA-RAW.
\end{itemize}

\begin{table}[H]
\centering  
\caption{MAP with the optimal number of dimensions $k$. Bold values are best.} 
\label{TIReucalpha1}
\begin{tabular}{lcccccccc}    
&\multicolumn{2}{c}{BBCNews} &\multicolumn{2}{c}{BBCSport}&\multicolumn{2}{c}{20 Newsgroups}&\multicolumn{2}{c}{Wilhelmus} \\
&$k$&MAP&$k$&MAP&$k$&MAP&$k$&MAP\\
\hline 
RAW &&0.358&&0.394 &&0.339&& 0.489\\
LSA-RAW  &6 & 0.652& 9 & 0.625 & 12 & 0.510&24&0.492\\
LSA-NROWL1 &5&0.733&6&0.721&10&0.565&16& 0.470\\
LSA-NROWL2 &5&0.738&5&0.748&4&0.636& 13&0.482\\
LSA-TFIDF &10&0.669&9&0.668&12&0.512&19&0.521\\
CA&4&\textbf{0.829}&4&\textbf{0.785}&4&\textbf{0.722}&6&\textbf{0.599}\\
\hline 
\end{tabular}  
\end{table} 

\subsubsection{MAP as a function of the weighting exponent $\alpha$ for LSA compared with MAP for CA under varying numbers of dimensions}\label{Sub: constantk}

In Section \ref{Subsub: standardalpha}, we found that CA outperforms the four versions of LSA in terms of MAP, where LSA had the usual weighting exponent $\alpha = 1$. In this section, we study whether the performance of LSA-RAW improves when we vary $\alpha$. 

\begin{figure}[h]
    \centering
     \begin{subfigure}[b]{0.45\linewidth}
         \centering
         \includegraphics[width=1\textwidth]{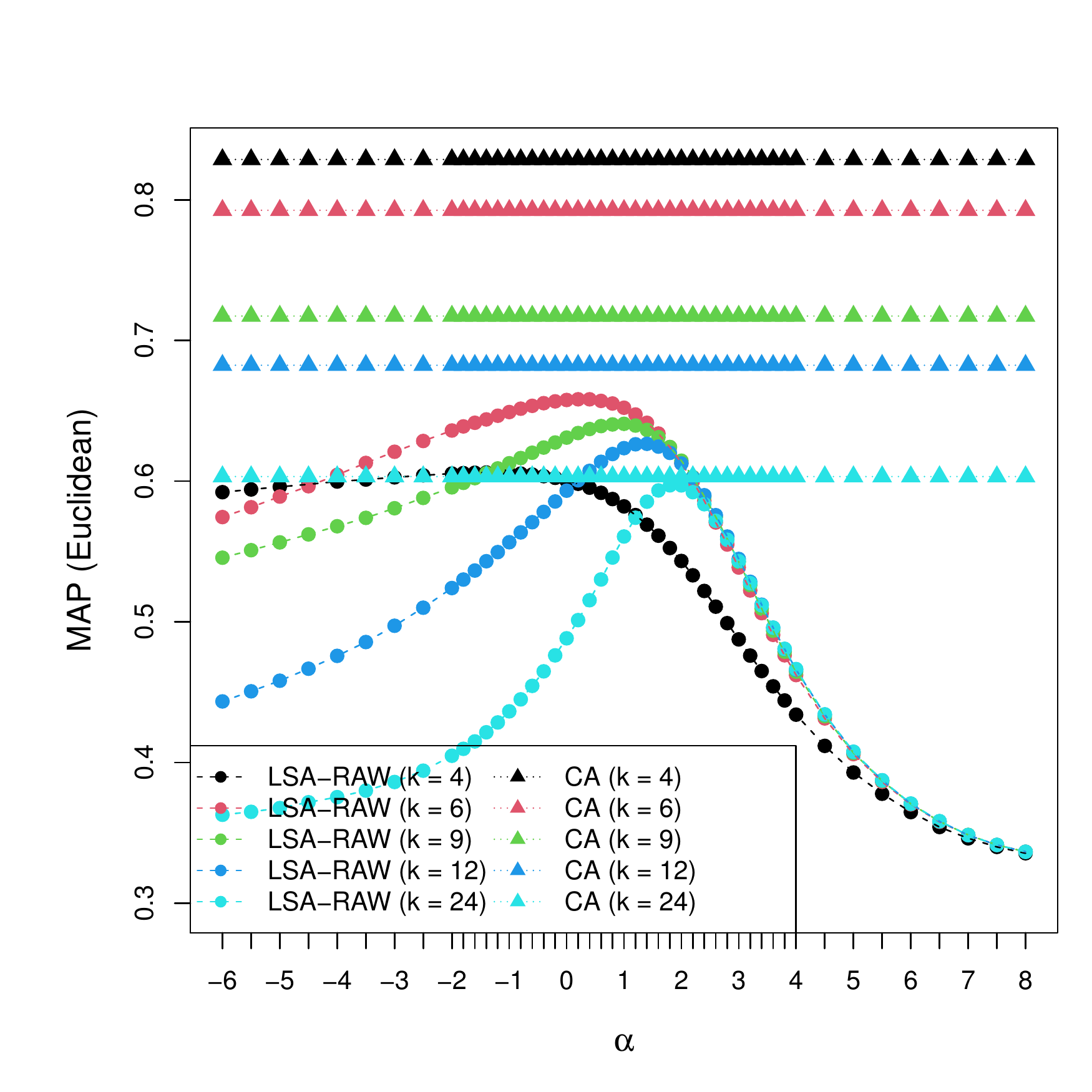}
         \caption{BBCNews}\label{F: eucstandarddBBCnews}
         \end{subfigure}
      \begin{subfigure}[b]{0.45\linewidth}
         \centering
         \includegraphics[width=1\textwidth]{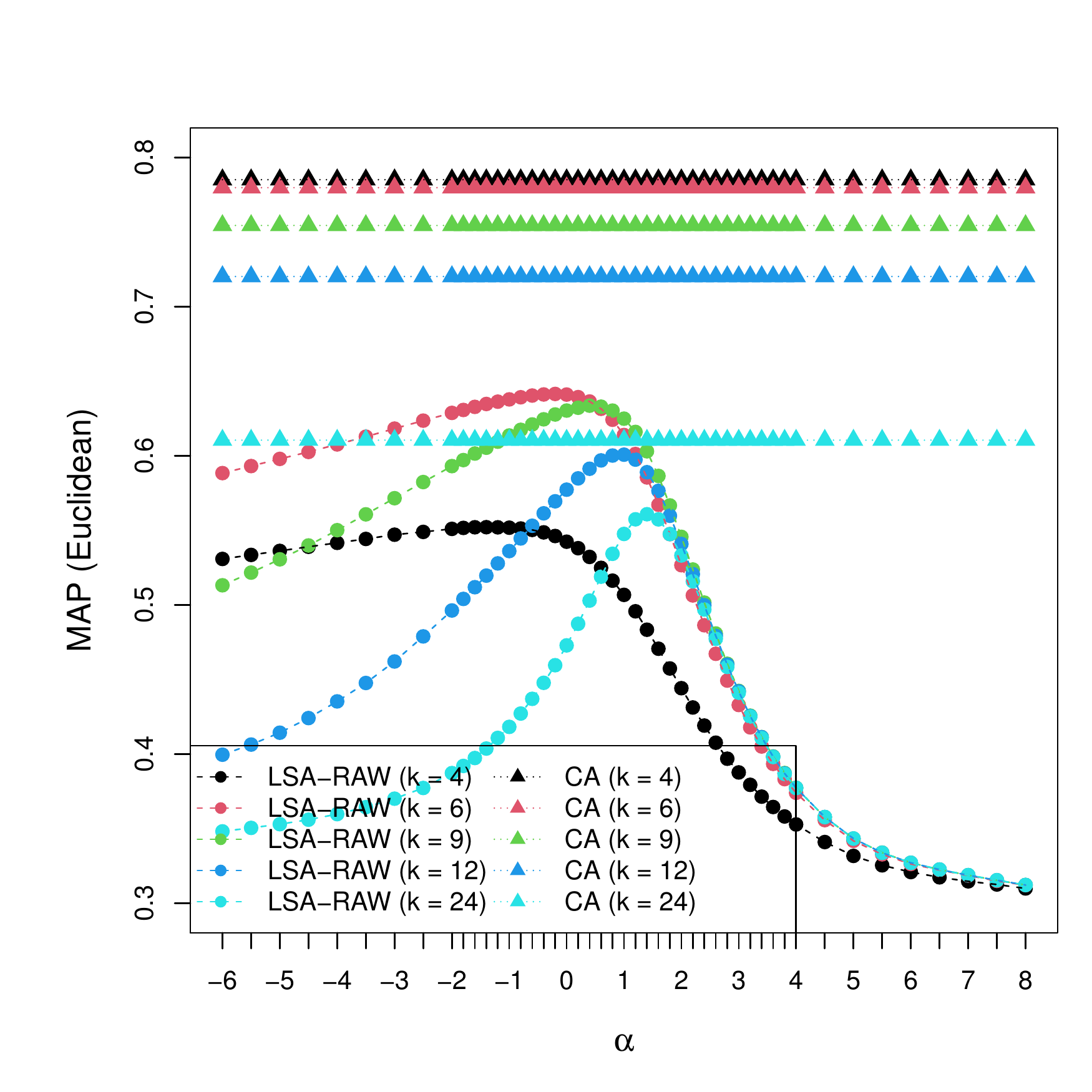}
         \caption{BBCSport}\label{F: eucstandarddBBCsport}
         \end{subfigure}
      \begin{subfigure}[b]{0.45\linewidth}
         \centering
         \includegraphics[width=1\textwidth]{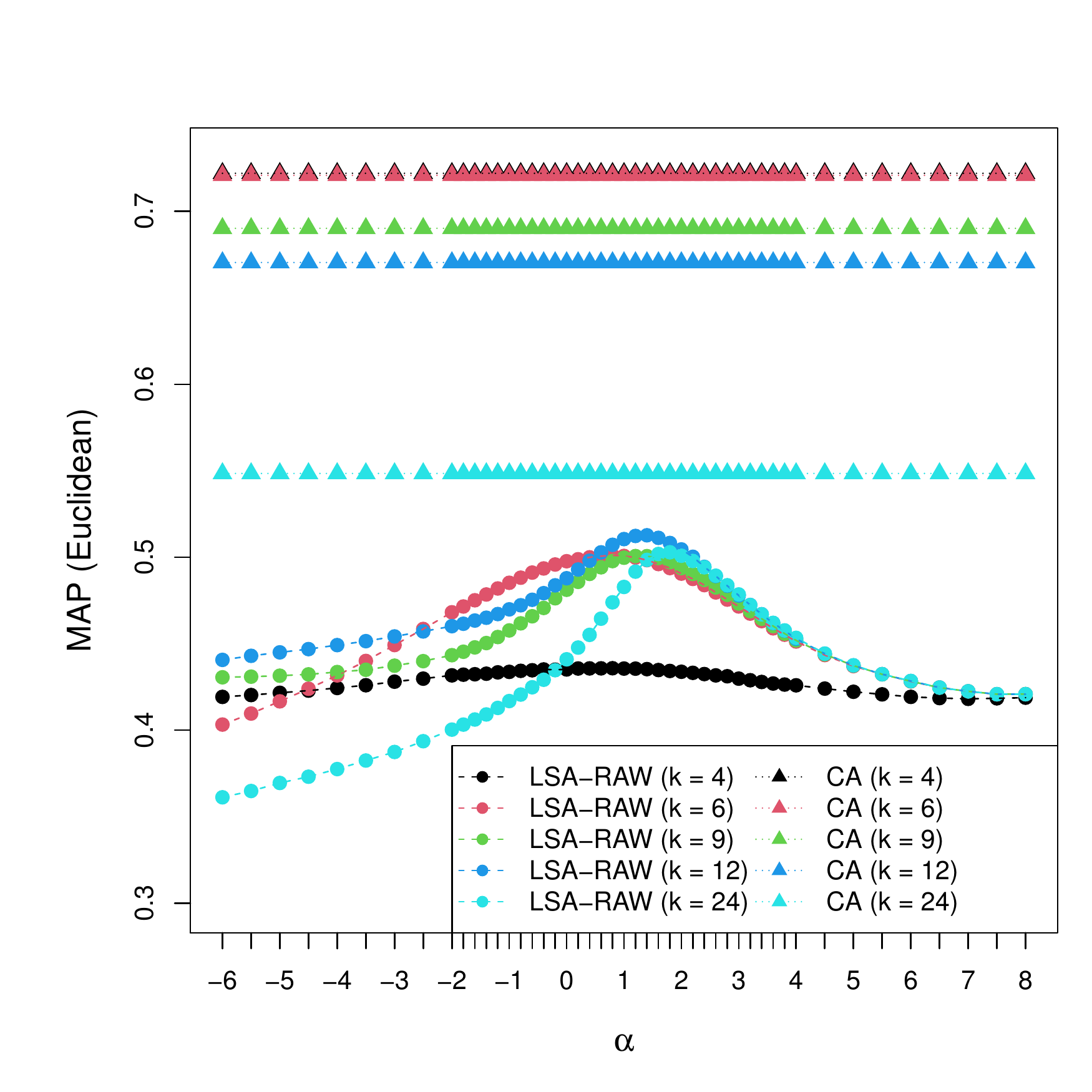}
         \caption{20 Newsgroups}\label{F: eucstandardd20newsgroups}
         \end{subfigure}
      \begin{subfigure}[b]{0.45\linewidth}
         \centering
         \includegraphics[width=1\textwidth]{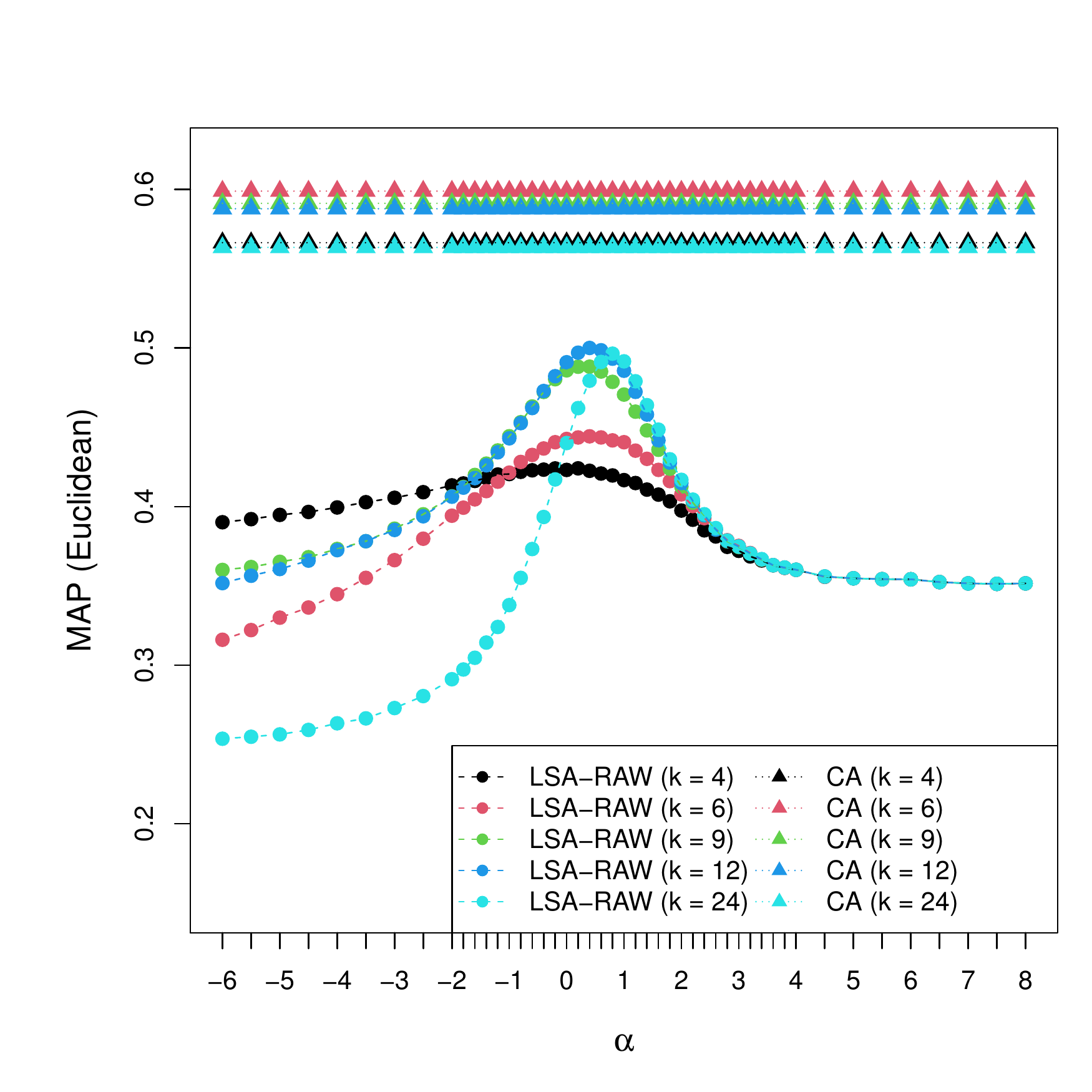}
         \caption{Wilhelmus}\label{F: eucstandarddwilhelmus}
         \end{subfigure}
    \caption{MAP as a function of $\alpha$ for LSA-RAW and MAP for CA under varying $k$.}
    \label{F: eucstandardd}
\end{figure}

Figure~\ref{F: eucstandardd} shows MAP as a function of $\alpha$ for LSA-RAW with the number of dimensions $k = 4, 6, 9, 12, \text{and }24$. For comparison, we also report the MAP values for CA found in Section \ref{Subsub: standardalpha} under these dimensions. 
We choose these values of $k$ because these dimensions are optimal for LSA-RAW and CA in Table~\ref{TIReucalpha1}. Table~\ref{TIReucdim10} shows the optimal $\alpha$ and corresponding MAP, which is a condensed version of Figure~\ref{F: eucstandardd}. We conclude the following from Figure~\ref{F: eucstandardd} and Table~\ref{TIReucdim10}:

\begin{itemize}
\item Although the performance of LSA-RAW improves by varying $\alpha$, CA still outperforms LSA-RAW. 
\item For LSA-RAW, the overall MAP first increases and then decreases as a function of $\alpha$. This means that varying $\alpha$ can potentially improve the performance of LSA-RAW. 
\item The increase in MAP is minor. Consider, for example, the BBCNews dataset. In Section~\ref{Subsub: standardalpha}, we found that the MAP was optimal with a value of 0.652 for $\alpha = 1$, when $k = 6$. Table~\ref{TIReucdim10} shows that for $\alpha = 0.2$, the MAP increases to 0.658. Apparently, for 6 dimensions, when $\alpha = 0.2$, the information - noise ratio is optimal in terms of MAP. For $\alpha = 0.2$, the distances on later dimensions (of the 6 dimensions) are increased and those on initial dimensions are reduced. This means that, with $\alpha = 0.2$, the impact of the initial dimensions affected most by the margins is reduced. This is consistent with the results of \citet{bullinaria2012extracting}, which indicates that reducing the initial dimensions improves performance.
\item Moreover, the optimal $\alpha$ for LSA-RAW is data dependent and generally increases with $k$. This replicates  results of \citet{caron2001experiments}. As the number of dimensions varies, the change in the optimal $\alpha$ is the result of the information - noise ratio for the specific number of dimensions studied. For example, for the BBCNews dataset, the optimal number of dimensions is 6; for larger numbers of dimensions, the optimal $\alpha$ increases. An increasing $\alpha$ indicates that distances at earlier dimensions are more important for information retrieval, and therefore, the role of the later dimensions is played down.
\end{itemize}

\begin{table}[H]
\centering  
\caption{MAP with the optimal weighting exponent $\alpha$ for LSA-RAW and MAP for CA under $k = 4, 6, 9, 12, \text{and }24$. Bold values are best.} 
\label{TIReucdim10}
\begin{tabular}{lcccccccc}    
&\multicolumn{2}{c}{BBCNews} &\multicolumn{2}{c}{BBCSport}&\multicolumn{2}{c}{20 Newsgroups}&\multicolumn{2}{c}{Wilhelmus} \\
&$\alpha$&MAP&$\alpha$&MAP&$\alpha$&MAP&$\alpha$&MAP\\
\hline 
LSA-RAW ($k = 4$) &-1.4& 0.606&-1.4&  0.552& 0.8&0.436 & 0.2 &0.424 \\
LSA-RAW ($k = 6$) &0.2& 0.658&-0.2& 0.642& 0.8&0.501 & 0.4 &0.444 \\
LSA-RAW ($k = 9$) &1& 0.641&0.4& 0.634&1.2&0.501 & 0.4&0.488 \\
LSA-RAW ($k = 12$) &1.4& 0.627&1& 0.601& 1.4&0.513 & 0.4 &0.500\\
LSA-RAW ($k = 24$) &1.8&0.597&1.4& 0.561& 1.8& 0.503 & 0.8 &0.496 \\
CA ($k = 4$) & &  \textbf{0.829}& & \textbf{0.785}&  &\textbf{0.722} &   &0.566 \\
CA ($k = 6$) & &  0.793& & 0.780&  &0.721 &   &\textbf{0.599} \\
CA ($k = 9$) & &  0.717& & 0.755&  &0.690 &   &0.591 \\
CA ($k = 12$) & & 0.682& & 0.720&  &0.670 &   &0.588\\
CA ($k = 24$) & & 0.603& & 0.611&  &0.548 &  & 0.563\\
\hline 
\end{tabular}  
\end{table}

\subsection{Adjusting CA using weighting}\label{Sub: improvingca}

\subsubsection{Weighting the elements of the raw document-term matrix for CA}\label{Subsub: caweightingscheme}

\begin{figure}[h]
    \centering
       \begin{subfigure}[b]{0.45\linewidth}
         \centering
         \includegraphics[width=1\textwidth]{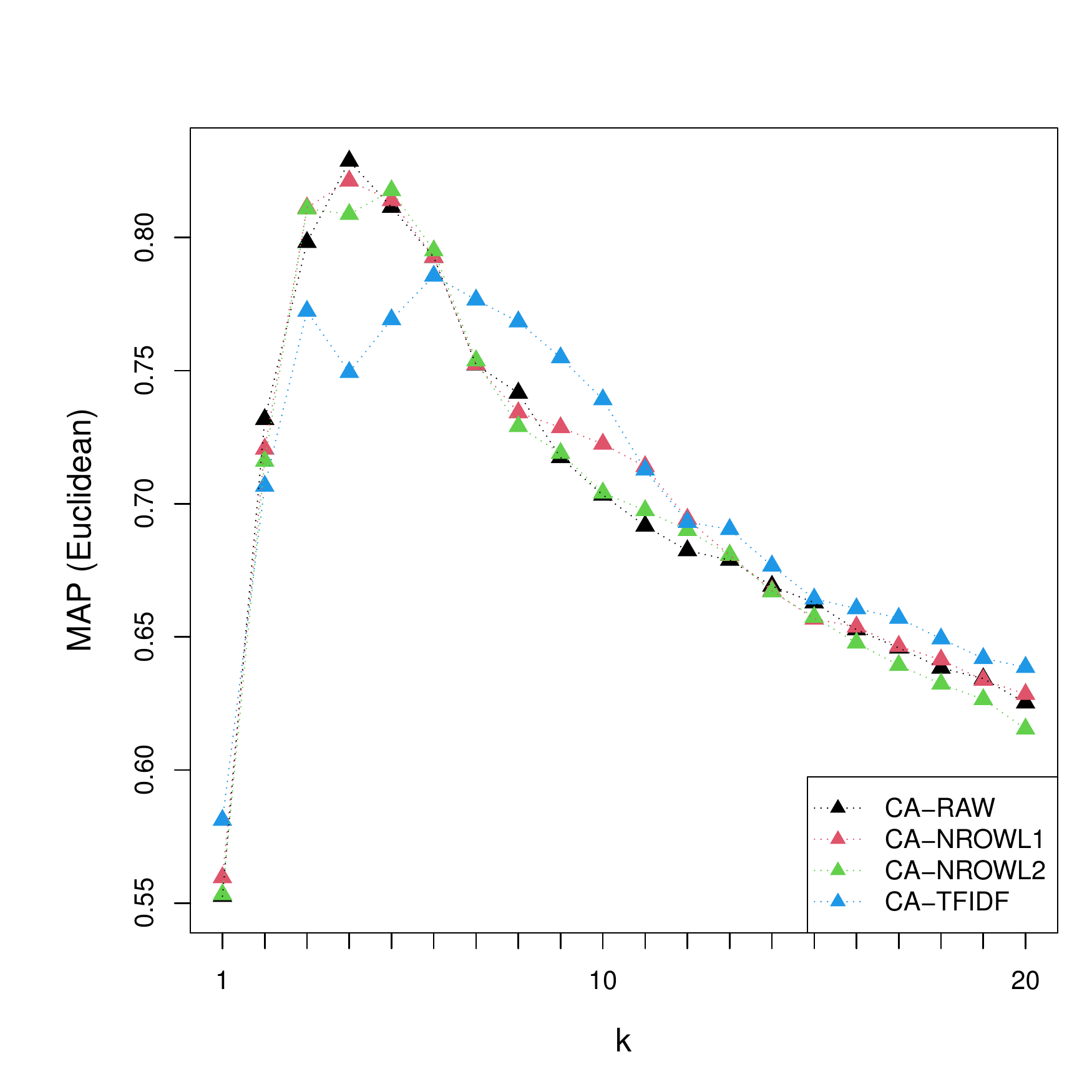}
         \caption{$\alpha = 1$ (BBCNews)}\label{F: eucstandardpBBCnewscaweischdim1to20}
         \end{subfigure}
      \begin{subfigure}[b]{0.45\linewidth}
         \centering
         \includegraphics[width=1\textwidth]{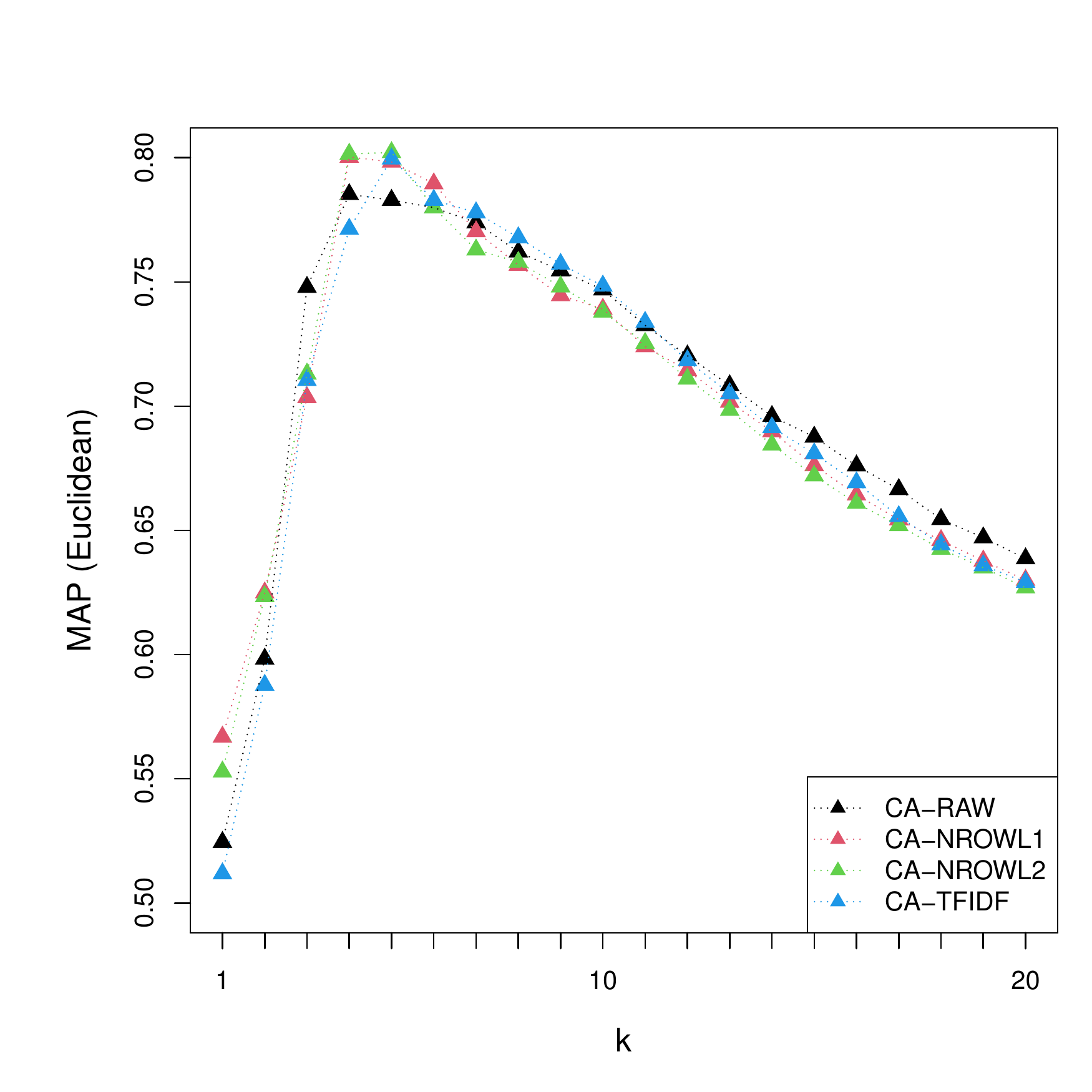}
         \caption{$\alpha = 1$ (BBCSport)}\label{F: eucstandardpBBCsportcaweischdim1to20}
         \end{subfigure}
     \begin{subfigure}[b]{0.45\linewidth}
         \centering
         \includegraphics[width=1\textwidth]{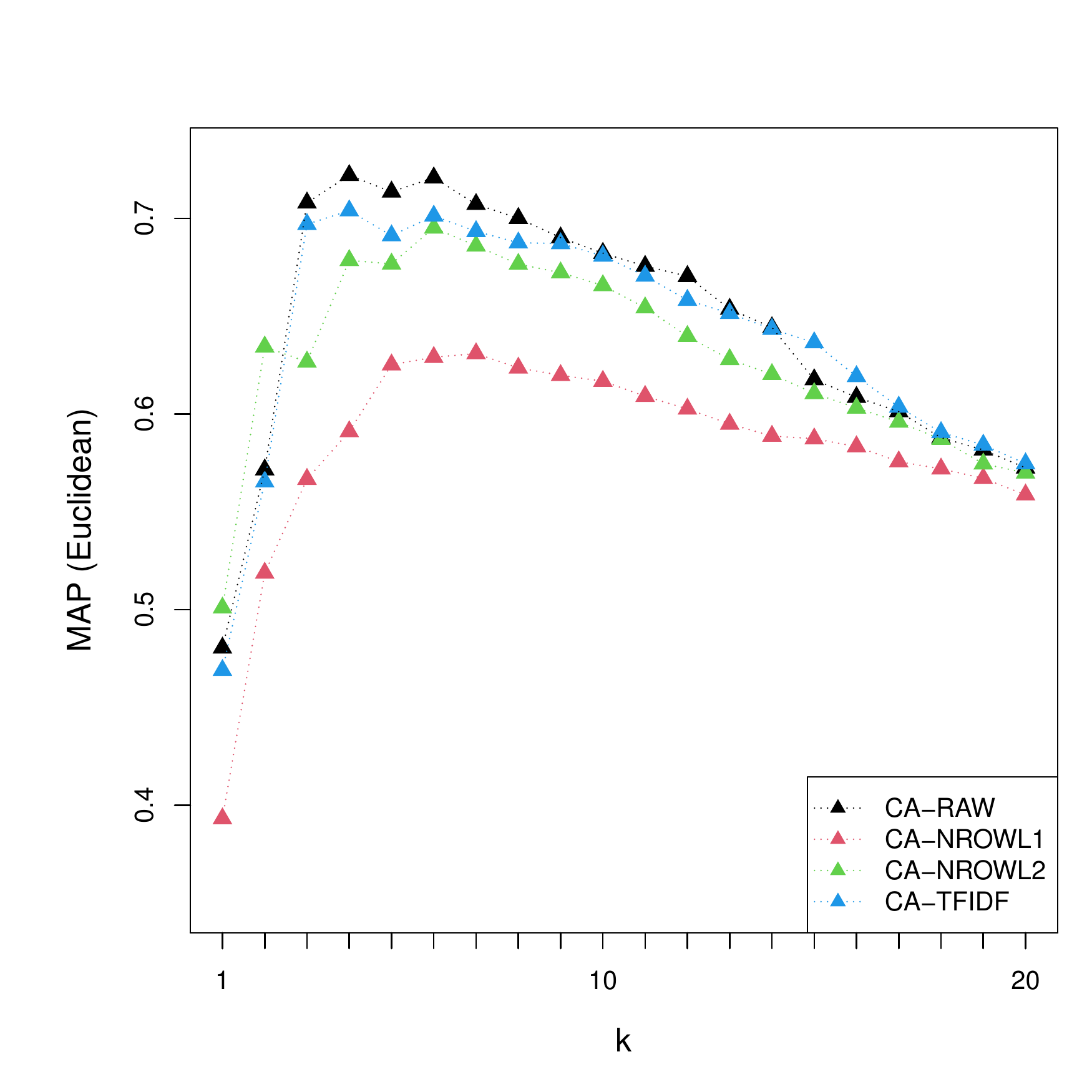}
         \caption{$\alpha = 1$ (20 Newsgroups)}\label{F: eucstandardp20newsgroupscaweischdim1to20}
         \end{subfigure}
      \begin{subfigure}[b]{0.45\linewidth}
         \centering
         \includegraphics[width=1\textwidth]{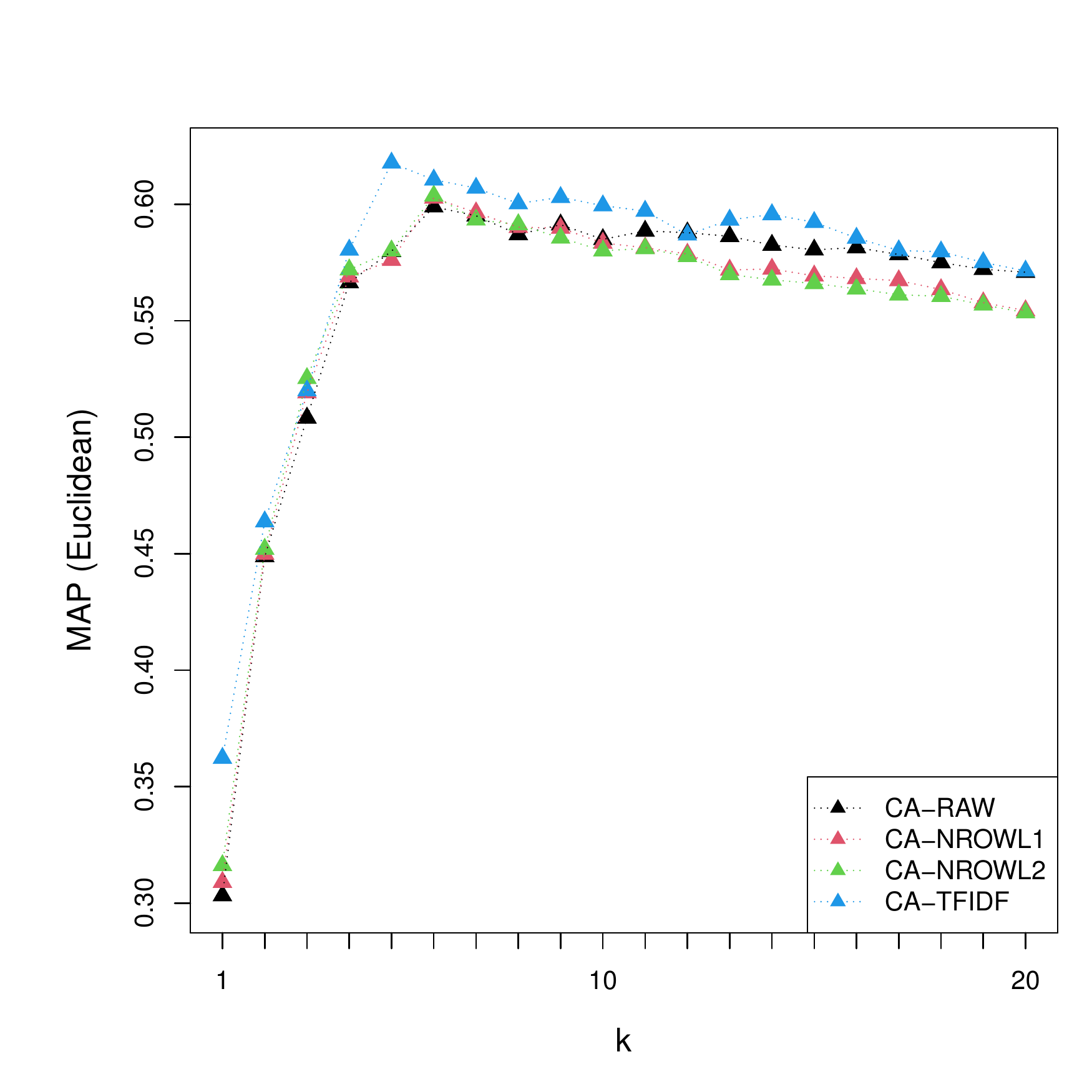}
         \caption{$\alpha = 1$ (Wilhelmus)}\label{F: eucstandardpwilhelmuscaweischdim1to20}
         \end{subfigure}
    \caption{MAP as a function of the number of dimensions $k$ for the four versions of CA under standard coordinates.}
    \label{F: eucstandardpcaweischdim1to20}
\end{figure}

Weighting the elements of the raw document-term matrix is an effective way to improve the performance of LSA for information retrieval. Here, we explore whether this holds for CA. Similar to Figure~\ref{F: eucstandardpdim1to20}, Figure~\ref{F: eucstandardpcaweischdim1to20} shows MAP as a function of $k$ for different weighting schemes of CA. CA in Figure~\ref{F: eucstandardpdim1to20} is referred to as CA-RAW in Figure~\ref{F: eucstandardpcaweischdim1to20}; for CA/CA-RAW, the results in these two figures are identical. For the four versions 
 of CA, Table~\ref{TIReucalpha1caweisch} shows the dimensionality for which the optimal MAP is reached, as well as the MAP value. We conclude the following from Figure~\ref{F: eucstandardpcaweischdim1to20} and Table~\ref{TIReucalpha1caweisch}:
\begin{itemize}
\item Overall, the weighting of the elements of the raw matrix sometimes improves the performance of CA, but these improvements over CA-RAW are small and data dependent.
\end{itemize}
Relative to LSA, it is harder to improve the performance of CA in information retrieval by weighting the elements of the raw matrix because (1) the MAP of CA-RAW is already relatively high, and (2) CA-RAW has weighted the elements of the raw document-term matrix as it is an integral part of this technique (Equation~(\ref{CA})).

\begin{table}[H]
\centering  
\caption{MAP with the optimal number of dimensions $k$ for the four versions of CA. Bold values are best.} 
\label{TIReucalpha1caweisch}
\begin{tabular}{lcccccccc}    
&\multicolumn{2}{c}{BBCNews} &\multicolumn{2}{c}{BBCSport}&\multicolumn{2}{c}{20 Newsgroups}&\multicolumn{2}{c}{Wilhelmus} \\
&$k$&MAP&$k$&MAP&$k$&MAP&$k$&MAP\\
\hline 
CA-RAW&4&\textbf{0.829}&4&0.785&4&\textbf{0.722}&6&0.599\\
CA-NROWL1&4&0.821&4&0.800&7&0.631&6&0.603\\
CA-NROWL2&5&0.818&5&\textbf{0.802}&6&0.695&6&0.604\\
CA-TFIDF&6&0.786&5&0.800&4&0.704&5&\textbf{0.618}\\
\hline 
\end{tabular}  
\end{table} 

\subsubsection{MAP as a function of the weighting exponent $\alpha$ for CA}\label{Subsub: caweightingexponent}

\begin{figure}[h]
    \centering
     \begin{subfigure}[b]{0.45\linewidth}
         \centering
         \includegraphics[width=1\textwidth]{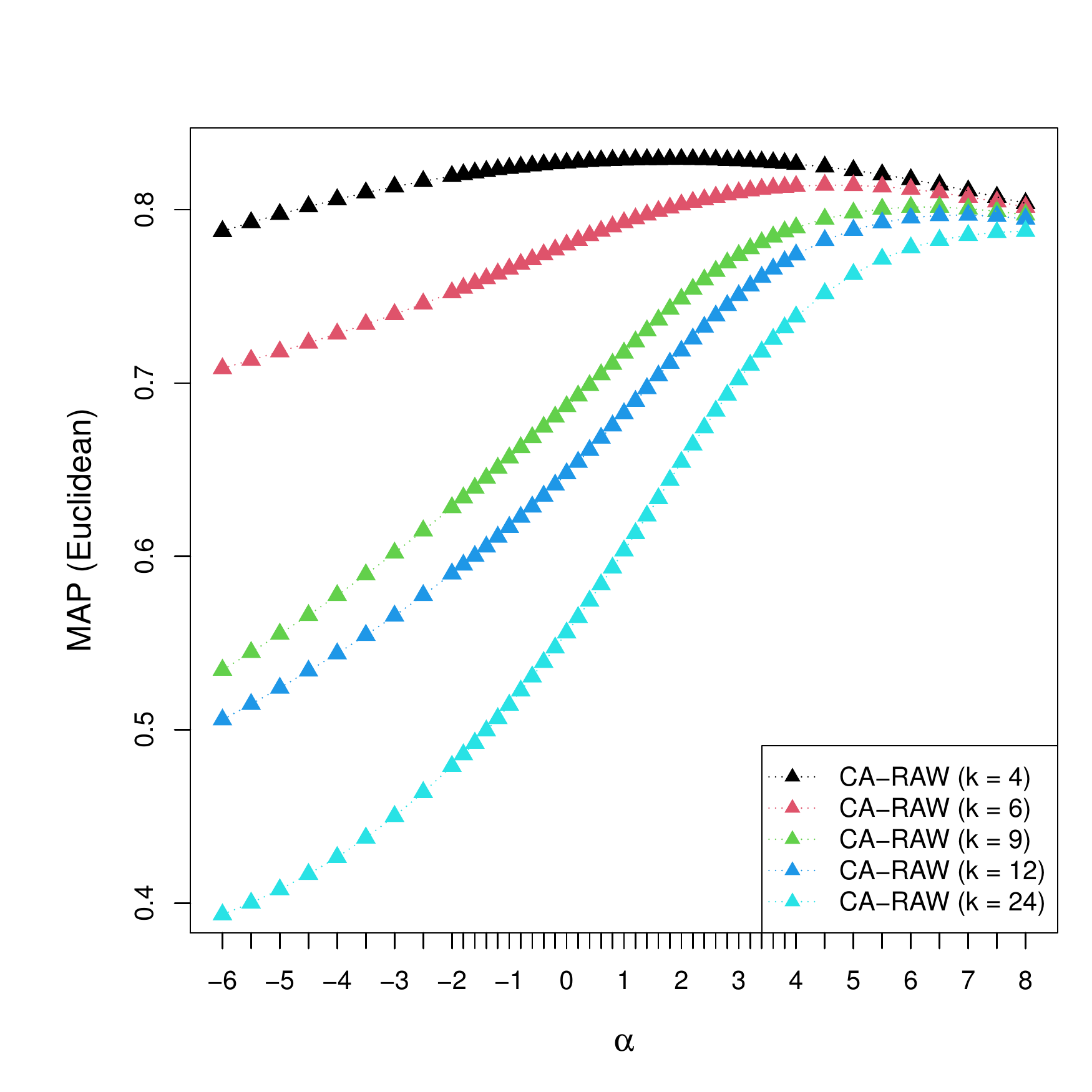}
         \caption{BBCNews}\label{F: eucstandarddBBCnewscaweiexp}
         \end{subfigure}
      \begin{subfigure}[b]{0.45\linewidth}
         \centering
         \includegraphics[width=1\textwidth]{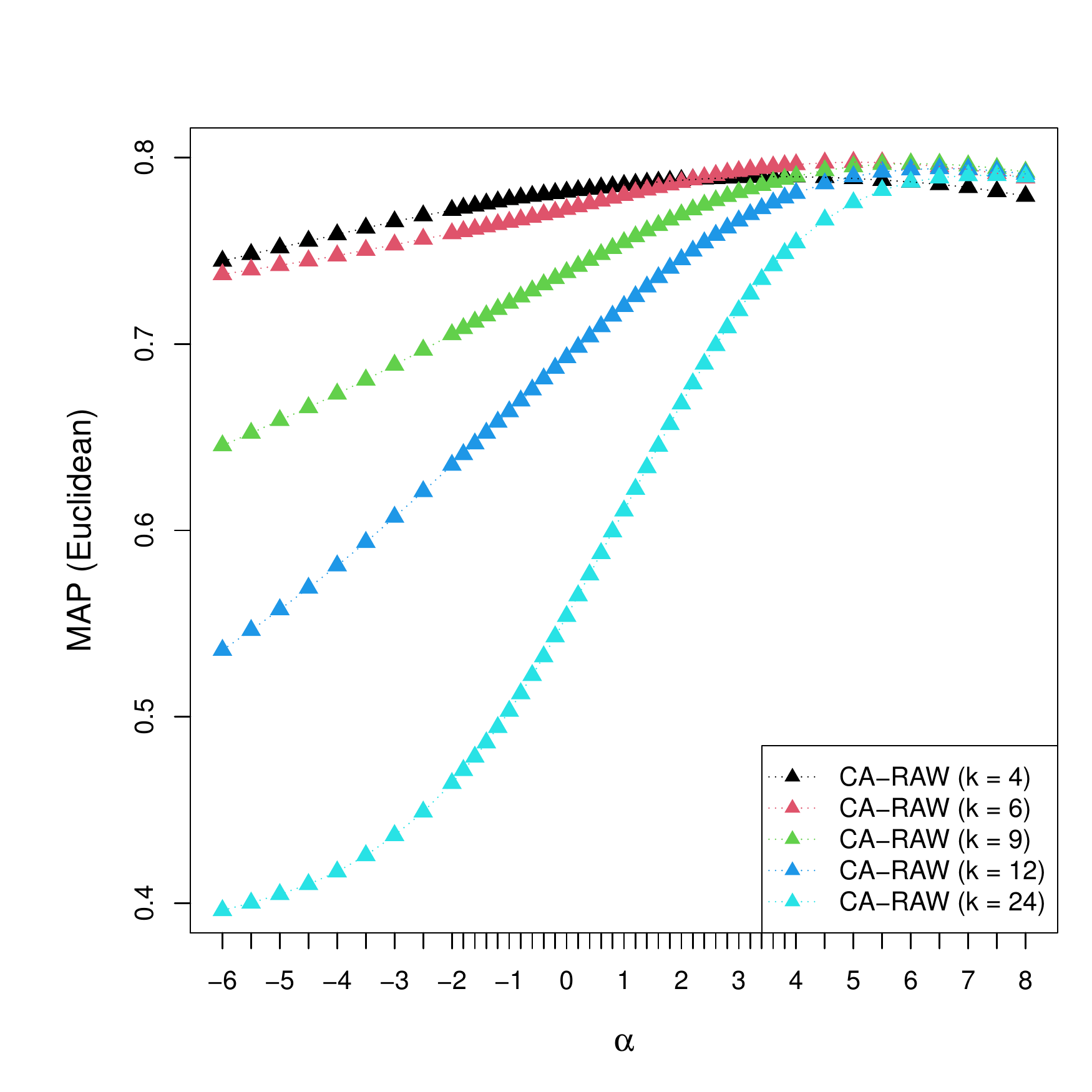}
         \caption{BBCSport}\label{F: eucstandarddBBCsportcaweiexp}
         \end{subfigure}
      \begin{subfigure}[b]{0.45\linewidth}
         \centering
         \includegraphics[width=1\textwidth]{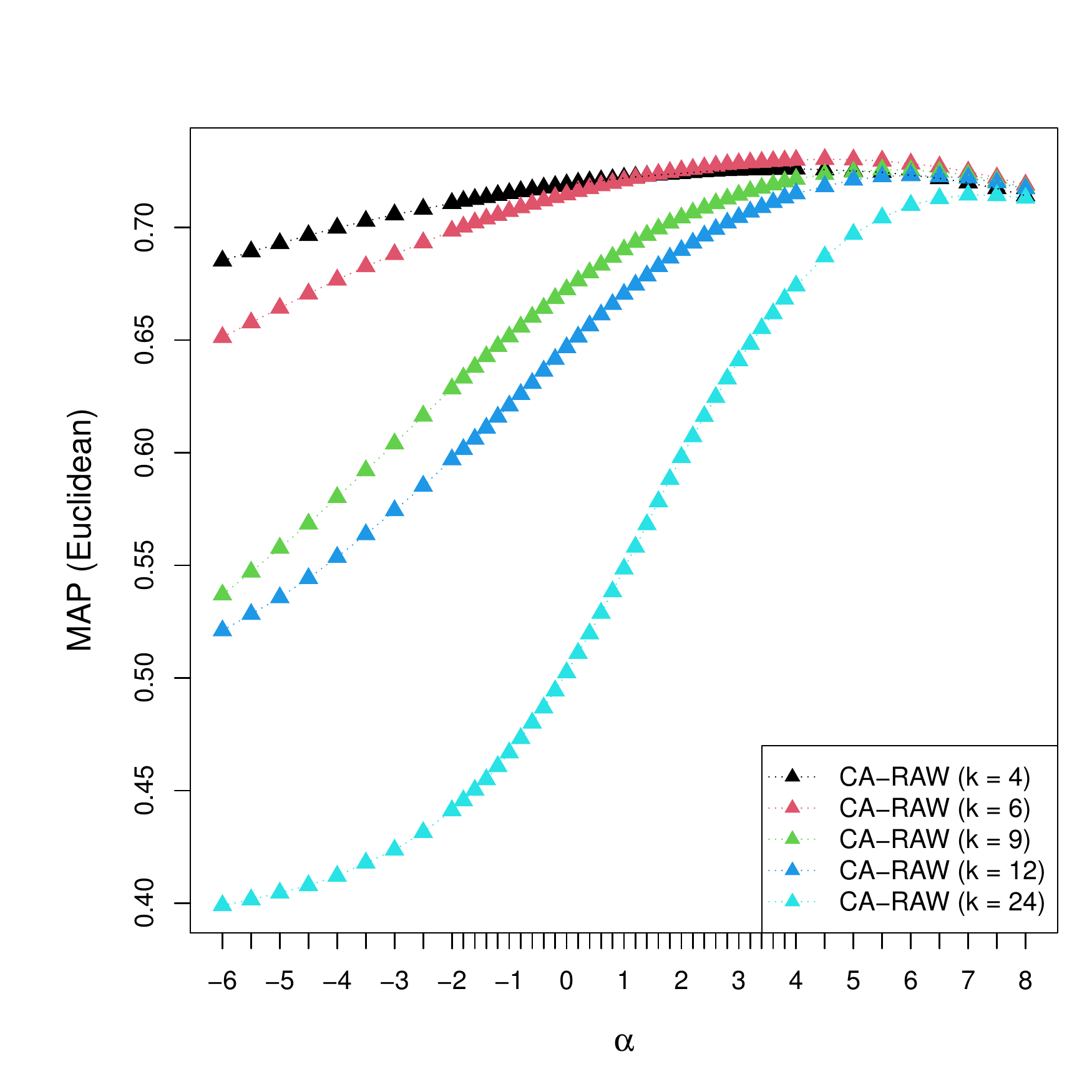}
         \caption{20 Newsgroups}\label{F: eucstandardd20newsgroupscaweiexp}
         \end{subfigure}
      \begin{subfigure}[b]{0.45\linewidth}
         \centering
         \includegraphics[width=1\textwidth]{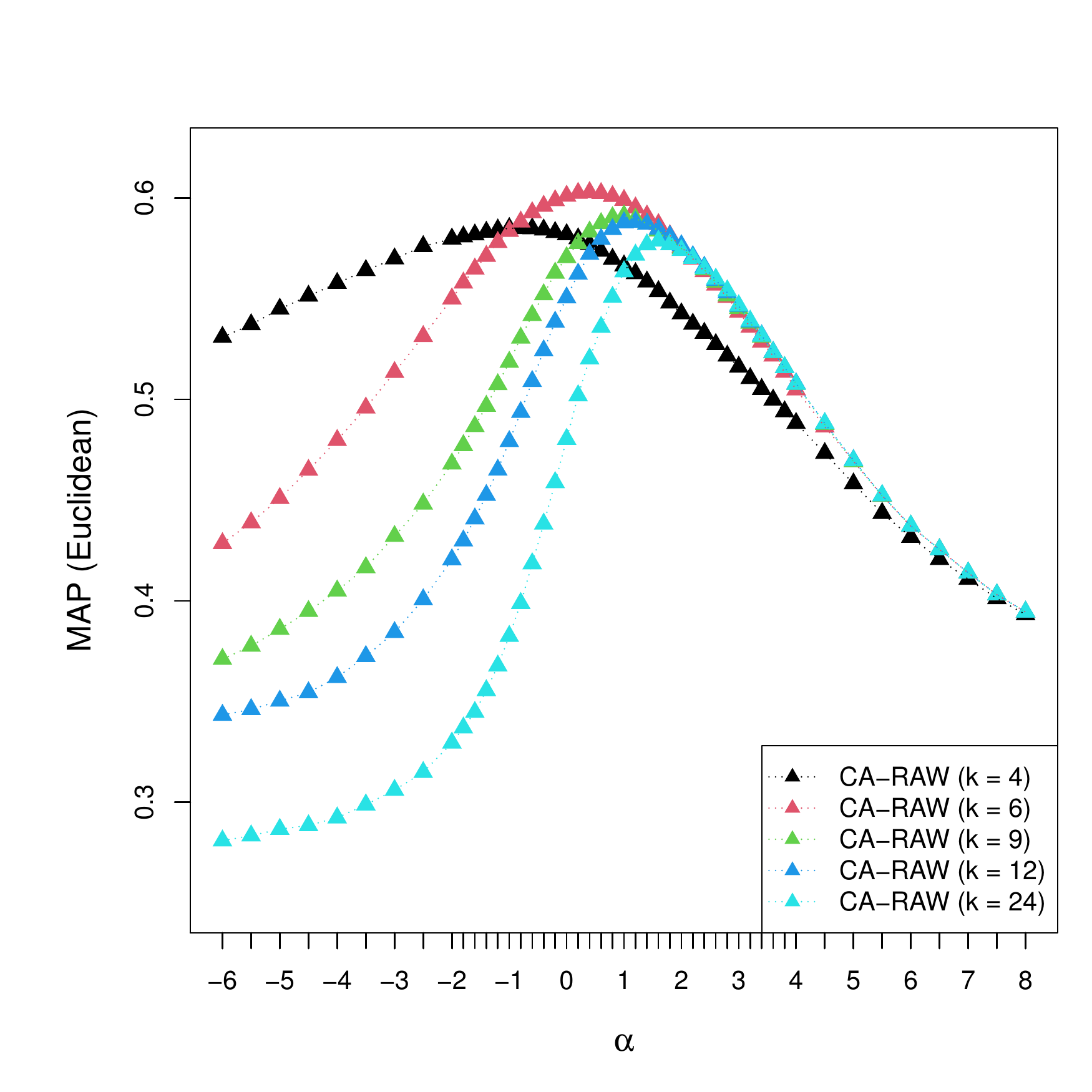}
         \caption{Wilhelmus}\label{F: eucstandarddwilhelmuscaweiexp}
         \end{subfigure}
    \caption{ MAP as a function of $\alpha$ for CA-RAW under various values of $k$.}
    \label{F: eucstandarddcaweiexp}
\end{figure}

In this section, we introduce CA with weighting exponent $\alpha$. Similar to Figure~\ref{F: eucstandardd},  Figure~\ref{F: eucstandarddcaweiexp} shows MAP as a function of $\alpha$ in CA-RAW for the number of dimensions $k = 4, 6, 9, 12, \text{and }24$. Table~\ref{TIReucdim10caweiexp} shows the optimal $\alpha$ and the corresponding MAP, which is a condensed version of Figure~\ref{F: eucstandarddcaweiexp}. We conclude the following from Figure~\ref{F: eucstandarddcaweiexp} and Table~\ref{TIReucdim10caweiexp}:

\begin{itemize}
\item For CA, the overall MAP first increases and then decreases as a function of $\alpha$. This means that varying $\alpha$ can potentially improve the performance of CA. 

\item The increase in MAP by adjusting $\alpha$ is data and dimension dependent.

\item If we compare the maxima in Table~\ref{TIReucdim10} with those in Table~\ref{TIReucdim10caweiexp}, there is hardly a noticeable increase. 
\end{itemize}

Comparing Table~\ref{TIReucdim10caweiexp} with part LSA-RAW of Table~\ref{TIReucdim10}, the optimal $\alpha$ for CA-RAW is almost always larger than LSA-RAW and is almost always larger than 1. That is, CA-RAW needs a larger $\alpha$ than LSA-RAW to obtain its maximum MAP. Thus, compared to LSA, CA improves by placing more emphasis on its initial dimensions. The important difference between LSA and CA is that LSA involves margins, and CA does not. Therefore, we infer that margins in LSA considerably contribute to the initial dimensions; however, they are irrelevant ("noise") for information retrieval. On the other hand, CA effectively eliminates this irrelevant information.

\begin{table}[H]
\centering  
\caption{MAP with the optimal $\alpha$ for CA-RAW under $k = 4, 6, 9, 12, \text{and }24$. Bold values are best.} 
\label{TIReucdim10caweiexp}
\begin{tabular}{lcccccccc}    
&\multicolumn{2}{c}{BBCNews} &\multicolumn{2}{c}{BBCSport}&\multicolumn{2}{c}{20 Newsgroups}&\multicolumn{2}{c}{Wilhelmus} \\
&$\alpha$&MAP&$\alpha$&MAP&$\alpha$&MAP&$\alpha$&MAP\\
\hline 
CA-RAW ($k = 4$) &2& \textbf{0.829} &3.6& 0.790& 4&0.726 & -1 &0.585 \\
CA-RAW ($k = 6$) &4.5& 0.814 &5& \textbf{0.798}& 4.5&\textbf{0.730} & 0.4&\textbf{0.603} \\
CA-RAW ($k = 9$) &6.5& 0.802&6& 0.797&5.5&0.726 & 1&0.591 \\
CA-RAW ($k = 12$) &7& 0.797&6.5& 0.794& 6&0.723 & 1.2 &0.588\\
CA-RAW ($k = 24$) &8&0.788&7.5& 0.791& 7& 0.715 & 1.6 & 0.579\\
\hline 
\end{tabular}  
\end{table} 

\subsection{Exploring MAP as a function of $\alpha$ under the optimal number of dimensions for LSA and CA}\label{Sub: optimalalpha}

\begin{figure}[h]
    \centering
       \begin{subfigure}[b]{0.45\linewidth}
         \centering
         \includegraphics[width=1\textwidth]{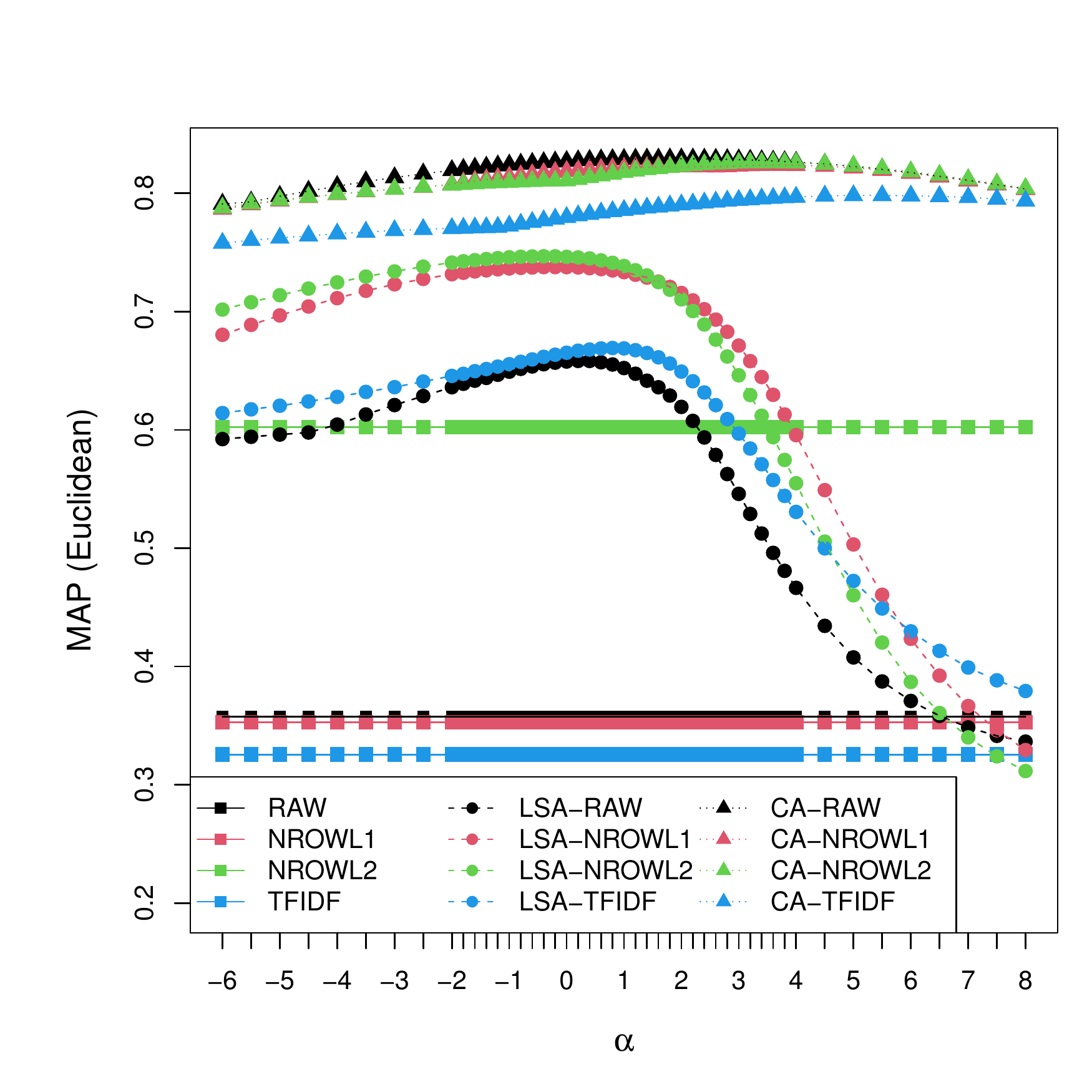}
         \caption{BBCNews}\label{F: eucoptimalMAPBBCnews}
         \end{subfigure}
      \begin{subfigure}[b]{0.45\linewidth}
         \centering
         \includegraphics[width=1\textwidth]{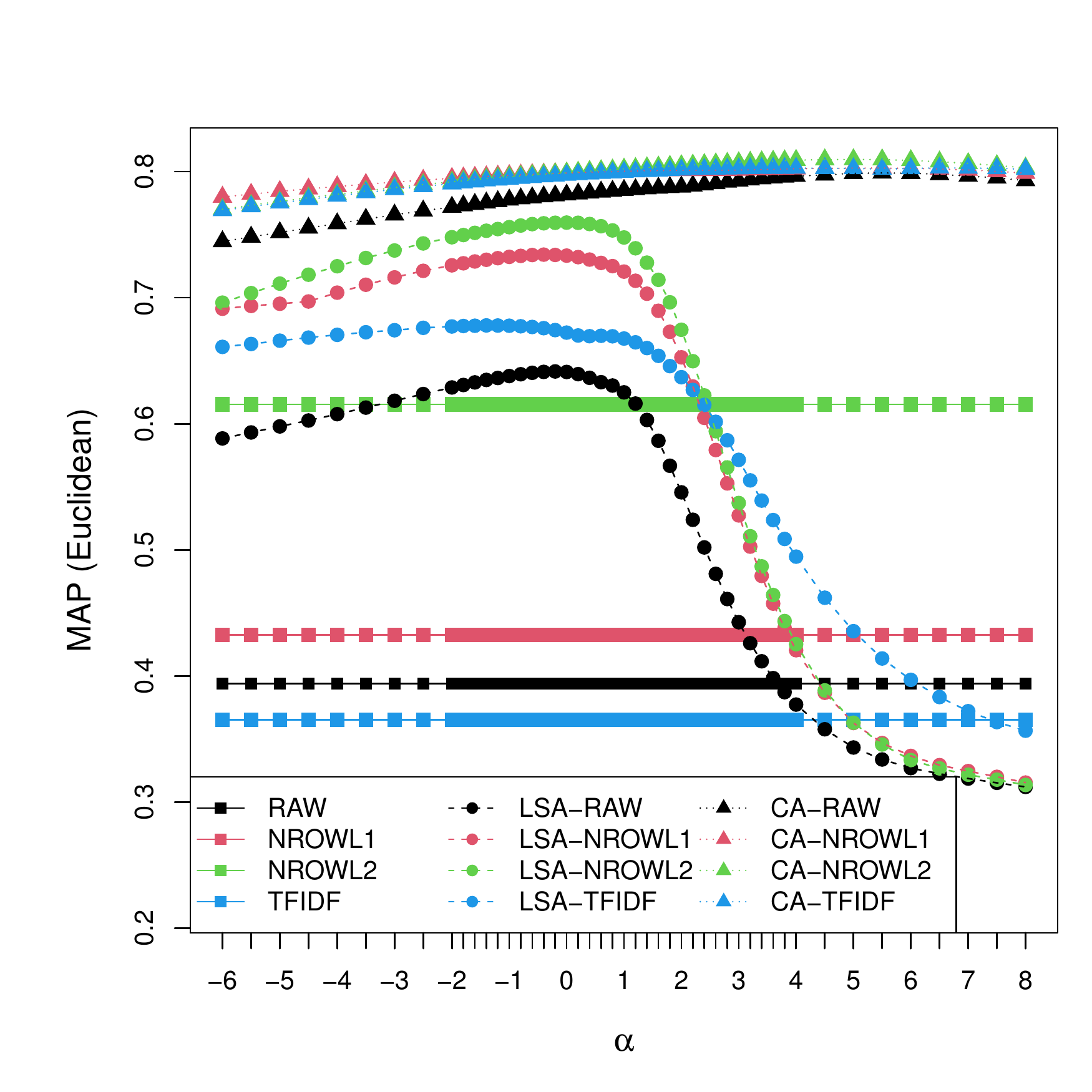}
         \caption{BBCSport}\label{F: eucoptimalMAPBBCsport}
         \end{subfigure}
        \begin{subfigure}[b]{0.45\linewidth}
         \centering
         \includegraphics[width=1\textwidth]{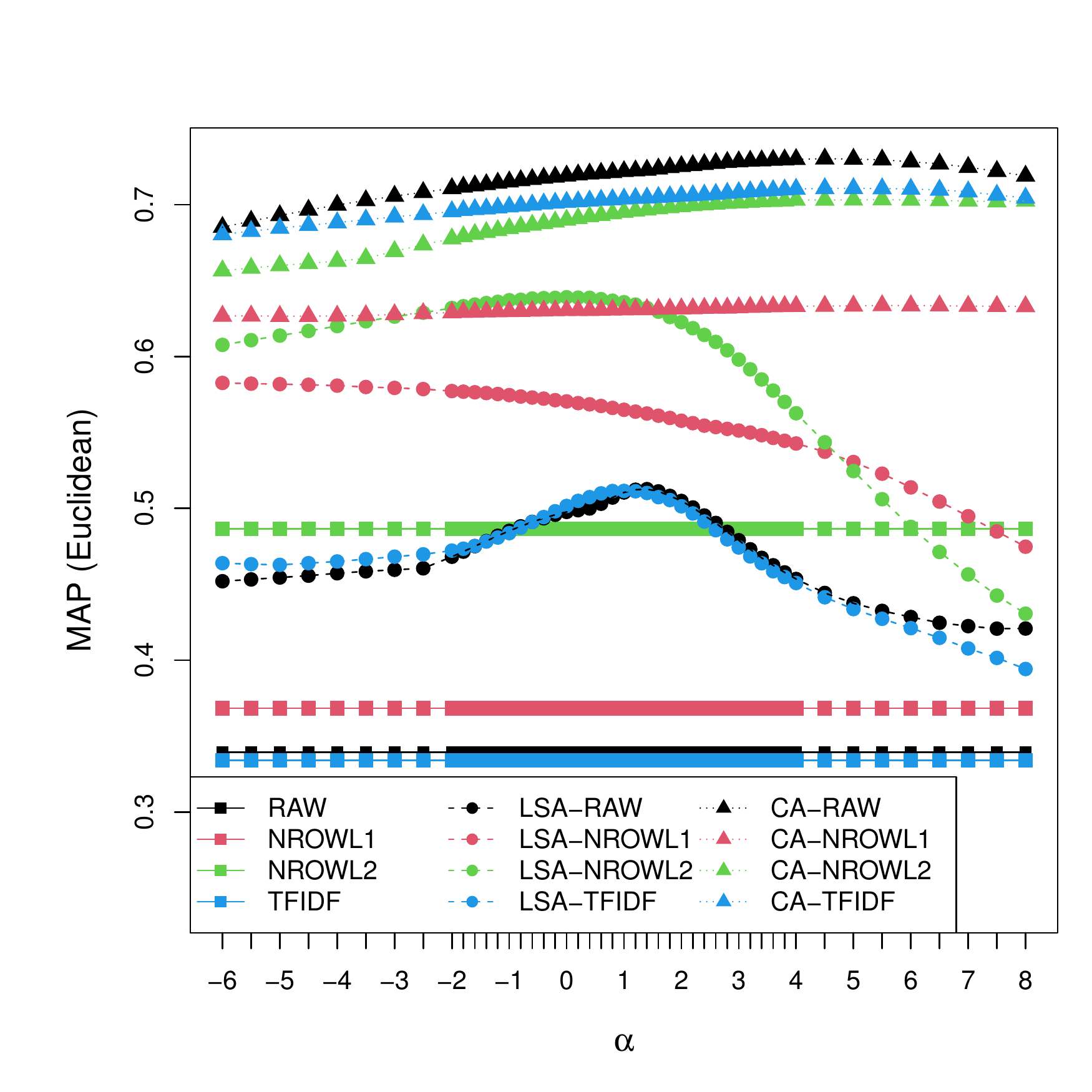}
         \caption{20 Newsgroups}\label{F: eucoptimalMAP20newsgroups}
         \end{subfigure}
        \begin{subfigure}[b]{0.45\linewidth}
         \centering
         \includegraphics[width=1\textwidth]{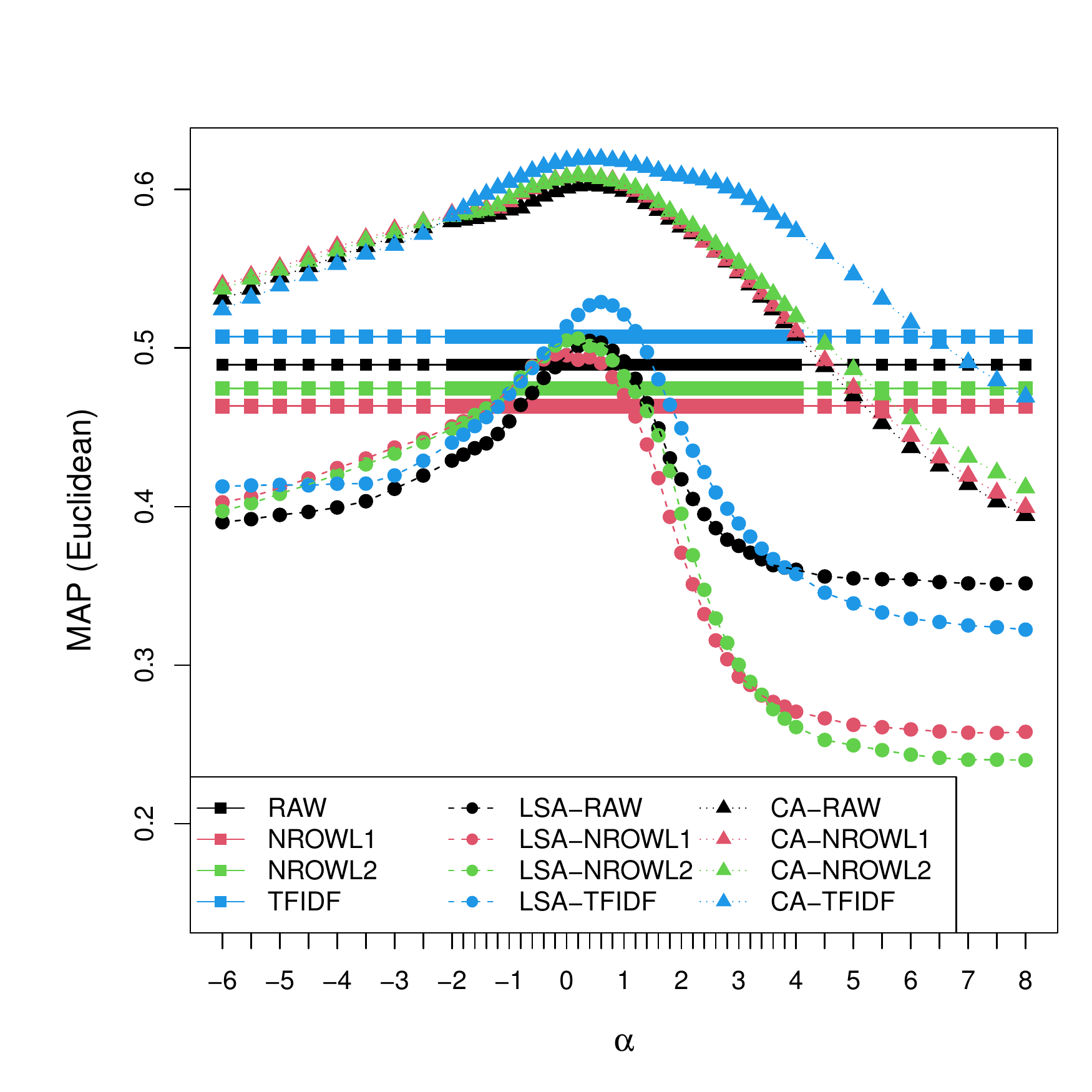}
         \caption{Wilhelmus}\label{F: eucoptimalMAPwilhelmus}
         \end{subfigure}
    \caption{MAP as a function of $\alpha$ under the optimal number of dimensions.}
    \label{F: eucoptimalMAP}
\end{figure}

Figure~\ref{F: eucoptimalMAP} shows the MAP as a function of $\alpha$ under the optimal number of dimensions. Similar to Section~\ref{Subsub: standardalpha} in the context of $\alpha = 1$, we can obtain the corresponding optimal $k$ (not shown in the figure) and the corresponding MAP (shown in the figure) for each $\alpha$. Table~\ref{TIReuc} shows the optimal $\alpha$, optimal $k$, and corresponding MAP, which is a condensed version of Figure~\ref{F: eucoptimalMAP}. Based on Figure~\ref{F: eucoptimalMAP} and Table~\ref{TIReuc}, we can see that 

\begin{itemize}
\item CA methods are always better than the LSA methods and term matching methods under the optimal $\alpha$ and optimal number of dimensions $k$. 
\item Weighting the elements of the raw document-term matrix under the optimal number of dimensions $k$ can improve the performance of CA; however, the improvements are small and data dependent.
\item Similar to dimension $k = 4, 6, 9, 12, \text{and }24$, MAP as a function of $\alpha$ under the optimal number of dimensions $k$ also first increases and then decreases. Thus, adjusting $\alpha$ can potentially improve the performance of LSA and CA. 
\item For different datasets, the optimal $\alpha$ under the optimal number of dimensions $k$ is very different. In constrast to LSA, CA needs a greater $\alpha$ to reach the optimal performance under the optimal number of dimensions $k$. This illustrates that CA places more emphasis on initial dimensions than LSA.
\end{itemize}

\begin{table}[htbp]
\centering  
\caption{MAP under the optimal $\alpha$ and optimal dimension $k$. Bold values are best within group; underlined values are best overall.} 
\label{TIReuc}
\begin{tabular}{ccccccccccccc}    
&\multicolumn{3}{c}{BBCNews} &\multicolumn{3}{c}{BBCSport}&\multicolumn{3}{c}{20 Newsgroups}&\multicolumn{3}{c}{Wilhelmus} \\
&$\alpha$&$k$&MAP&$\alpha$&$k$&MAP&$\alpha$&$k$&MAP&$\alpha$&$k$&MAP\\
\hline 
RAW & &&0.358& &&0.394 &&&0.339&&& 0.489\\
LSA-RAW &0.2 &6 & 0.658& -0.2&6 & 0.642 & 1.4 &12 & 0.513&0.4&13&0.505\\
CA-RAW&2&4&\underline{\textbf{0.829}}&5.5&7&\textbf{0.799}&4.5&6&\underline{\textbf{0.730}}&0.4&6&\textbf{0.603}\\
\hdashline
NROWL1& &&0.353& & &0.433 & &&0.368&&&0.463 \\
LSA-NROWL1 &-0.4&5&0.738&-0.4&5&0.734&-6&10&0.583&-0.2&8& 0.496\\
CA-NROWL1&3.6&5&\textbf{0.824}&5.5&5&\textbf{0.803}&5.5&7&\textbf{0.634}&0.2&6&\textbf{0.609}\\
\hdashline
NROWL2 & &&0.602 & & &0.615 && &0.486 &&&0.474 \\
LSA-NROWL2 &-0.4&5&0.747&0&5&0.760&0&4&0.639& 0.2&10&0.506\\
CA-NROWL2&3.6&5&\textbf{0.826}&5&5&\underline{\textbf{0.810}}&5.5&6&\textbf{0.703}&0.2&6&\textbf{0.609}\\
\hdashline
TFIDF & && 0.326 && &0.365& & &0.334 &&& 0.507 \\
LSA-TFIDF &0.8&10&0.669&-1.4&6&0.678&1&12&0.512&0.6&16&0.529\\
CA-TFIDF&5&6&\textbf{0.798}&6.5&7&\textbf{0.803}&5&6&\textbf{0.711}&0.6&5&\underline{\textbf{0.619}}\\
\hline 
\end{tabular}  
\end{table} 

\section{Results for dot similarity and cosine similarity}\label{S: resultsdotcos}

In Section~\ref{S: resultseuc}, we presented the results where Euclidean distance was used as a measure of similarity. Here, for comparison, we provide results for dot similarity and cosine similarity. To save space, we only show tables corresponding to the Euclidean distance (Table~\ref{TIReucalpha1}), dot similarity (Table~\ref{TIRdotalpha1}) and cosine similarity (Table~\ref{TIRcosalpha1}). Other tables and figures for dot similarity and cosine similarity are presented in the supplementary materials.

The results for both dot similarity and cosine similarity lead to conclusions that match those for Euclidean distance. However, cosine similarity leads to a better performance in terms of MAP than Euclidean distance and dot similarity. We displayed the results for Euclidean distance in Section~\ref{S: resultseuc} because (1) it is more easily interpretable in the context of adjusting weighting exponent $\alpha$: as $\alpha$ increases, Euclidean distances between row points (column points) on initial dimensions increase relative to the later dimensions; and (2) in the literature, the Euclidean distance is the preferred way to interpret CA (in fact, we have never seen an interpretation of CA in terms of cosine or dot similarity).

\begin{table}[H]
\centering  
\caption{MAP with the optimal number of dimensions $k$ about dot similarity. Bold values are best.} 
\label{TIRdotalpha1}
\begin{tabular}{lcccccccc}    
&\multicolumn{2}{c}{BBCNews} &\multicolumn{2}{c}{BBCSport}&\multicolumn{2}{c}{20 Newsgroups}&\multicolumn{2}{c}{Wilhelmus} \\
&$k$&MAP&$k$&MAP&$k$&MAP&$k$&MAP\\
\hline 
RAW &&0.543& &0.534&&0.454&&0.342\\
LSA-RAW  &14 &0.580&15 & 0.536 &  36 & 0.488&90& 0.343\\
LSA-NROWL1 &5&0.751&7& 0.663&14&0.601& 80& 0.404\\
LSA-NROWL2 &5&0.744&6&0.693&4&0.672& 34&0.475\\
LSA-TFIDF &32&0.613&22&0.651&60&0.490&100&0.421\\
CA&4&\textbf{0.896}&7&\textbf{0.865}& 6&\textbf{0.788}&12&\textbf{0.642}\\
\hline 
\end{tabular}  
\end{table} 

\begin{table}[H]
\centering  
\caption{MAP with the optimal number of dimensions $k$ about cosine similarity. Bold values are best.} 
\label{TIRcosalpha1}
\begin{tabular}{lcccccccc}    
&\multicolumn{2}{c}{BBCNews} &\multicolumn{2}{c}{BBCSport}&\multicolumn{2}{c}{20 Newsgroups}&\multicolumn{2}{c}{Wilhelmus} \\
&$k$&MAP&$k$&MAP&$k$&MAP&$k$&MAP\\
\hline 
RAW & &0.602& &0.615& &0.487&& 0.474\\
LSA-RAW  &6 &0.779&9 & 0.766 &  12 & 0.654&22& 0.484\\
LSA-NROWL1 &5&0.817&6& 0.775&12&0.651&13& 0.481\\
LSA-NROWL2 &5&0.825&5&0.795& 4&0.745& 13&0.482\\
LSA-TFIDF &10&0.796&9&0.819&12&0.665&19&0.531\\
CA&4&\textbf{0.902}&7&\textbf{0.878}& 6&\textbf{0.794}&12&\textbf{0.652}\\
\hline 
\end{tabular}  
\end{table}

\section{Conclusions and discussions}\label{S: condis}

We compared four versions of LSA with CA and found that CA always performs better than LSA in terms of MAP. Then, we compared LSA-RAW as a function of weighting exponent $\alpha$ with CA under a range of the numbers of dimensions. Even though LSA is improved by choosing an appropriate value for $\alpha$, CA always performed better than LSA.

Next, we applied different weighting elements of raw document-term matrix to CA. We found that weighting elements of the raw matrix sometimes improves the performance of CA, but improvements over CA-RAW are small and data dependent. Then, we adjusted the weighting exponents $\alpha$ in CA. For CA, as a function of $\alpha$, MAP first increases and then decreases. Adjusting the weighting exponent $\alpha$ can potentially improve the performance of CA. However, the increased performance obtained by adjusting $\alpha$ is data and dimension dependent. 

Using the standard coordinates of $\alpha = 1$, for LSA, the Euclidean distances between the rows of coordinates approximate the Euclidean distances between the rows of the decomposed matrix. For CA, the Euclidean distances between the rows of coordinates approximate the $\chi^2-$distances between the rows of the decomposed matrix. $\alpha < 1$ gives less emphasis to the initial dimensions relative to the standard coordinates. Conversely, $\alpha > 1$ gives more emphasis to the initial dimensions relative to the standard coordinates. The optimal $\alpha$ for CA is almost always larger than that for LSA and is almost always larger than 1.

Finally, we studied MAP as a function of $\alpha$ under the optimal number of dimensions. Again, CA performs better than LSA. Although the optimal $\alpha$ under the optimal number of dimensions is data dependent, the optimal $\alpha$ of CA is usually considerably larger than that of LSA.

\citet{bullinaria2012extracting} argued that the initial dimensions in LSA tend not to contribute the most useful information about semantics and tend to be contaminated by "noise". The above mentioned results indicate that CA places more emphasis on the initial dimensions than LSA. The major difference between LSA and CA is that LSA involves margins but CA does not \citep{qi2021comparison}. Thus, we infer that margins considerably contribute to the initial dimensions in LSA. These margins are irrelevant for information retrieval. The CA effectively eliminates this irrelevant information.

In this paper, we focused on the performances of CA and LSA using Euclidean distances. We also performed identical experiments for dot similarity and cosine similarity. Both have nearly identical results with the Euclidean distance. Cosine similarity performs better than the Euclidean distance and dot similarity. We focus on Euclidean distance in the paper because (1) it is more easily interpretable in the context of adjusting $\alpha$:  as $\alpha$ increases, the Euclidean distances between row points (column
points) on the initial dimensions increase relative to the later dimensions; (2) for CA, dot similarity and cosine similarity have never been used before, and therefore, by focusing on Euclidean distances, the results fit better into the existing literature. 

Based on experimental results and analysis, we have the following three suggestions for practical guidance: 
\begin{enumerate}
\item Use CA instead of LSA; use CA for visualizing data.
\item If information retrieval is the key issue, use cosine similarity instead of Euclidean distance and dot similarity for calculating MAP.  
\item If optimal performance in terms of MAP is not of key importance, there is no need to weight the elements of raw document-term matrix for CA and optimize the performance over $\alpha$ for CA to saving time. Otherwise, these two weightings may be considered potential approaches for improving the performance of CA.
\end{enumerate}

There exist various applications of LSA. In future work, it will be interesting to attempt to improve these applications by replacing LSA with CA.

\section*{Acknowledgments}

Author Qianqian Qi is supported by the China Scholarship Council.

\bibliography{references.bib}

\appendixqq

\appendixqqsection{Euclidean distance}\label{SS: Euclidean distance}

\begin{figure}[H]
    \centering
       \begin{subfigure}[b]{0.45\linewidth}
         \centering
         \includegraphics[width=1\textwidth]{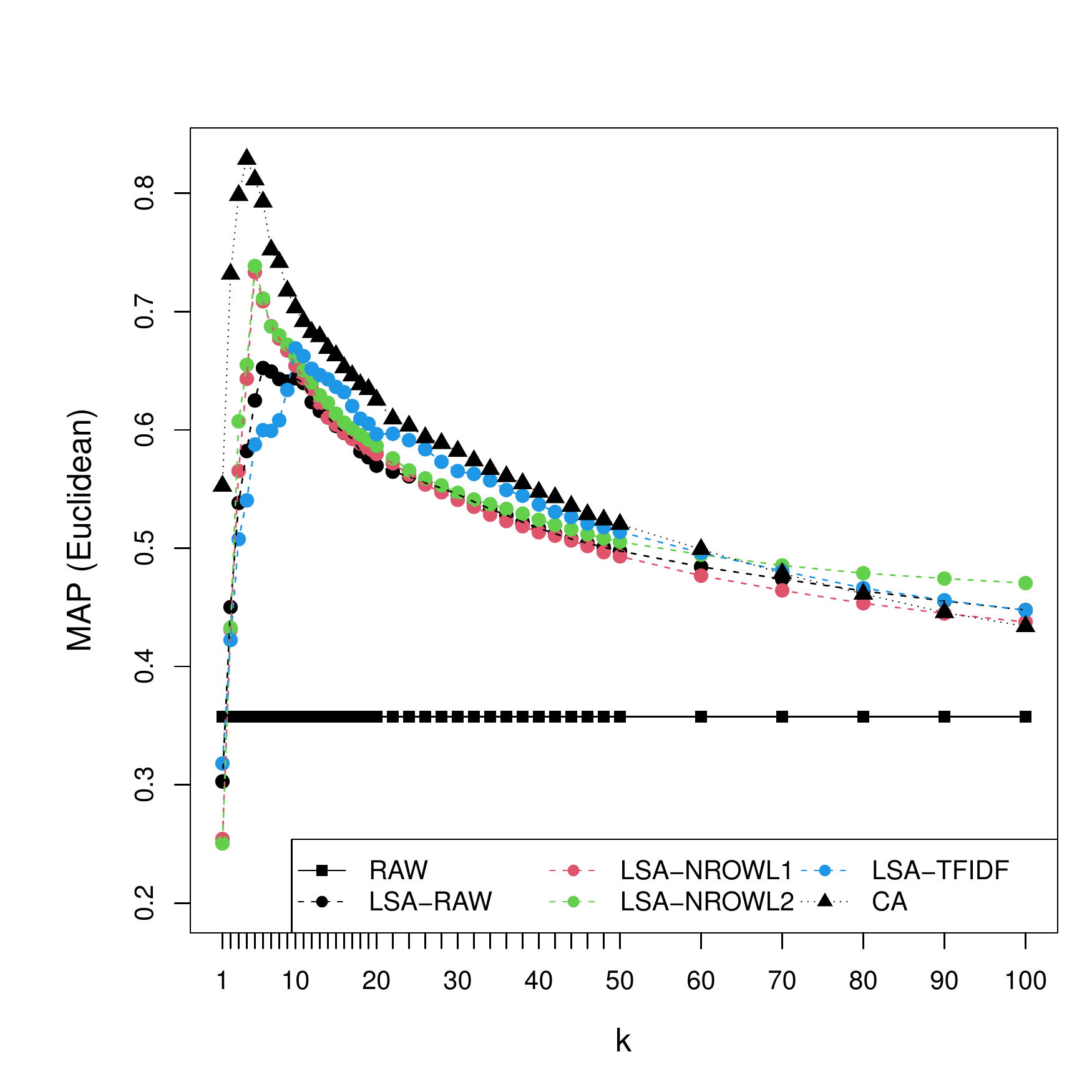}
         \caption{$\alpha = 1$ (BBCNews)}\label{F: eucstandardpBBCnews}
         \end{subfigure}
      \begin{subfigure}[b]{0.45\linewidth}
         \centering
         \includegraphics[width=1\textwidth]{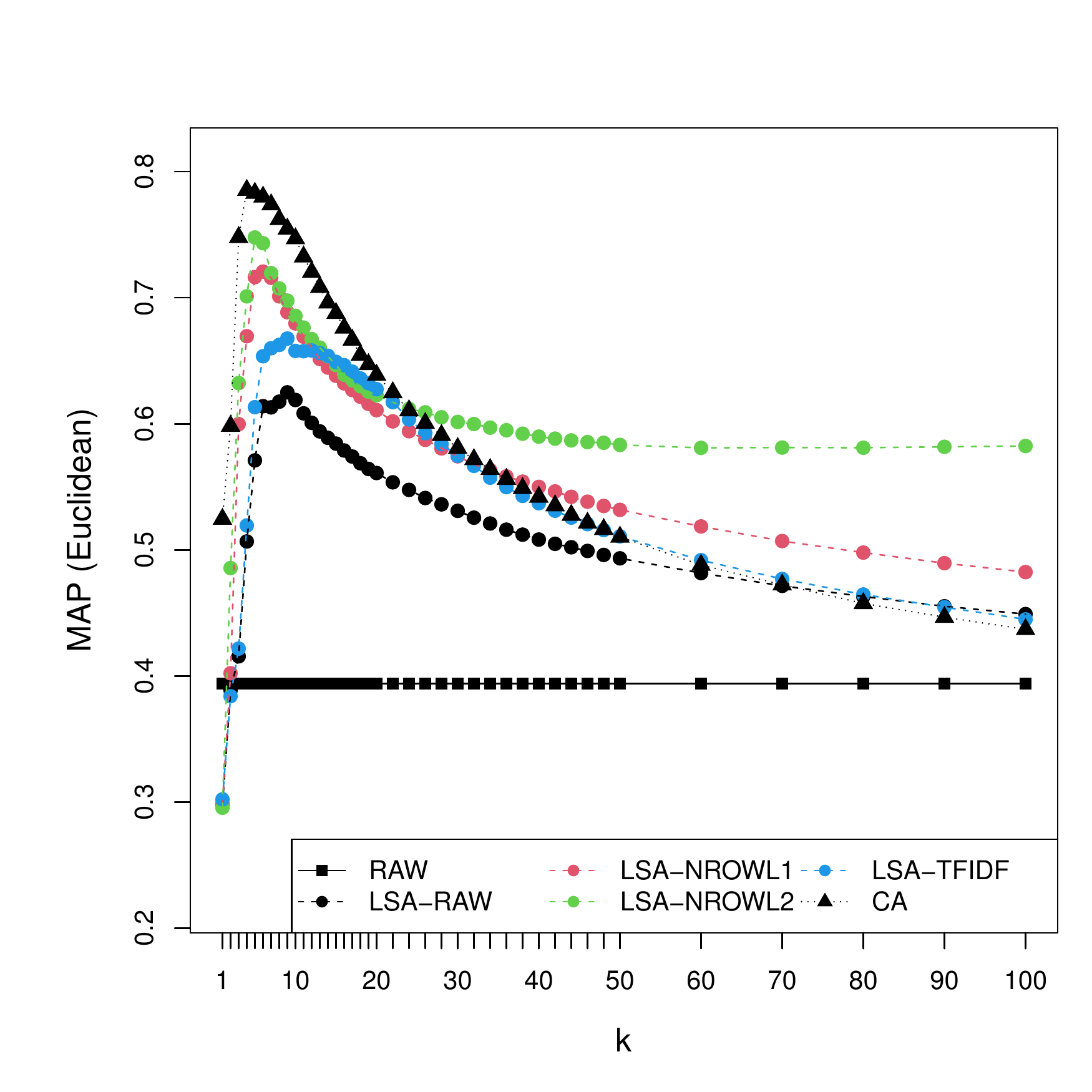}
         \caption{$\alpha = 1$ (BBCSport)}\label{F: eucstandardpBBCsport}
         \end{subfigure}
     \begin{subfigure}[b]{0.45\linewidth}
         \centering
         \includegraphics[width=1\textwidth]{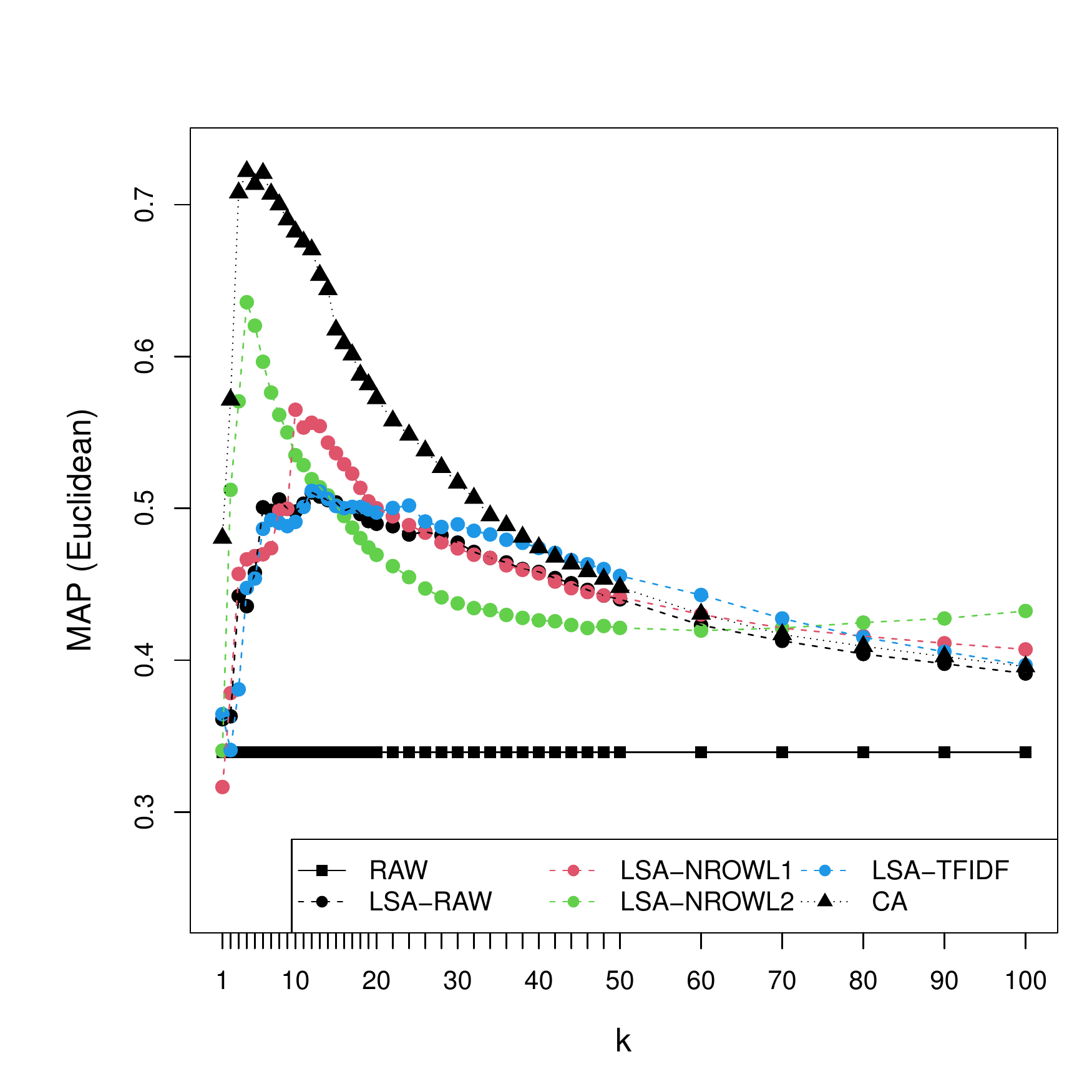}
         \caption{$\alpha = 1$ (20 Newsgroups)}\label{F: eucstandardp20newsgroups}
         \end{subfigure}
      \begin{subfigure}[b]{0.45\linewidth}
         \centering
         \includegraphics[width=1\textwidth]{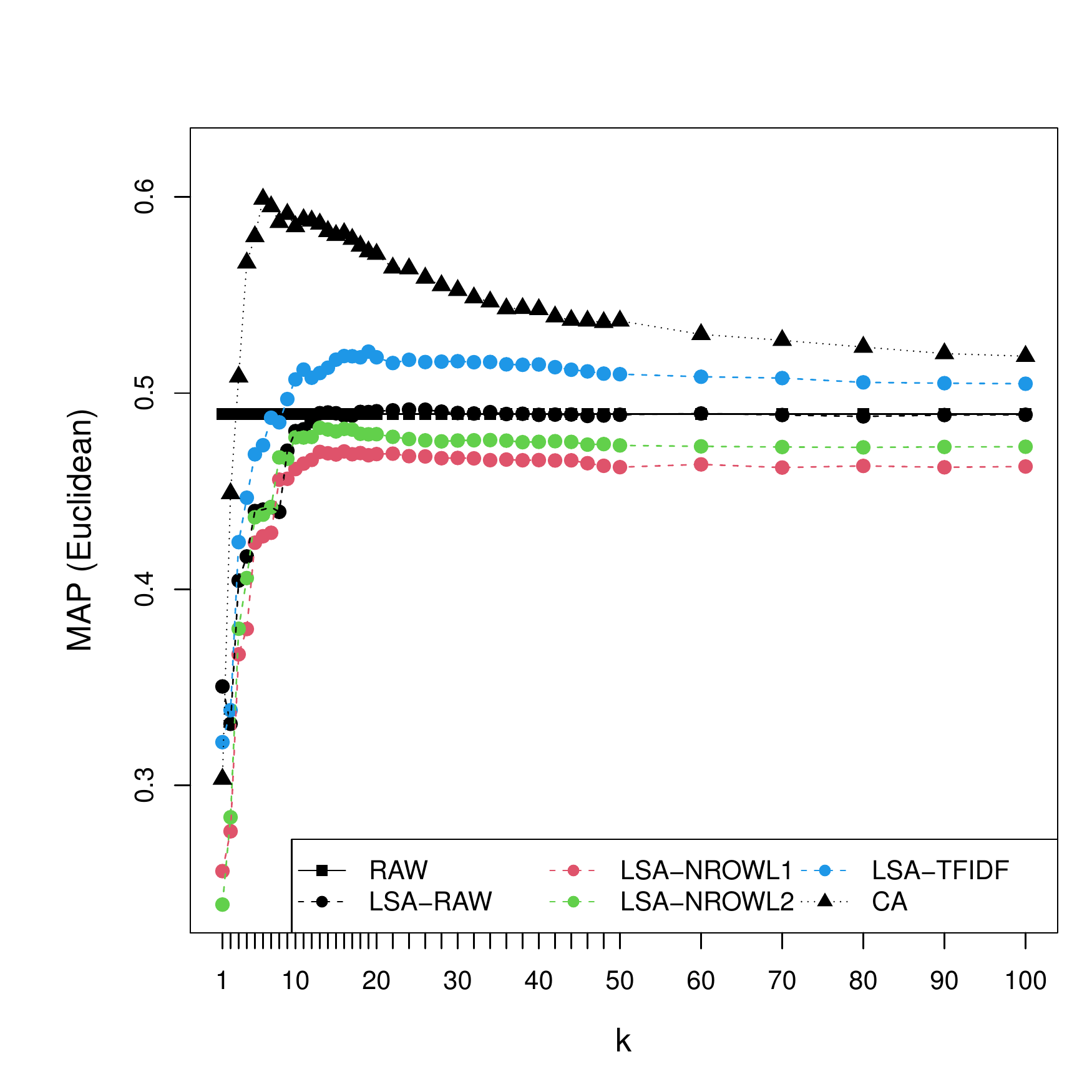}
         \caption{$\alpha = 1$ (Wilhelmus)}\label{F: eucstandardpwilhelmus}
         \end{subfigure}
    \caption{MAP as a function of the number of dimensions $k$ under standard coordinates.}
    \label{F: eucstandardp}
\end{figure}

\begin{figure}[H]
    \centering
       \begin{subfigure}[b]{0.45\linewidth}
         \centering
         \includegraphics[width=1\textwidth]{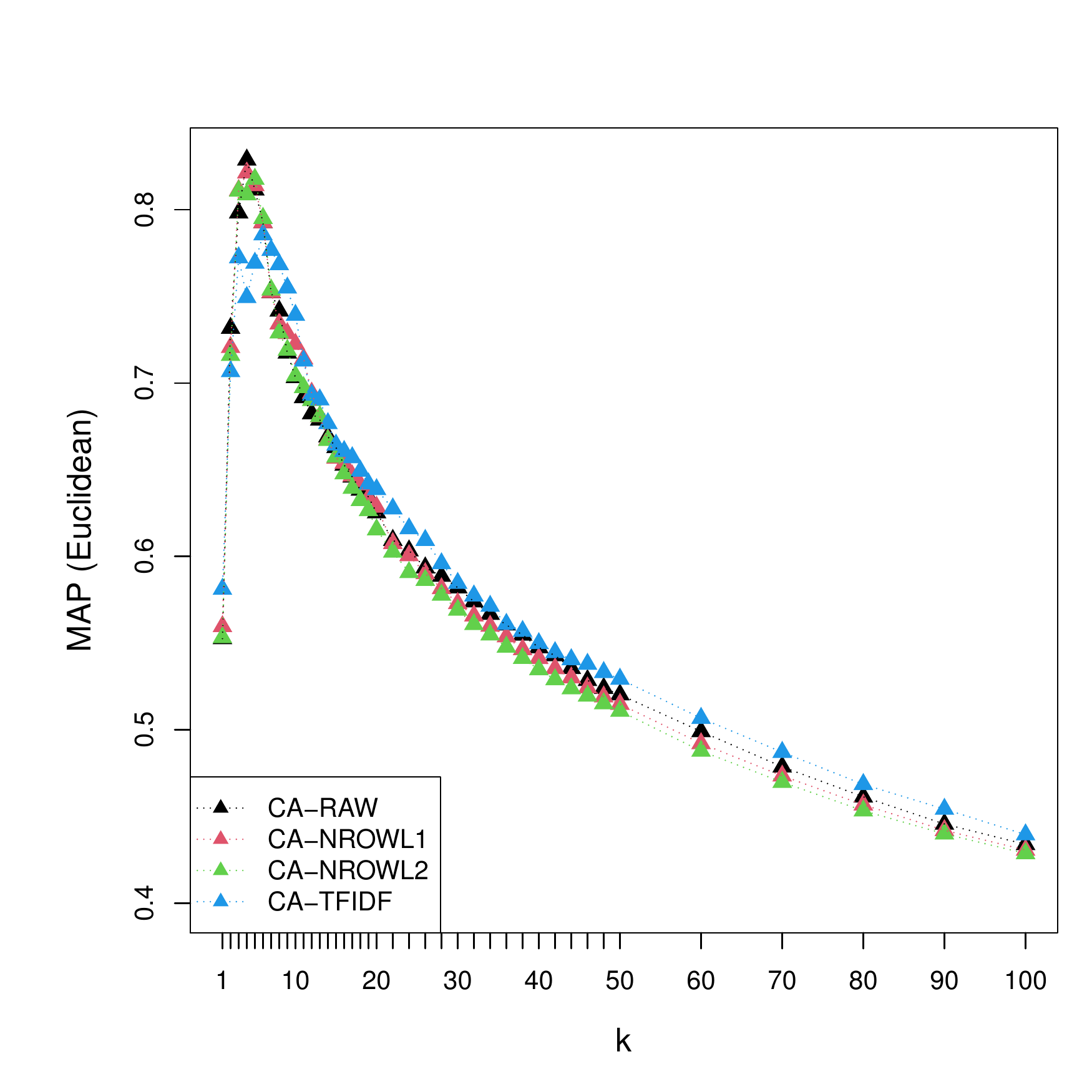}
         \caption{$\alpha = 1$ (BBCNews)}\label{F: eucstandardpBBCnewscaweisch}
         \end{subfigure}
      \begin{subfigure}[b]{0.45\linewidth}
         \centering
         \includegraphics[width=1\textwidth]{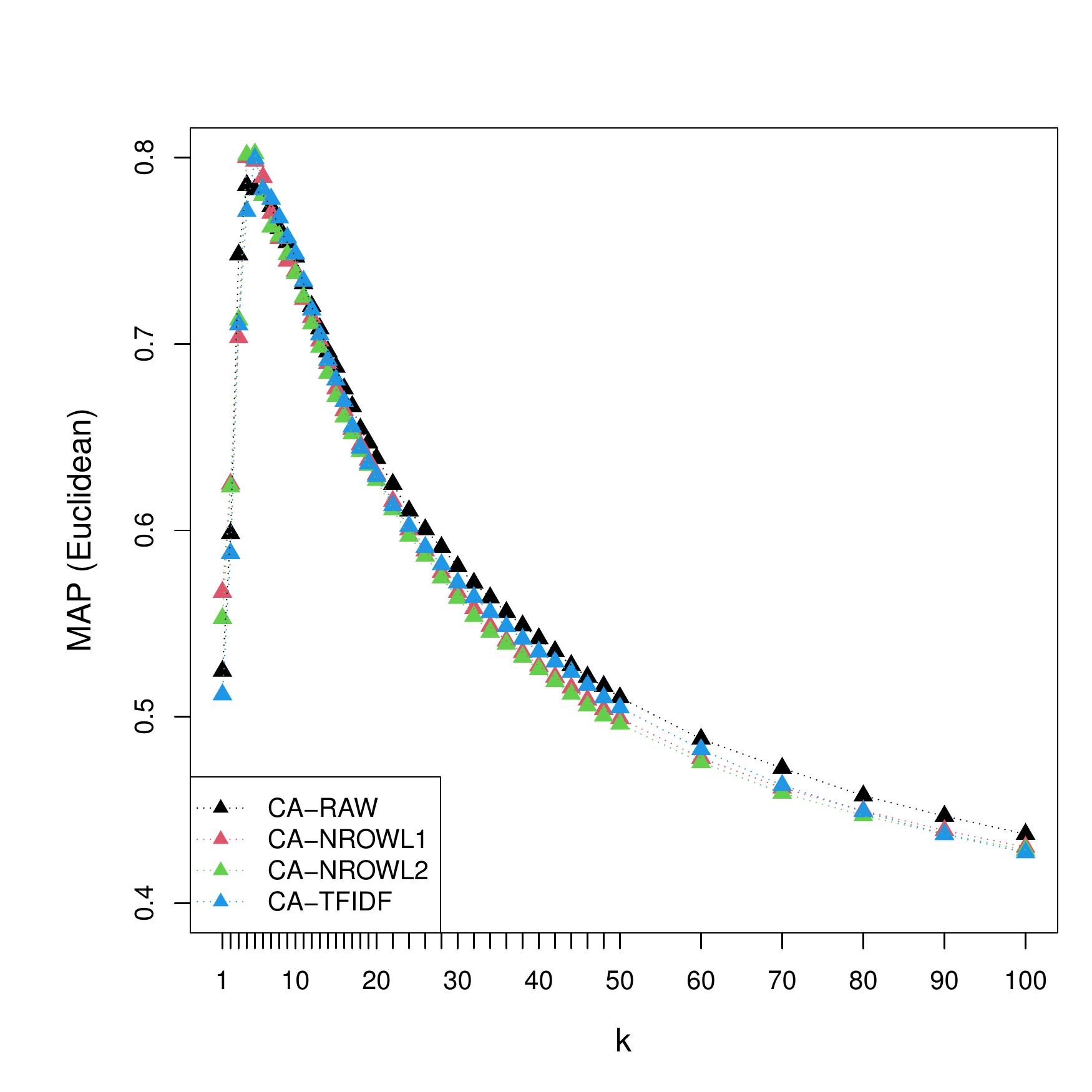}
         \caption{$\alpha = 1$ (BBCSport)}\label{F: eucstandardpBBCsportcaweisch}
         \end{subfigure}
     \begin{subfigure}[b]{0.45\linewidth}
         \centering
         \includegraphics[width=1\textwidth]{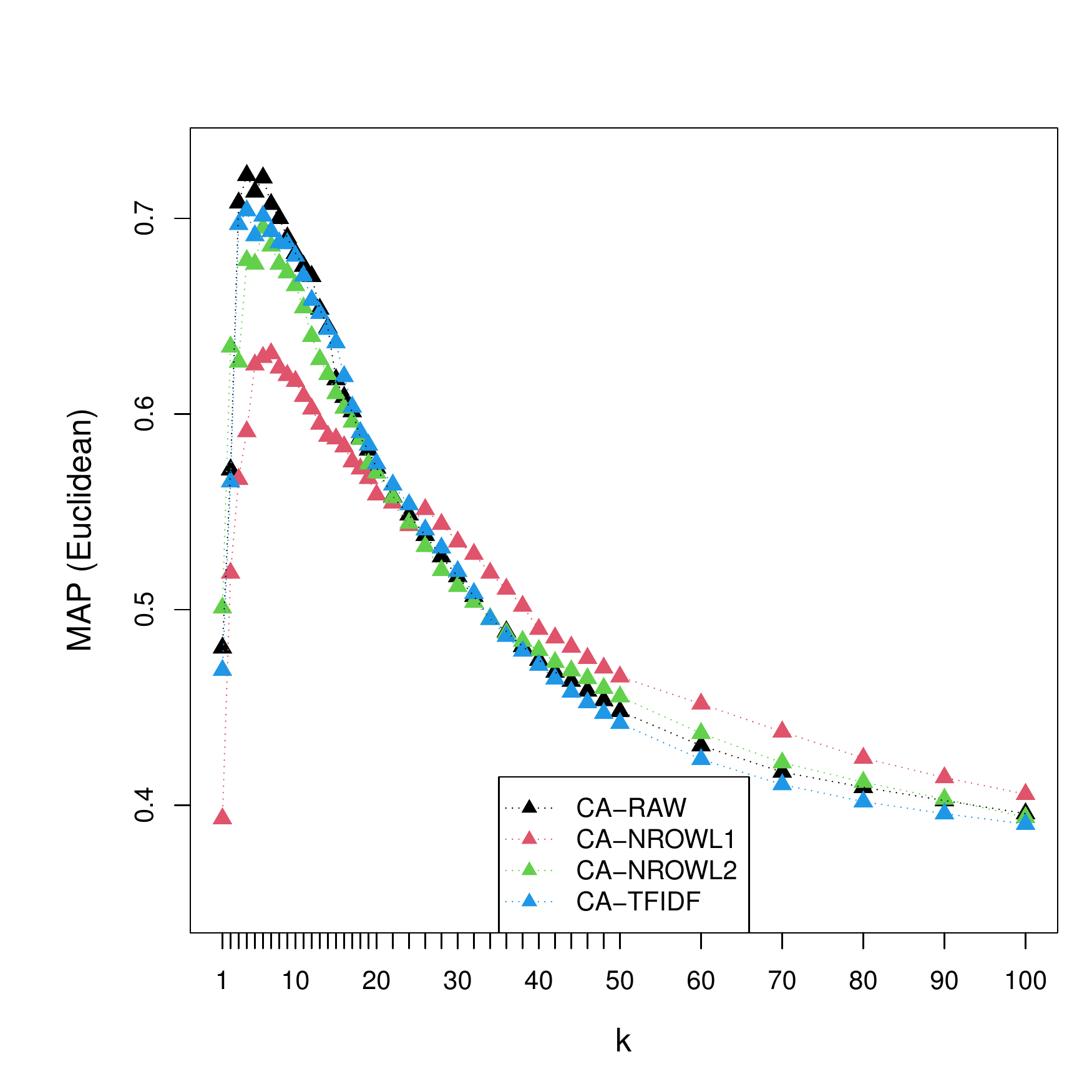}
         \caption{$\alpha = 1$ (20 Newsgroups)}\label{F: eucstandardp20newsgroupscaweisch}
         \end{subfigure}
      \begin{subfigure}[b]{0.45\linewidth}
         \centering
         \includegraphics[width=1\textwidth]{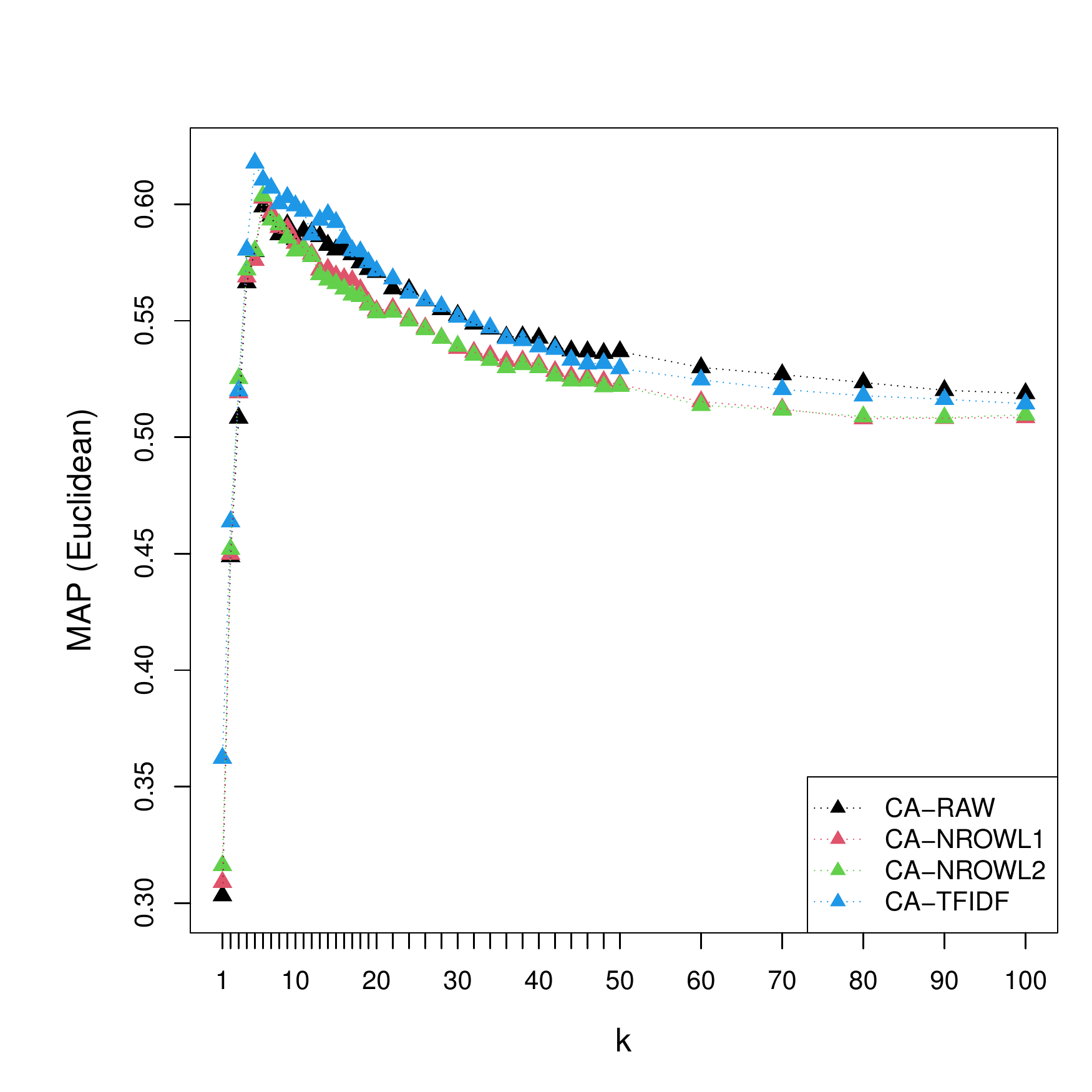}
         \caption{$\alpha = 1$ (Wilhelmus)}\label{F: eucstandardpwilhelmuscaweisch}
         \end{subfigure}
    \caption{MAP as a function of the number of dimensions $k$ for the four versions of CA under standard coordinates.}
    \label{F: eucstandardpcaweisch}
\end{figure}

\appendixqqsection{Dot similarity}\label{SS: dot similarity}

We performed identical experiments to the main paper, but using dot similarity, instead of Euclidean distance, as similarity measurement method. The results lead to matching conclusions as those for Euclidean distance used in the main paper.

\subsection{Comparing LSA and CA for information retrieval}\label{Sub: standardcadot}

\subsubsection{MAP as a function of the number of dimensions for four versions of LSA with standard weighting exponent $\alpha = 1$ and CA}\label{Subsub: standardalphadot}

\begin{figure}[H]
    \centering
       \begin{subfigure}[b]{0.45\linewidth}
         \centering
         \includegraphics[width=1\textwidth]{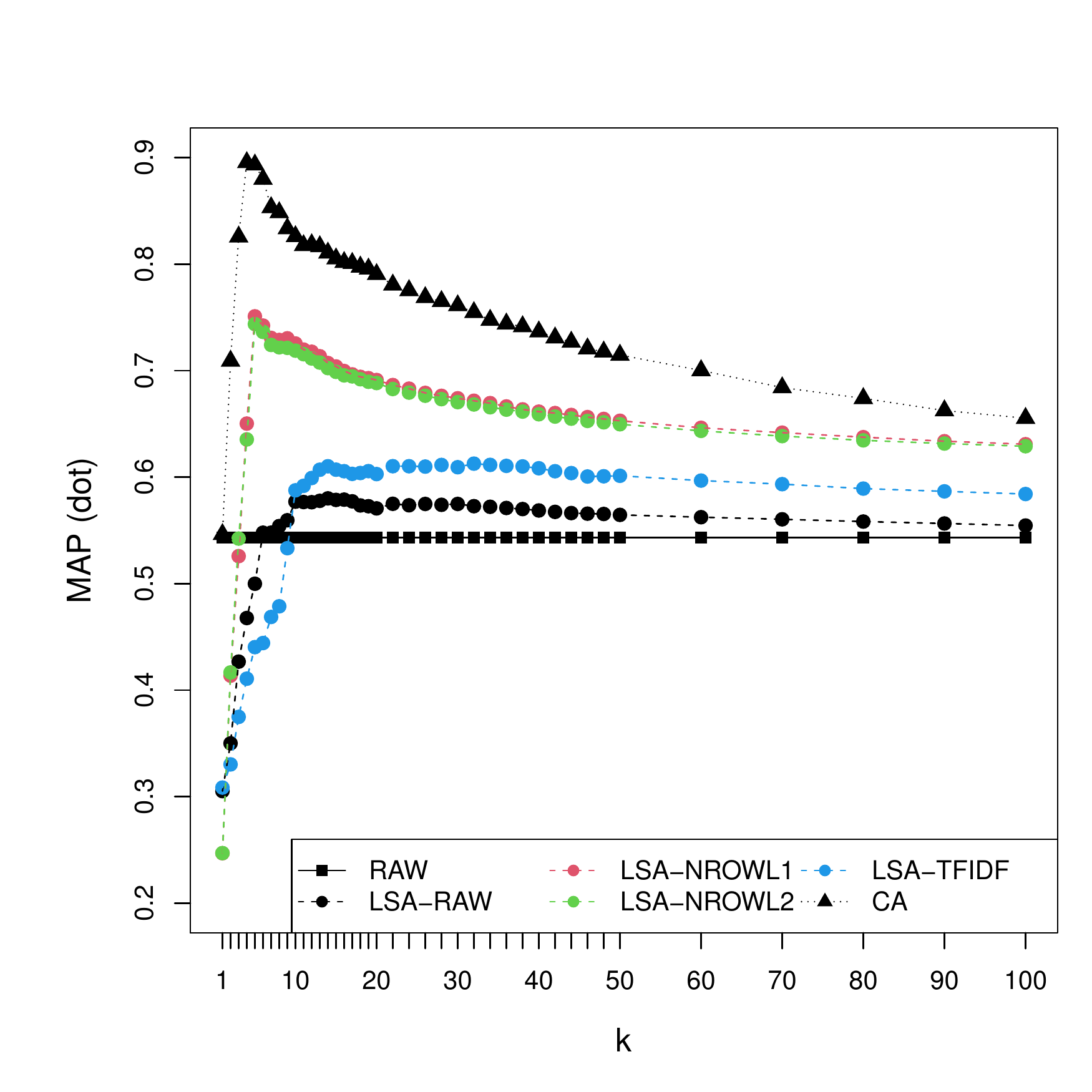}
        \caption{BBCNews}\label{F: dotstandardpBBCnews}
         \end{subfigure}
      \begin{subfigure}[b]{0.45\linewidth}
         \centering
         \includegraphics[width=1\textwidth]{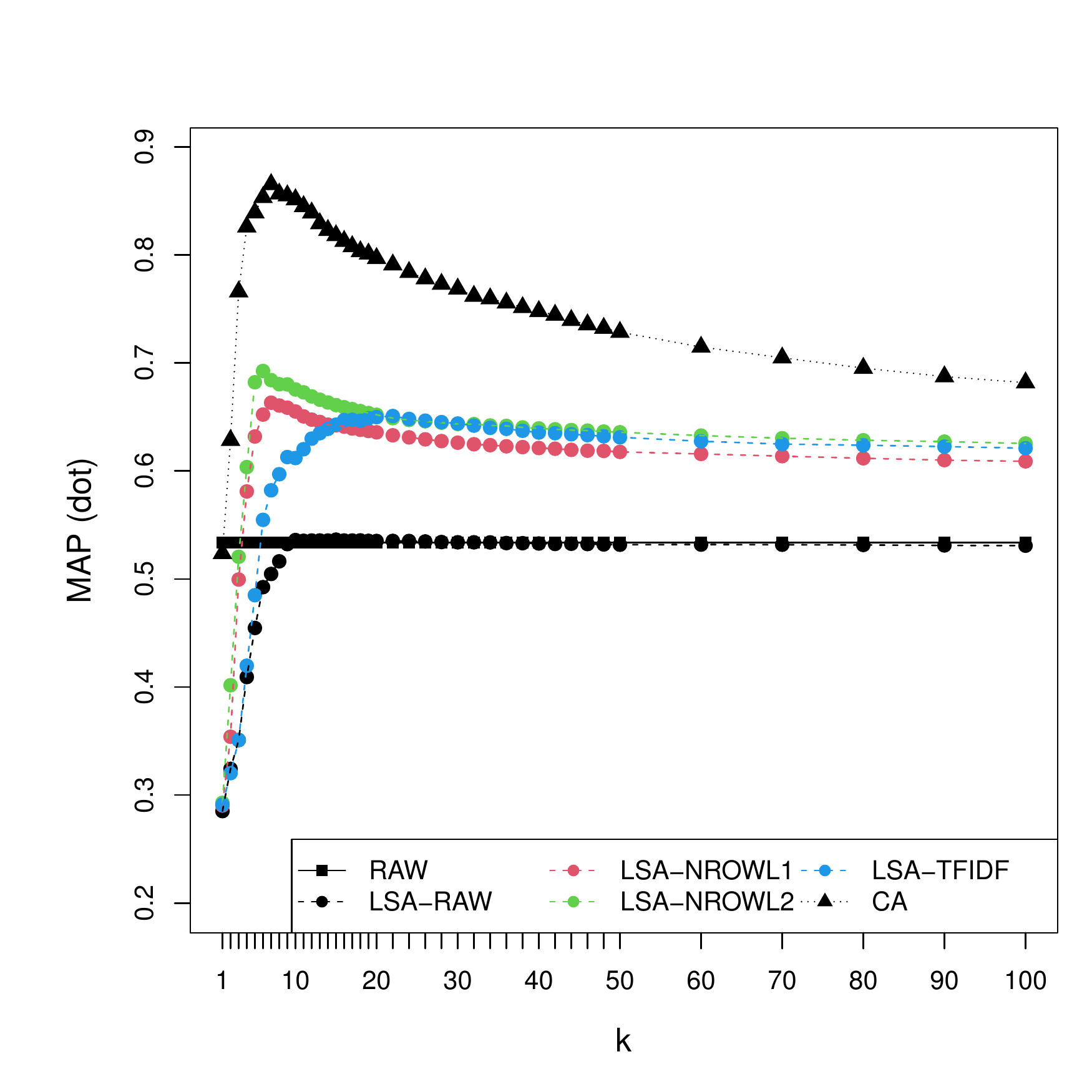}
         \caption{BBCSport}\label{F: dotstandardpBBCsport}
         \end{subfigure}
     \begin{subfigure}[b]{0.45\linewidth}
         \centering
         \includegraphics[width=1\textwidth]{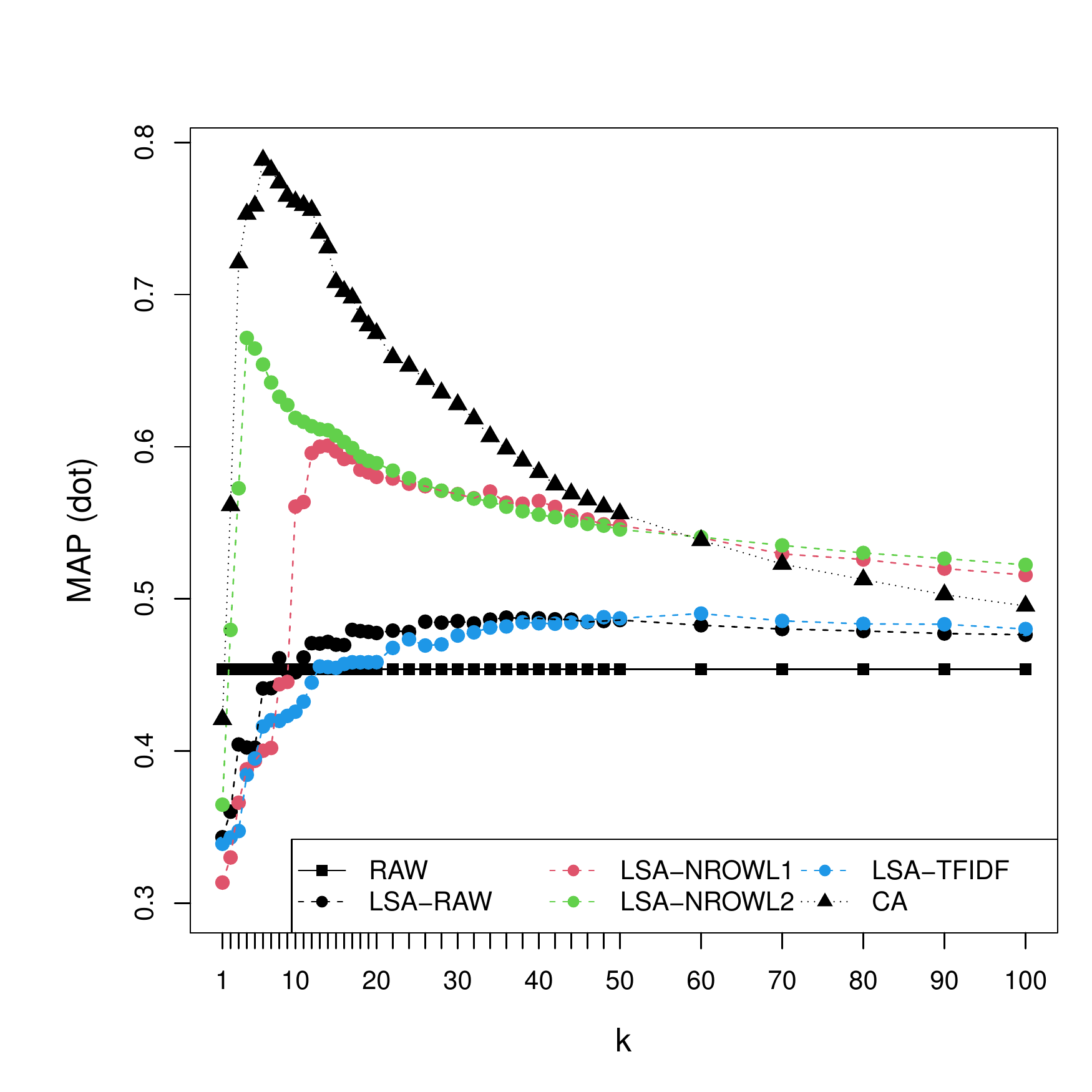}
        \caption{20 Newsgroups}\label{F: dotstandardp20newsgroups}
         \end{subfigure}
      \begin{subfigure}[b]{0.45\linewidth}
         \centering
         \includegraphics[width=1\textwidth]{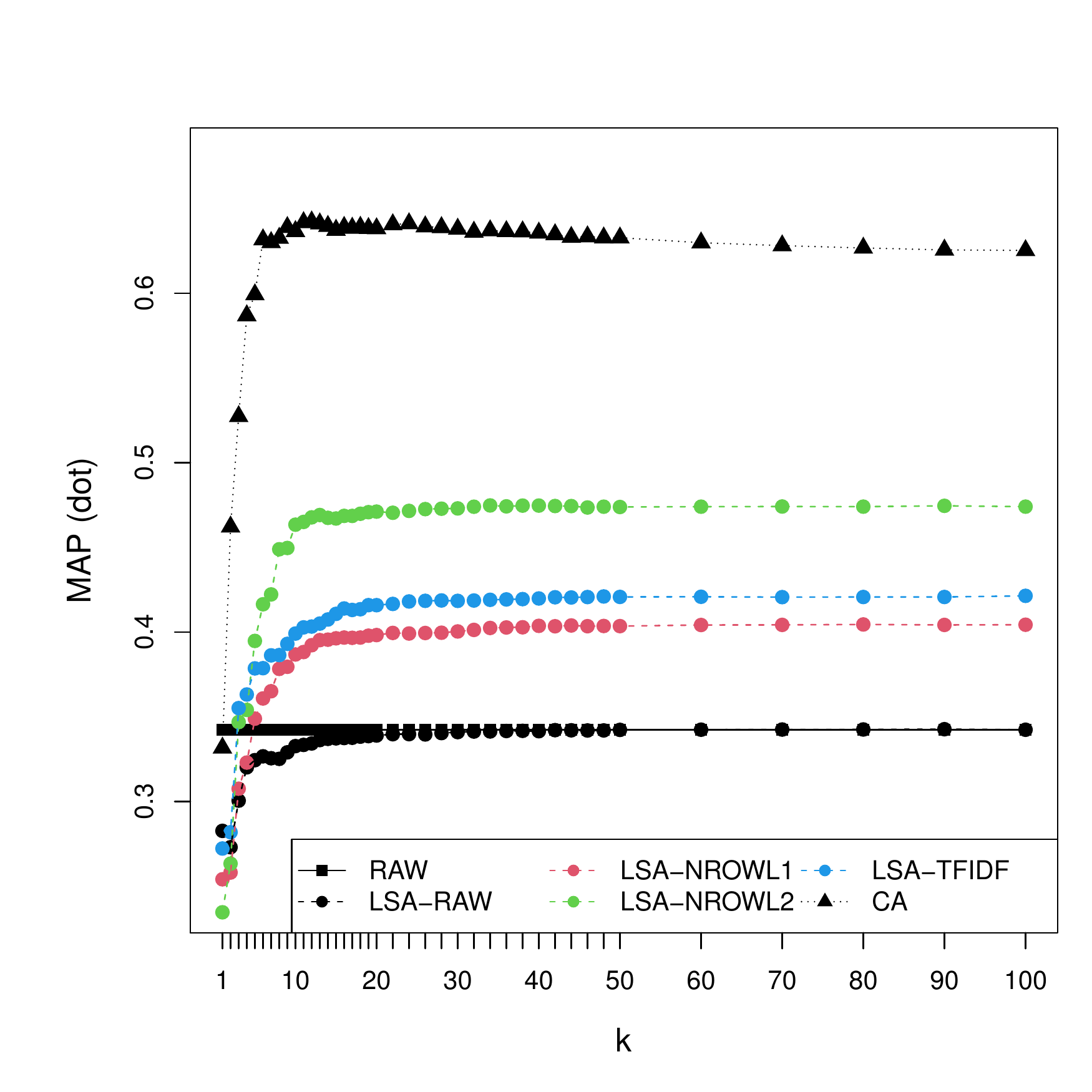}
         \caption{Wilhelmus}\label{F: dotstandardpwilhelmus}
         \end{subfigure}
    \caption{MAP as a function of the number of dimensions under standard coordinates.}
    \label{F: dotstandardp}
\end{figure}

\subsubsection{MAP as a function of the weighting exponent $\alpha$ about LSA and MAP about CA for various values of the number of dimensions}\label{Sub: constantkdot}

\begin{figure}[H]
    \centering
     \begin{subfigure}[b]{0.45\linewidth}
         \centering
         \includegraphics[width=1\textwidth]{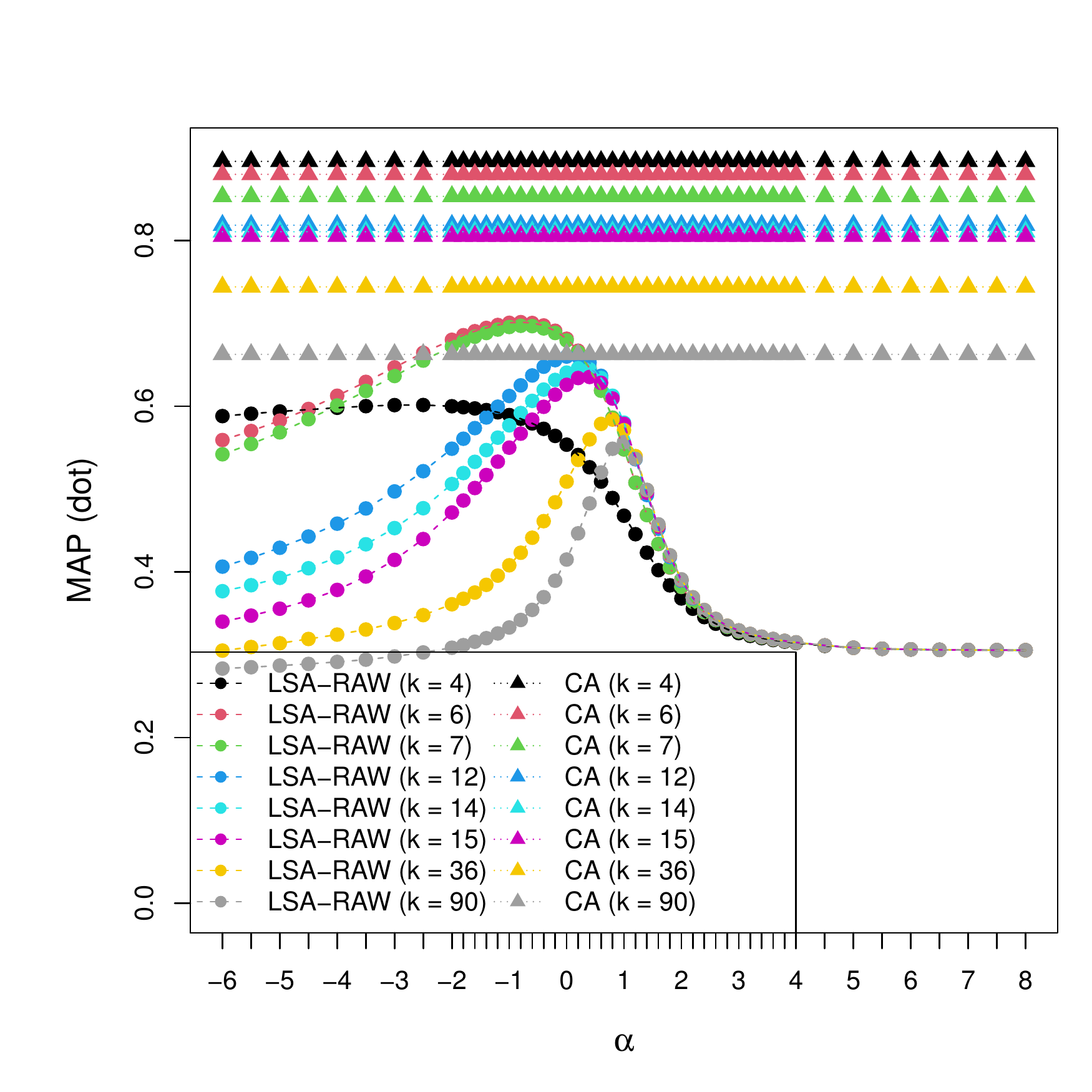}
         \caption{BBCNews}\label{F: dotstandarddBBCnews}
         \end{subfigure}
      \begin{subfigure}[b]{0.45\linewidth}
         \centering
         \includegraphics[width=1\textwidth]{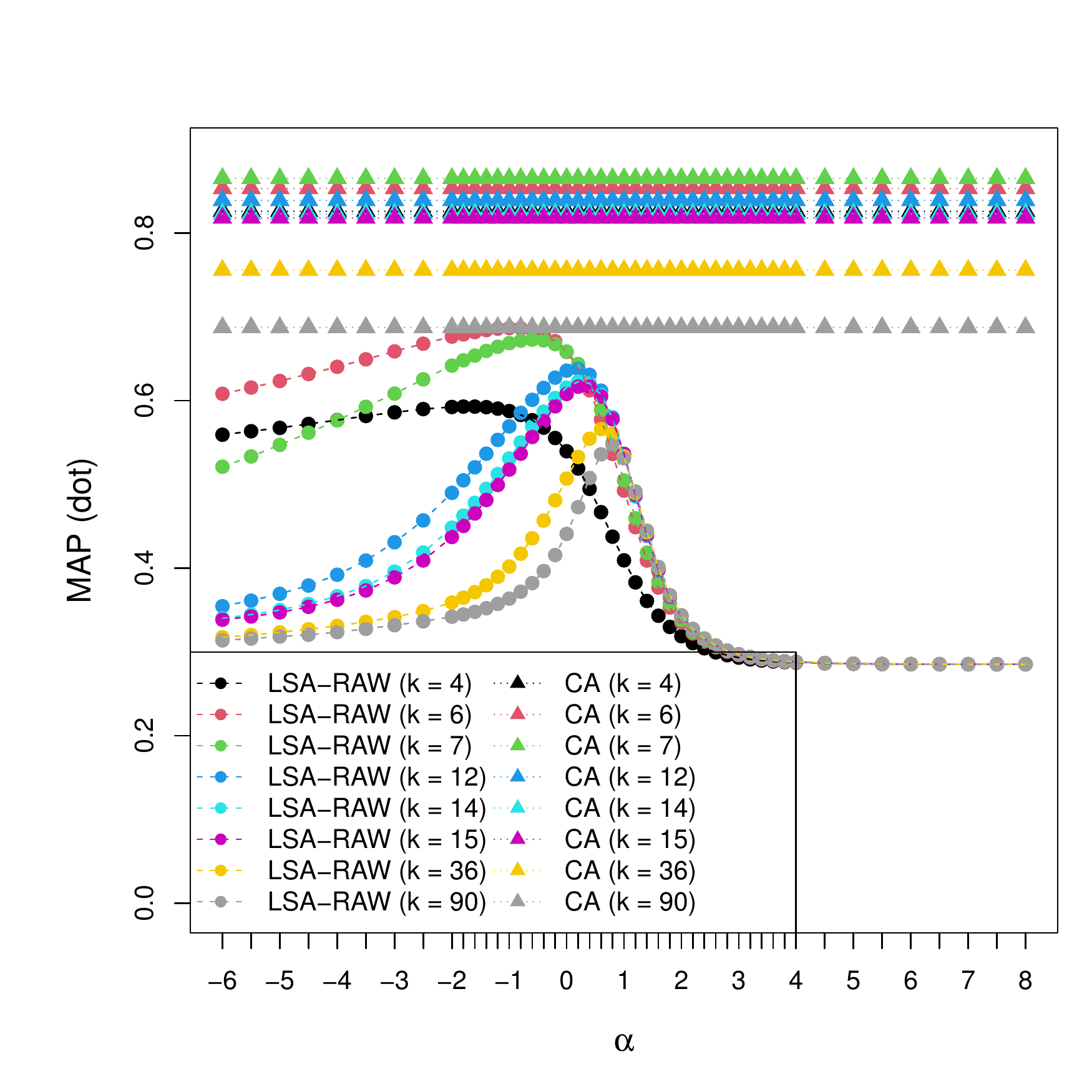}
         \caption{BBCSport}\label{F: dotstandarddBBCsport}
         \end{subfigure}
      \begin{subfigure}[b]{0.45\linewidth}
         \centering
         \includegraphics[width=1\textwidth]{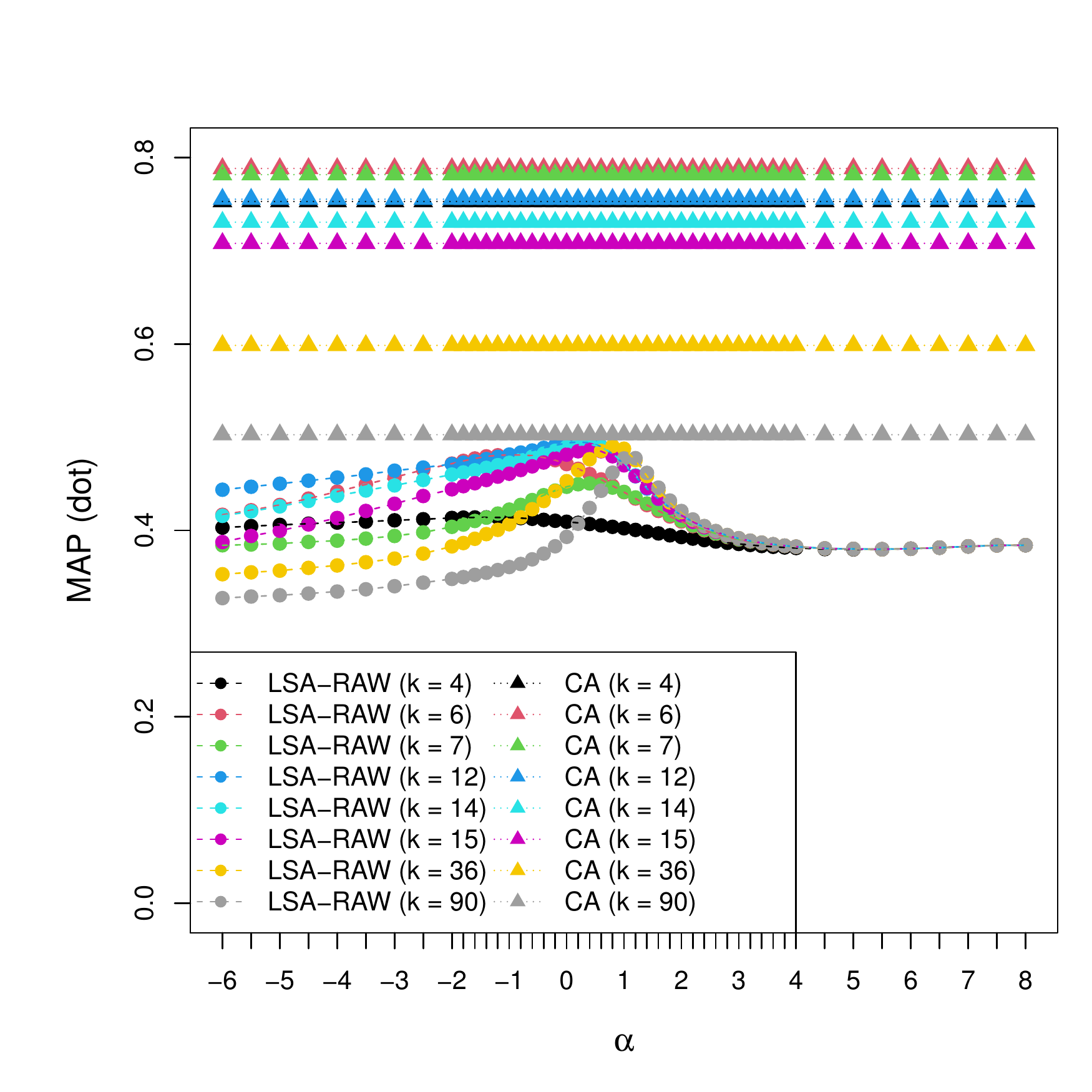}
         \caption{20 Newsgroups}\label{F: dotstandardd20newsgroups}
         \end{subfigure}
      \begin{subfigure}[b]{0.45\linewidth}
         \centering
         \includegraphics[width=1\textwidth]{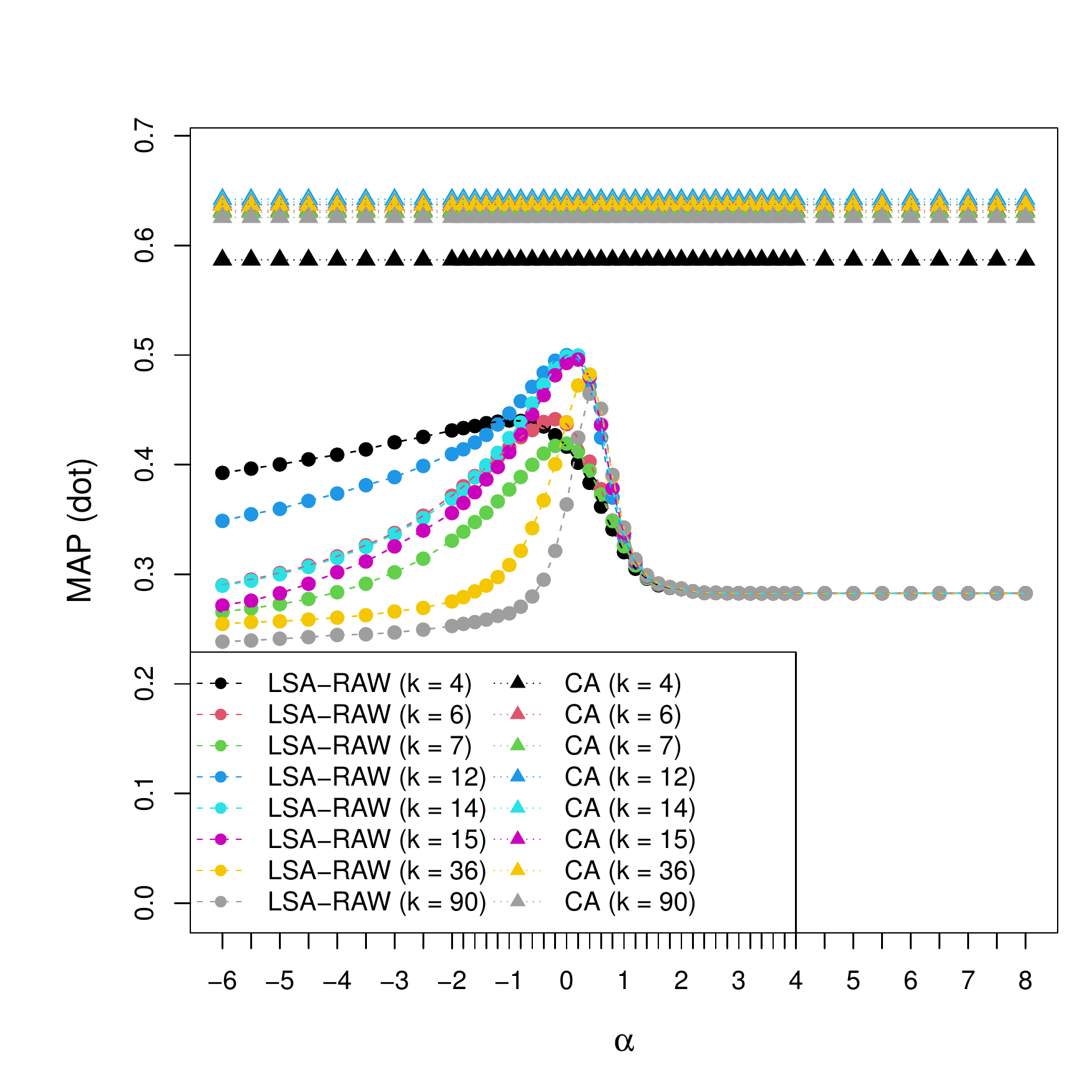}
         \caption{Wilhelmus}\label{F: dotstandarddwilhelmus}
         \end{subfigure}
    \caption{MAP as a function of $\alpha$ for LSA-RAW and MAP for CA under various values of $k$.}
    \label{F: dotstandardd}
\end{figure}

Figure~\ref{F: dotstandardd} shows MAP as a function of $\alpha$ about LSA-RAW and MAP about CA for the number of dimensions: $k = 4, 6, 7, 12, 14, 15, 36, \text{and }90$.

\begin{table}[H]
\centering  
\caption{MAP with the optimal weighting exponent $\alpha$ for LSA-RAW and MAP for CA under $k = 4, 6, 7, 12, 14, 15, 36, \text{and }90$ about dot similarity. Bold values are best.} 
\label{TIRdotdim10}
\begin{tabular}{ccccccccc}    
&\multicolumn{2}{c}{BBCNews} &\multicolumn{2}{c}{BBCSport}&\multicolumn{2}{c}{20 Newsgroups}&\multicolumn{2}{c}{Wilhelmus} \\
&$\alpha$&MAP&$\alpha$&MAP&$\alpha$&MAP&$\alpha$&MAP\\
\hline 
LSA-RAW ($k = 4$) &-2.5& 0.601&-1.8&  0.593& -1.4&0.414 & -1 &0.440 \\
LSA-RAW ($k = 6$) &-0.8& 0.701&-1& 0.687& -1&0.481 & -0.2 &0.441 \\
LSA-RAW ($k = 7$) &-0.8& 0.697&-0.6& 0.673&0.4&0.450 & 0&0.419 \\
LSA-RAW ($k = 12$) &0& 0.660&0.2&0.638& 0.2&0.495 & 0 &0.500\\
LSA-RAW ($k = 14$) &0.2&0.646&0.2& 0.623& 0.4& 0.492 & 0.2 &0.500 \\
LSA-RAW ($k = 15$) &0.4&0.635&0.4& 0.617& 0.4& 0.486 & 0.2 &0.496 \\
LSA-RAW ($k = 36$) &0.8&0.584& 0.6&  0.566& 0.8& 0.490 & 0.4 &0.482 \\
LSA-RAW ($k = 90$) &1&0.556&0.8& 0.547& 1.2& 0.478 & 0.4 &0.465 \\
CA ($k = 4$) & &  \textbf{0.896}& & 0.826&  &0.753 &   &0.587 \\
CA ($k = 6$) & &  0.880& & 0.853&  &\textbf{0.788} &   &0.632 \\
CA ($k = 7$) & &  0.853& & \textbf{0.865}&  &0.782 &   &0.630 \\
CA ($k = 12$) & & 0.818& & 0.839&  &0.756 &   &\textbf{0.642}\\
CA ($k = 14$) & & 0.811& & 0.823&  &0.731 &  & 0.640\\
CA ($k = 15$) & & 0.805& & 0.818&  &0.708 &  & 0.637\\
CA ($k = 36$) & & 0.744& & 0.756&  &0.599&  & 0.637\\
CA ($k = 90$) & & 0.663& & 0.687&  & 0.503 &  & 0.626\\
\hline 
\end{tabular}  
\end{table} 

\subsection{Improving performance of CA for information retrieval}\label{Sub: improvingcadot}

\subsubsection{Weighting scheme of raw document-term matrix for CA}\label{Subsub: caweightingschemedot}

\begin{figure}[H]
    \centering
       \begin{subfigure}[b]{0.45\linewidth}
         \centering
         \includegraphics[width=1\textwidth]{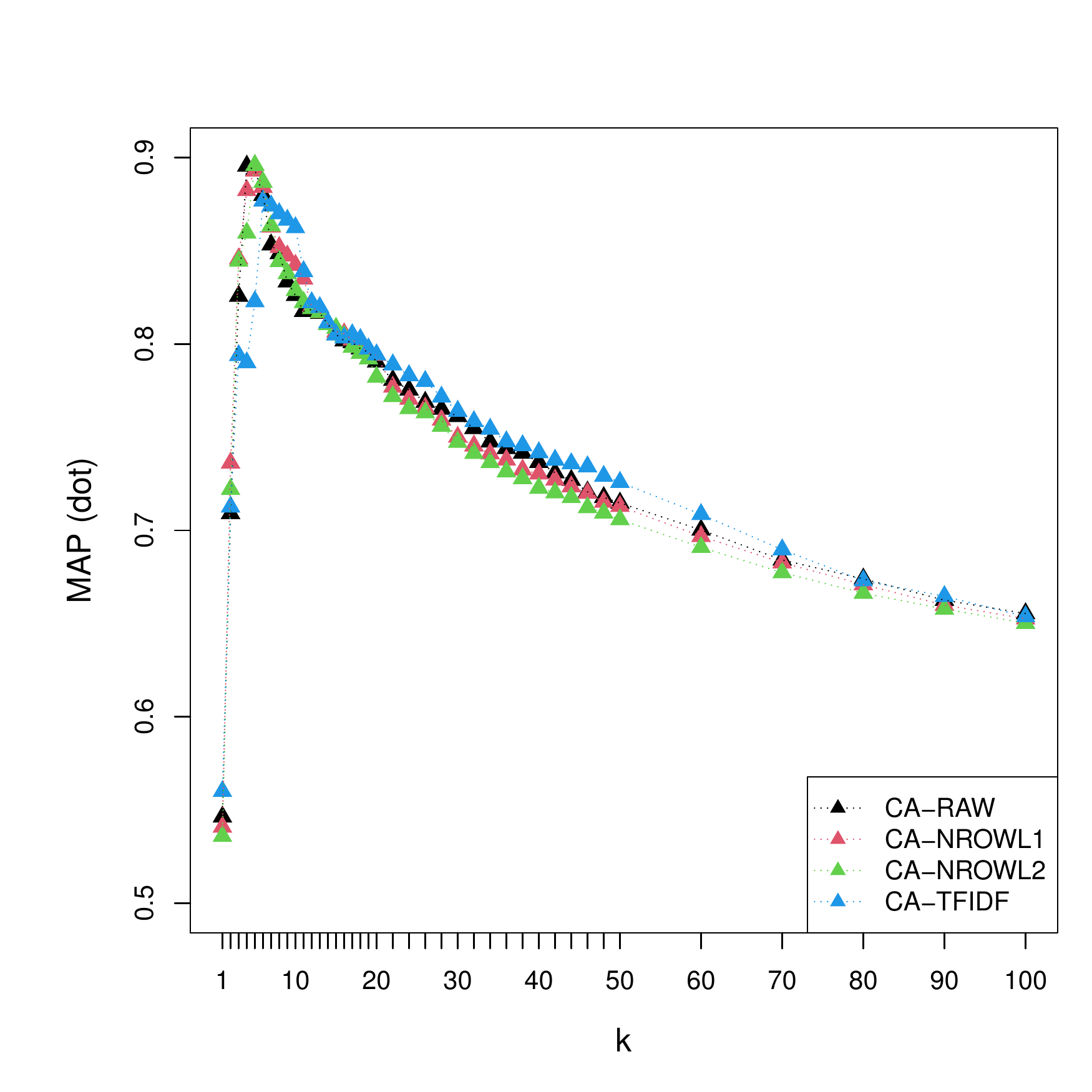}
         \caption{$\alpha = 1$ (BBCNews)}\label{F: dotstandardpBBCnewscaweisch}
         \end{subfigure}
      \begin{subfigure}[b]{0.45\linewidth}
         \centering
         \includegraphics[width=1\textwidth]{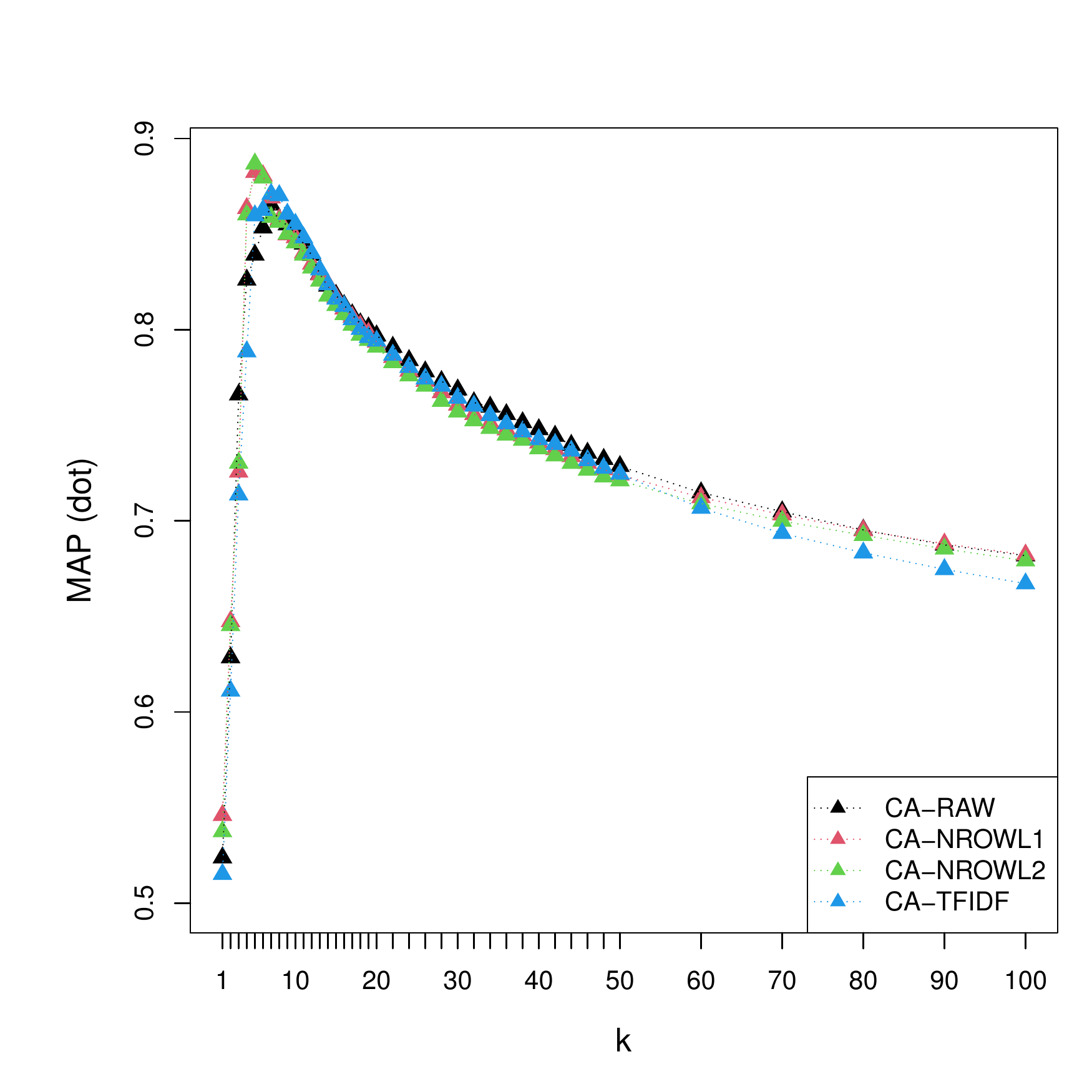}
         \caption{$\alpha = 1$ (BBCSport)}\label{F: dotstandardpBBCsportcaweisch}
         \end{subfigure}
     \begin{subfigure}[b]{0.45\linewidth}
         \centering
         \includegraphics[width=1\textwidth]{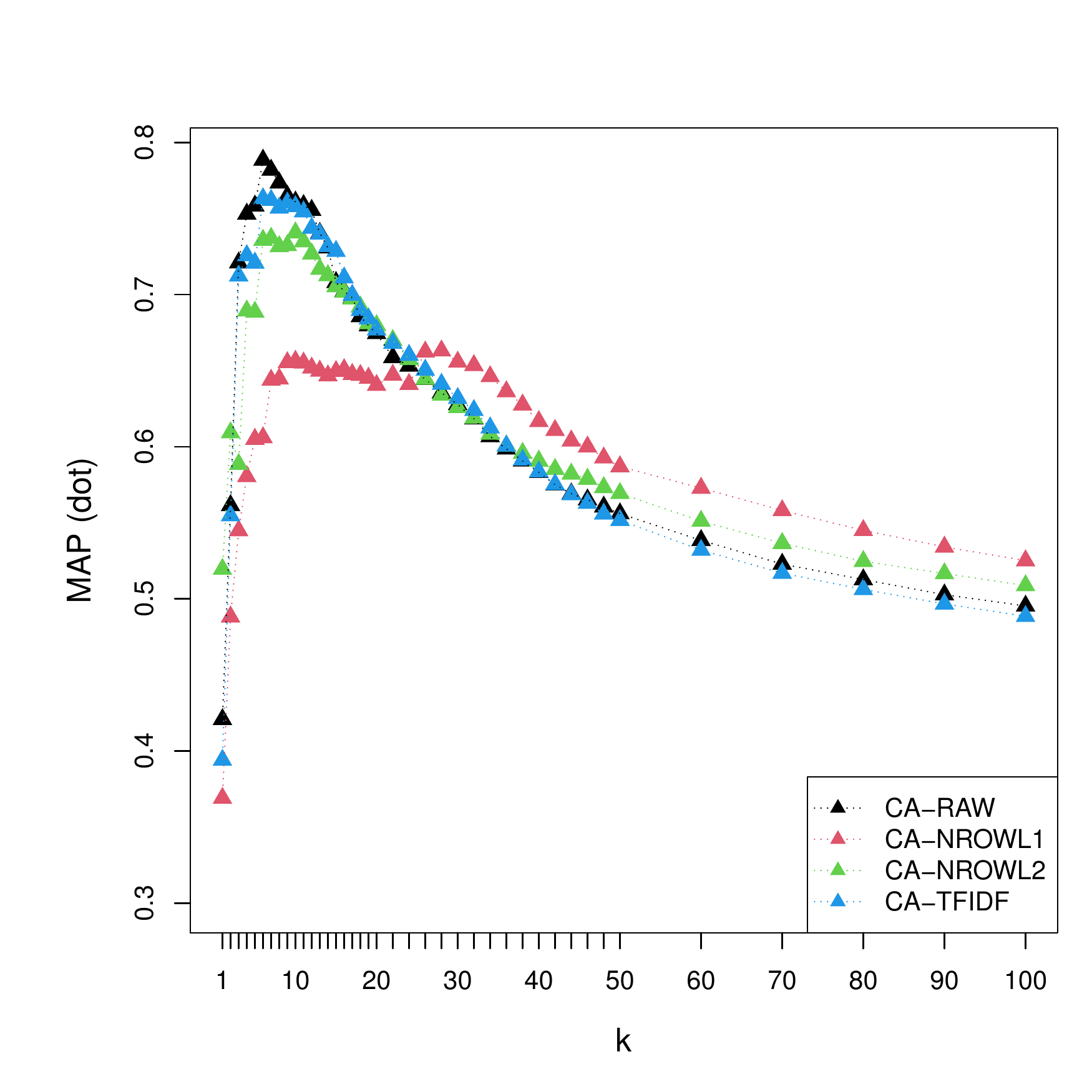}
         \caption{$\alpha = 1$ (20 Newsgroups)}\label{F: dotstandardp20newsgroupscaweisch}
         \end{subfigure}
      \begin{subfigure}[b]{0.45\linewidth}
         \centering
         \includegraphics[width=1\textwidth]{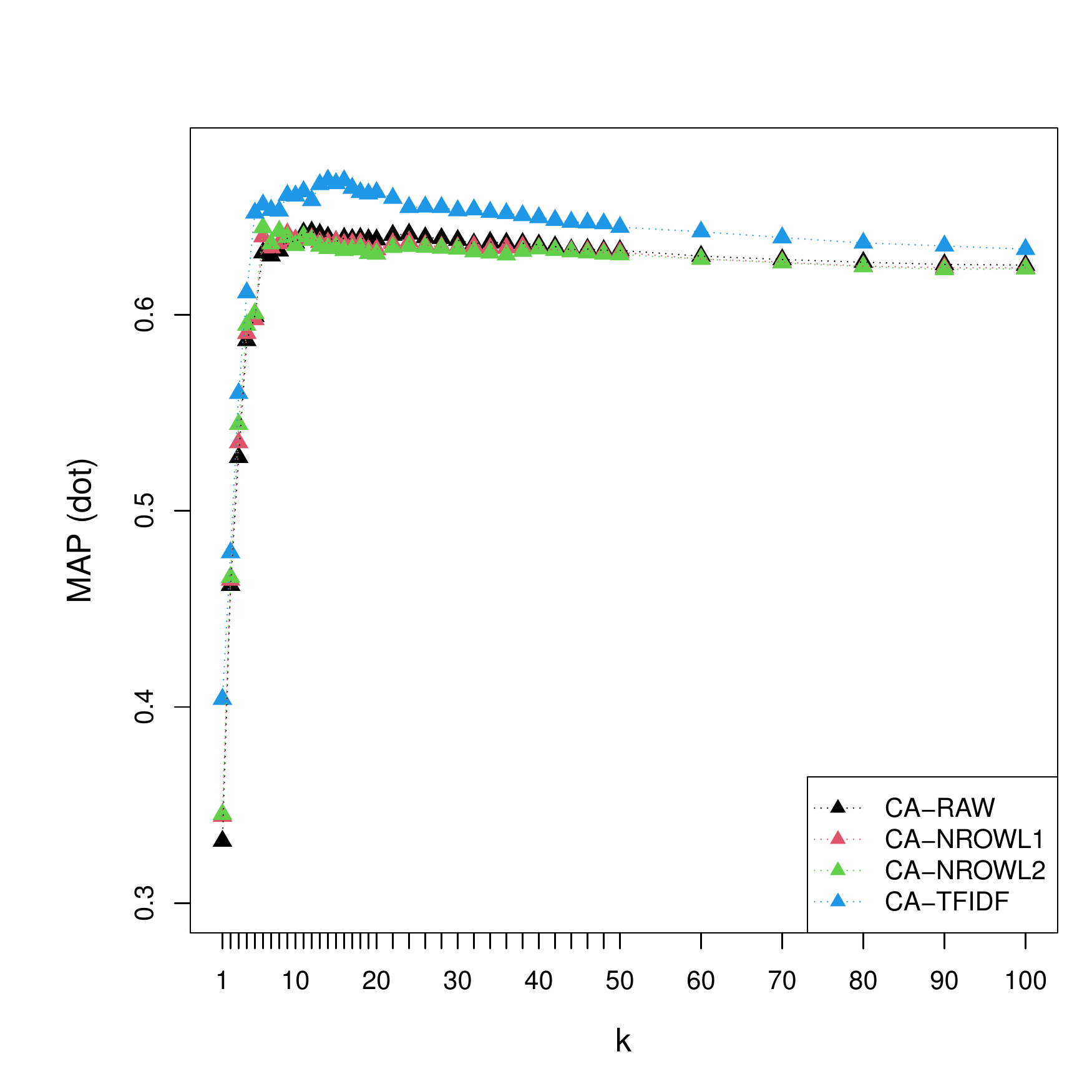}
         \caption{$\alpha = 1$ (Wilhelmus)}\label{F: dotstandardpwilhelmuscaweisch}
         \end{subfigure}
    \caption{MAP as a function of the number of dimensions $k$ for the four versions of CA under standard coordinates.}
    \label{F: dotstandardpcaweisch}
\end{figure}

\begin{table}[H]
\centering  
\caption{MAP with the optimal number of dimensions $k$ for the four versions of CA about dot similarity. Bold values are best.} 
\label{TIRdotalpha1caweisch}
\begin{tabular}{lcccccccc}    
&\multicolumn{2}{c}{BBCNews} &\multicolumn{2}{c}{BBCSport}&\multicolumn{2}{c}{20 Newsgroups}&\multicolumn{2}{c}{Wilhelmus} \\
&$k$&MAP&$k$&MAP&$k$&MAP&$k$&MAP\\
\hline 
CA-RAW&4&\textbf{0.896}&7&0.865& 6&\textbf{0.788}&12&0.642\\
CA-NROWL1&5&0.893&5&0.882&28&0.663&9&0.641\\
CA-NROWL2&5&\textbf{0.896}&5&\textbf{0.887}&10&0.741&6&0.644\\
CA-TFIDF&6&0.877&7&0.871&6&0.763&14&\textbf{0.669}\\
\hline 
\end{tabular}  
\end{table} 

\subsubsection{Weighting exponent $\alpha$ in CA}\label{Subsub: caweightingexponentdot}

\begin{figure}[H]
    \centering
     \begin{subfigure}[b]{0.45\linewidth}
         \centering
         \includegraphics[width=1\textwidth]{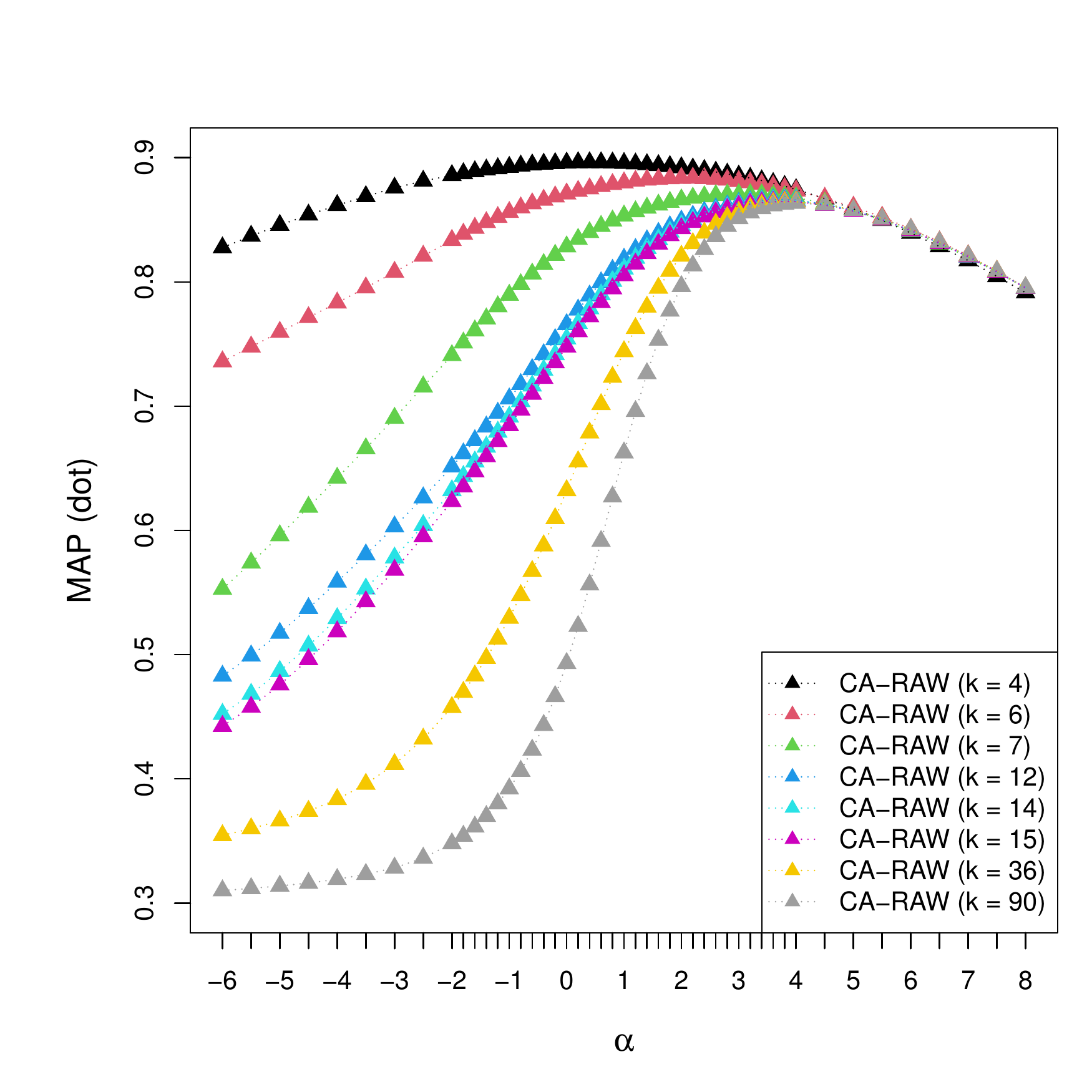}
         \caption{BBCNews}\label{F: dotstandarddBBCnewscaweiexp}
         \end{subfigure}
      \begin{subfigure}[b]{0.45\linewidth}
         \centering
         \includegraphics[width=1\textwidth]{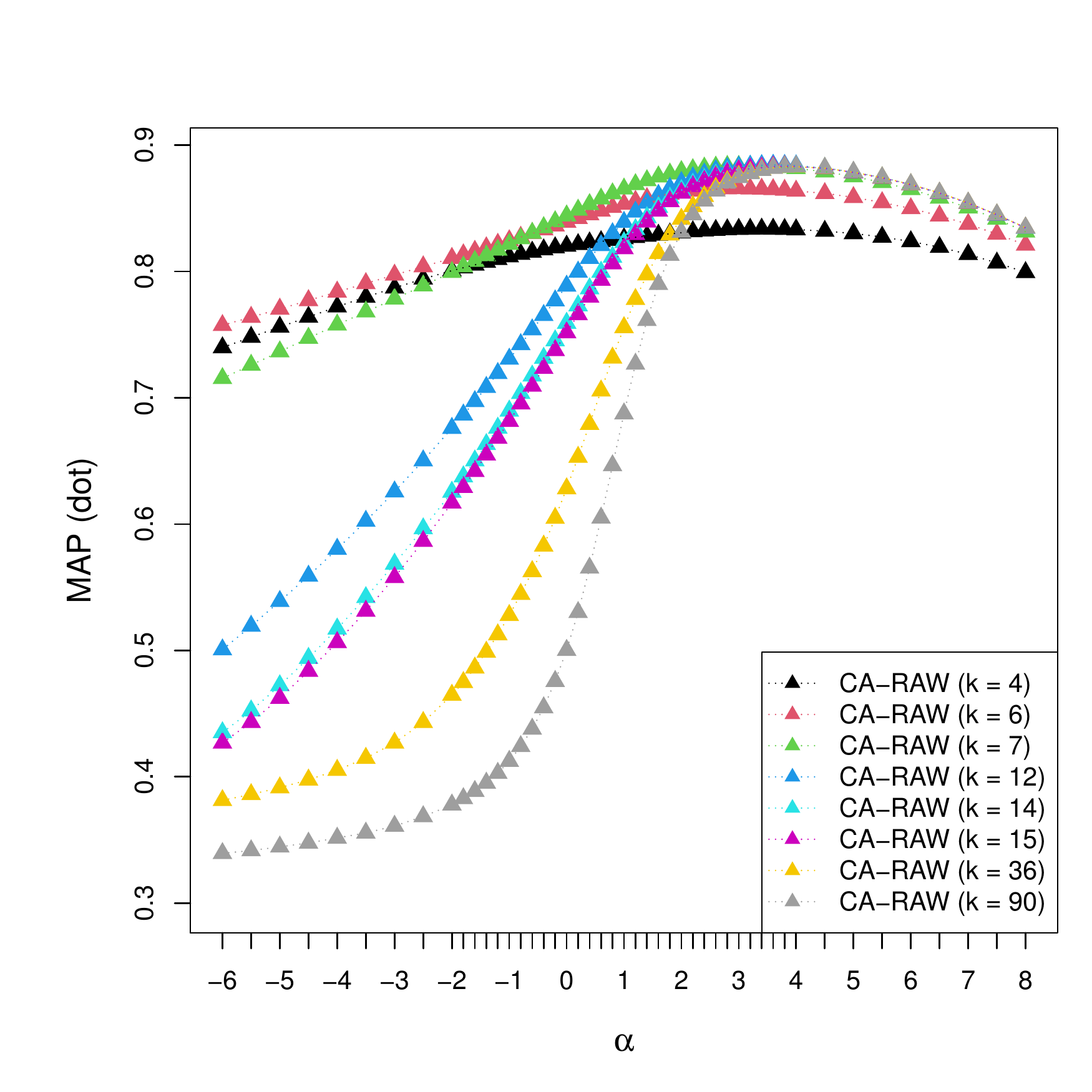}
         \caption{BBCSport}\label{F: dotstandarddBBCsportcaweiexp}
         \end{subfigure}
      \begin{subfigure}[b]{0.45\linewidth}
         \centering
         \includegraphics[width=1\textwidth]{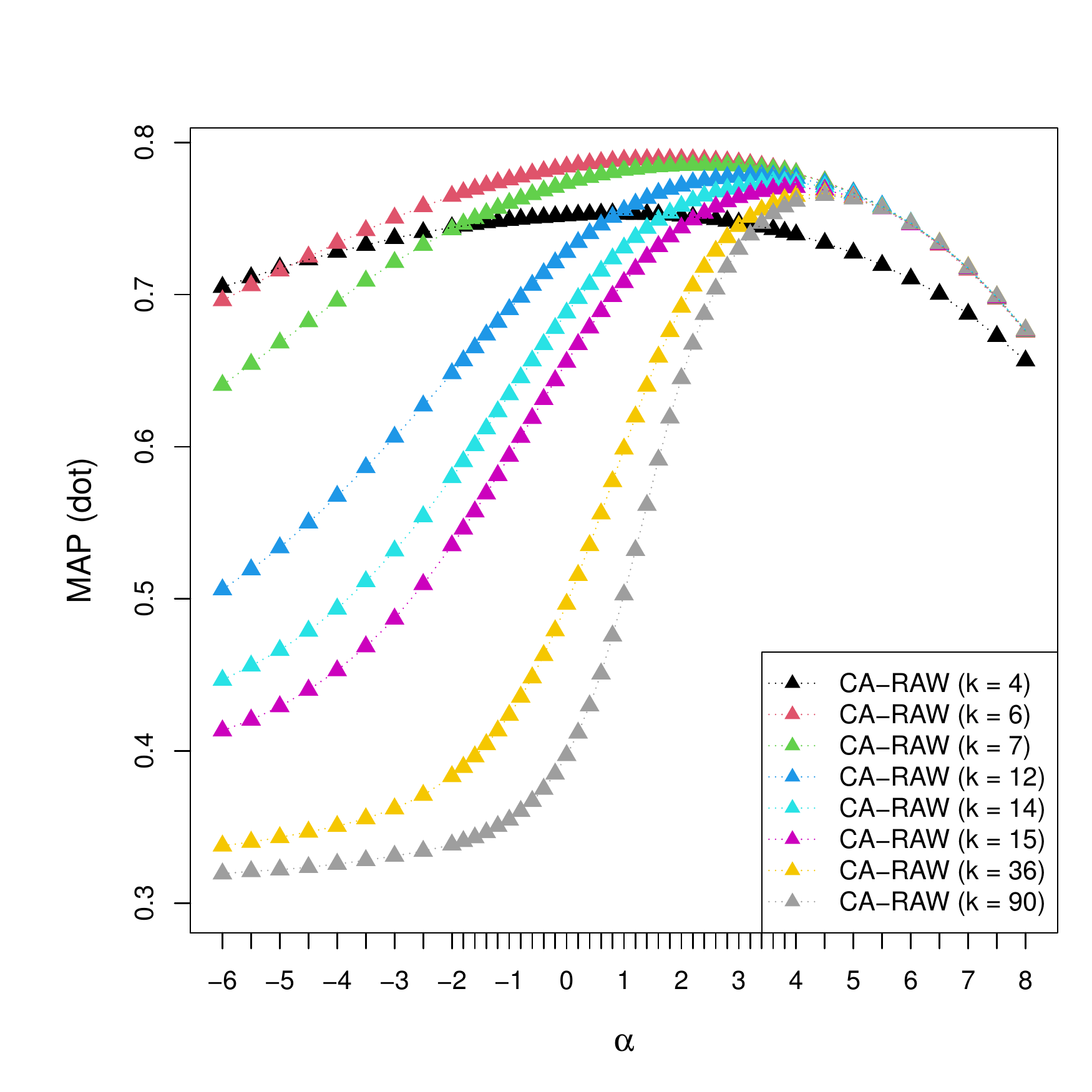}
         \caption{20 Newsgroups}\label{F: dotstandardd20newsgroupscaweiexp}
         \end{subfigure}
      \begin{subfigure}[b]{0.45\linewidth}
         \centering
         \includegraphics[width=1\textwidth]{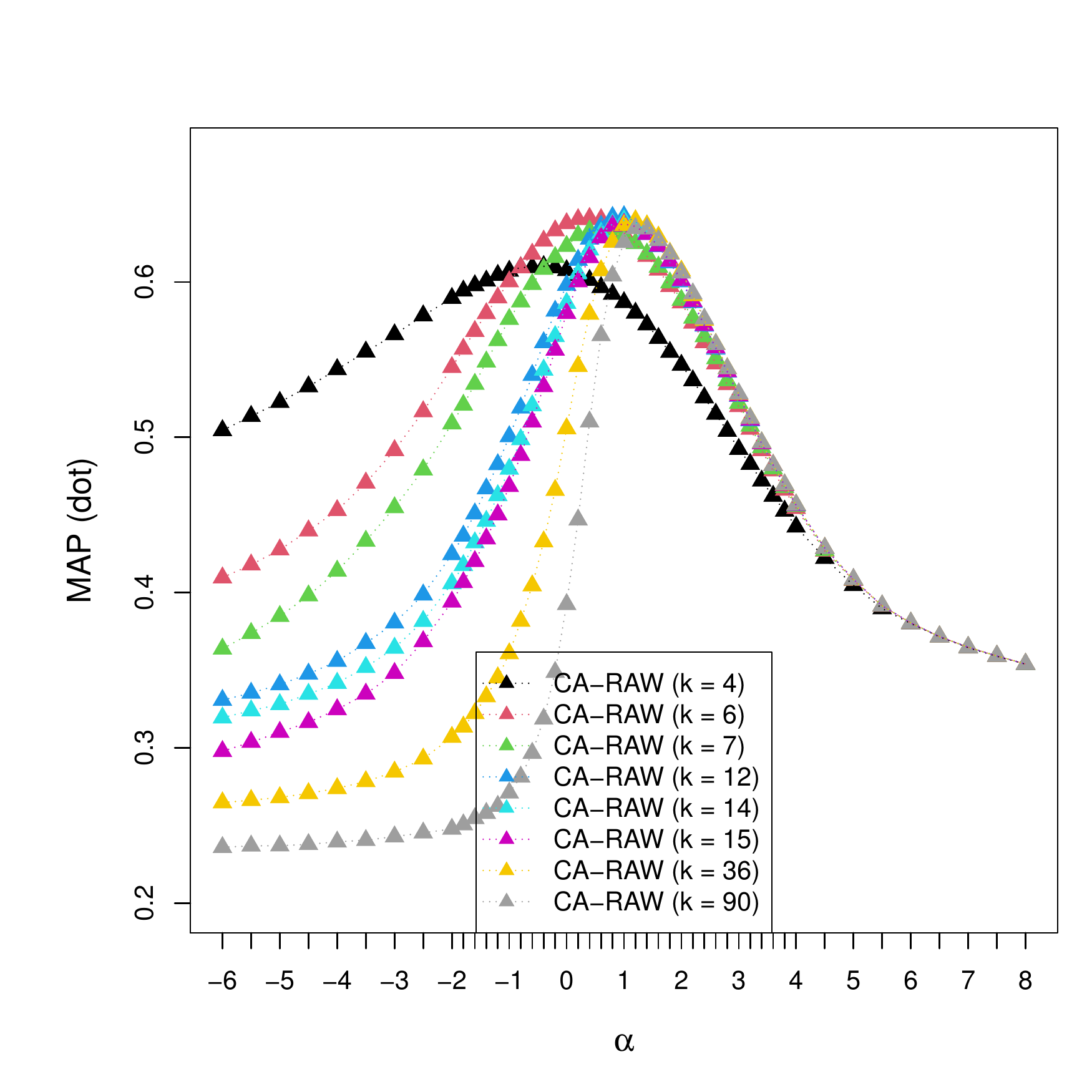}
         \caption{Wilhelmus}\label{F: dotstandarddwilhelmuscaweiexp}
         \end{subfigure}
    \caption{ MAP as a function of $\alpha$ for CA-RAW under various values of $k$.}
    \label{F: dotstandarddcaweiexp}
\end{figure}

\begin{table}[H]
\centering  
\caption{MAP with the optimal $\alpha$ for CA-RAW under $k = 4, 6, 7, 12, 14, 15, 36, \text{and }90$ about dot similarity. Bold values are best.} 
\label{TIRdotdim10caweiexp}
\begin{tabular}{ccccccccc}    
&\multicolumn{2}{c}{BBCNews} &\multicolumn{2}{c}{BBCSport}&\multicolumn{2}{c}{20 Newsgroups}&\multicolumn{2}{c}{Wilhelmus} \\
&$\alpha$&MAP&$\alpha$&MAP&$\alpha$&MAP&$\alpha$&MAP\\
\hline 
CA ($k = 4$) &0.4 &  \textbf{0.896}&3.4 & 0.834& 0.8 &0.753 &  -0.4 &0.610 \\
CA ($k = 6$) & 2.2&   0.884&3 &0.866& 1.6 &\textbf{0.789} &  0.4 &0.641 \\
CA ($k = 7$) & 3.2&  0.870&3 & 0.883&2.4  &0.785 & 0.6  &0.635 \\
CA ($k = 12$) & 3.8& 0.866&3.6 & \textbf{0.884}& 3.4 & 0.779& 1  &\textbf{0.642}\\
CA ($k = 14$) & 3.8& 0.866&3.8 & 0.883&3.8  & 0.774&  1& 0.640\\
CA ($k = 15$) &3.8 & 0.865&3.8 & 0.883&4  &0.771 & 1 & 0.637\\
CA ($k = 36$) & 4& 0.864&4 &0.883& 4.5 &0.767 &1.2  &0.640 \\
CA ($k = 90$) &4 & 0.863&4 & 0.883& 4.5 &0.765 & 1.2 &0.635 \\
\hline 
\end{tabular}  
\end{table} 

\subsection{Exploring MAP as a function of $\alpha$ under the optimal number of dimensions for LSA and CA}\label{Sub: optimalalphadot}

\begin{figure}[H]
    \centering
       \begin{subfigure}[b]{0.45\linewidth}
         \centering
         \includegraphics[width=1\textwidth]{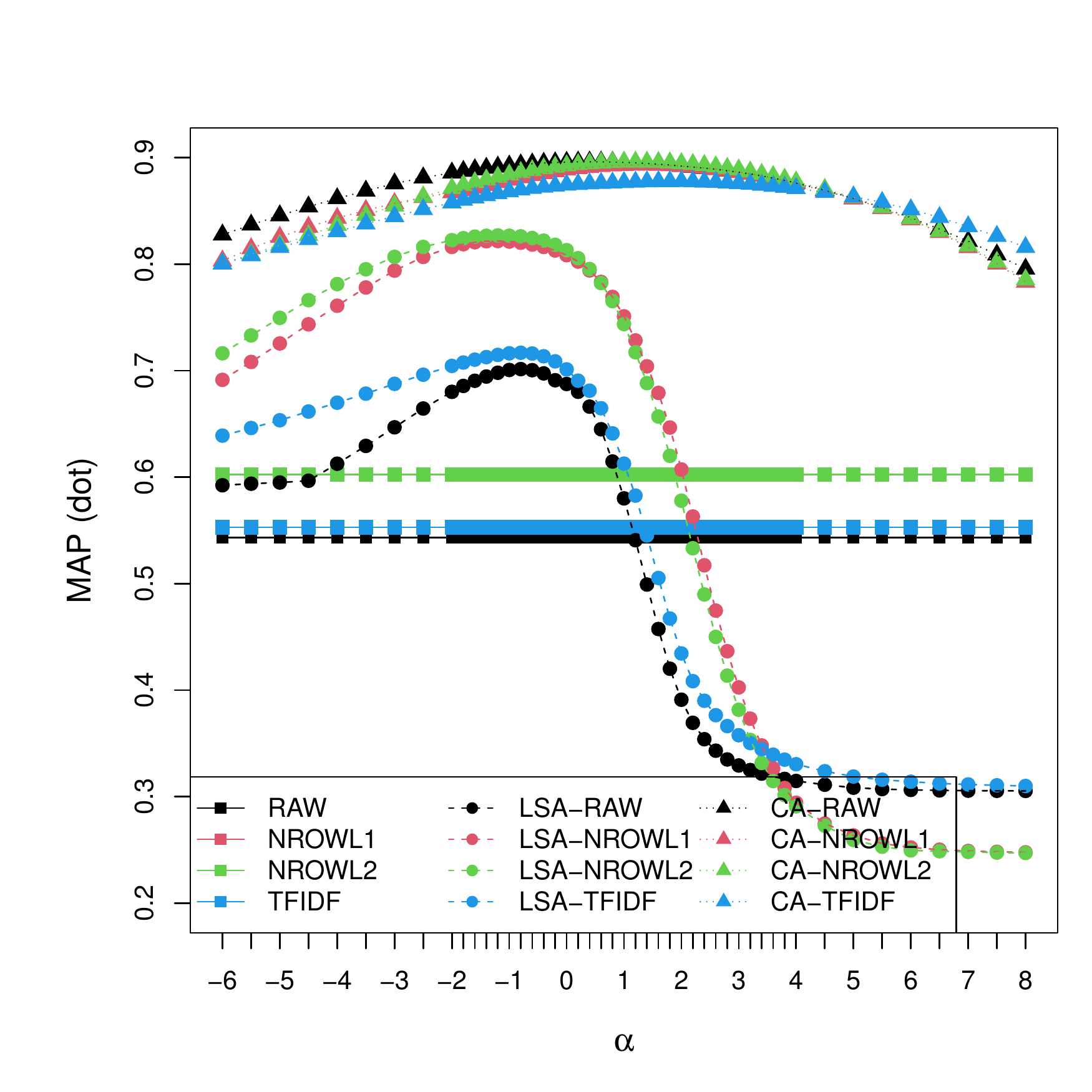}
         \caption{BBCNews}\label{F: dotoptimalMAPBBCnews}
         \end{subfigure}
      \begin{subfigure}[b]{0.45\linewidth}
         \centering
         \includegraphics[width=1\textwidth]{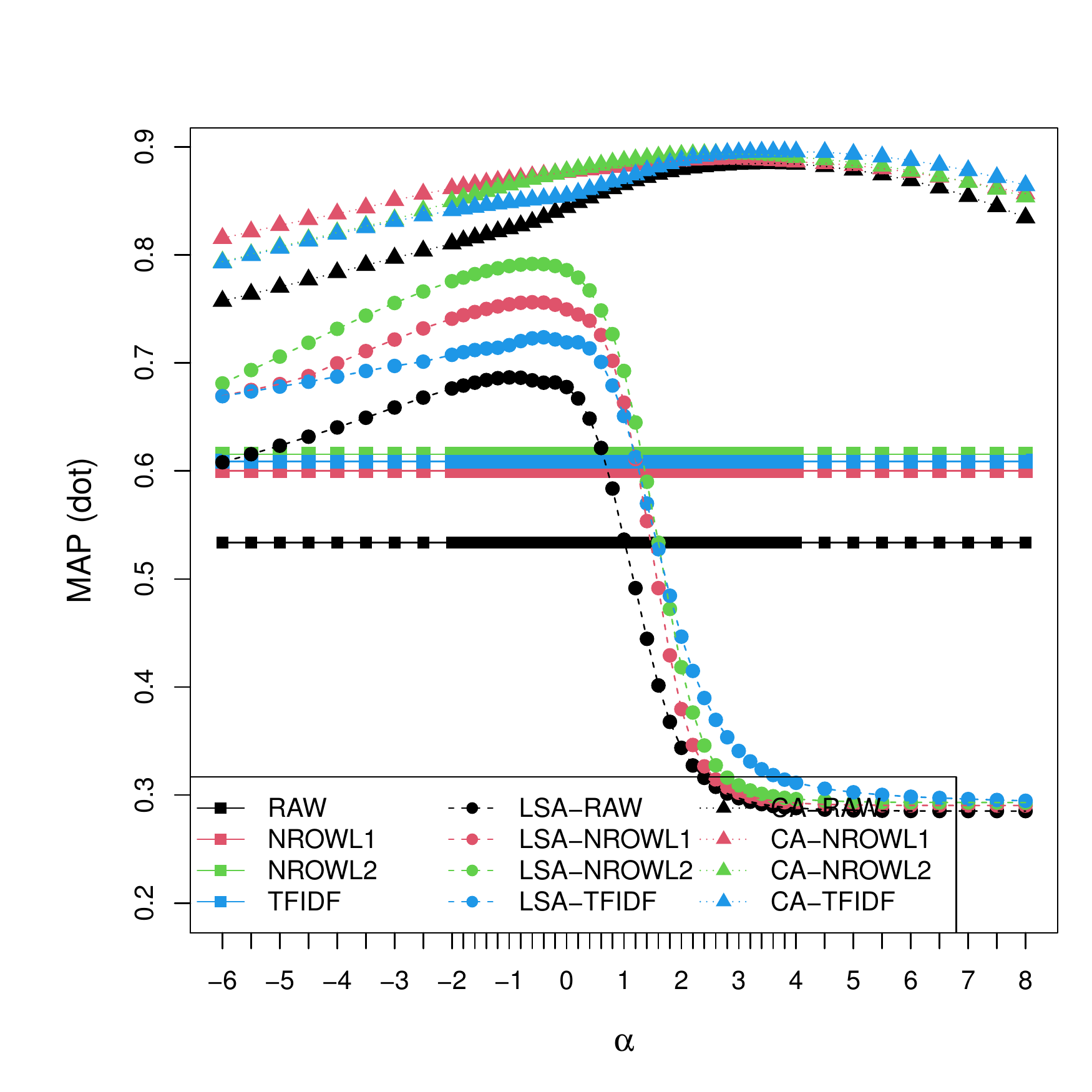}
         \caption{BBCSport}\label{F: dotoptimalMAPBBCsport}
         \end{subfigure}
        \begin{subfigure}[b]{0.45\linewidth}
         \centering
         \includegraphics[width=1\textwidth]{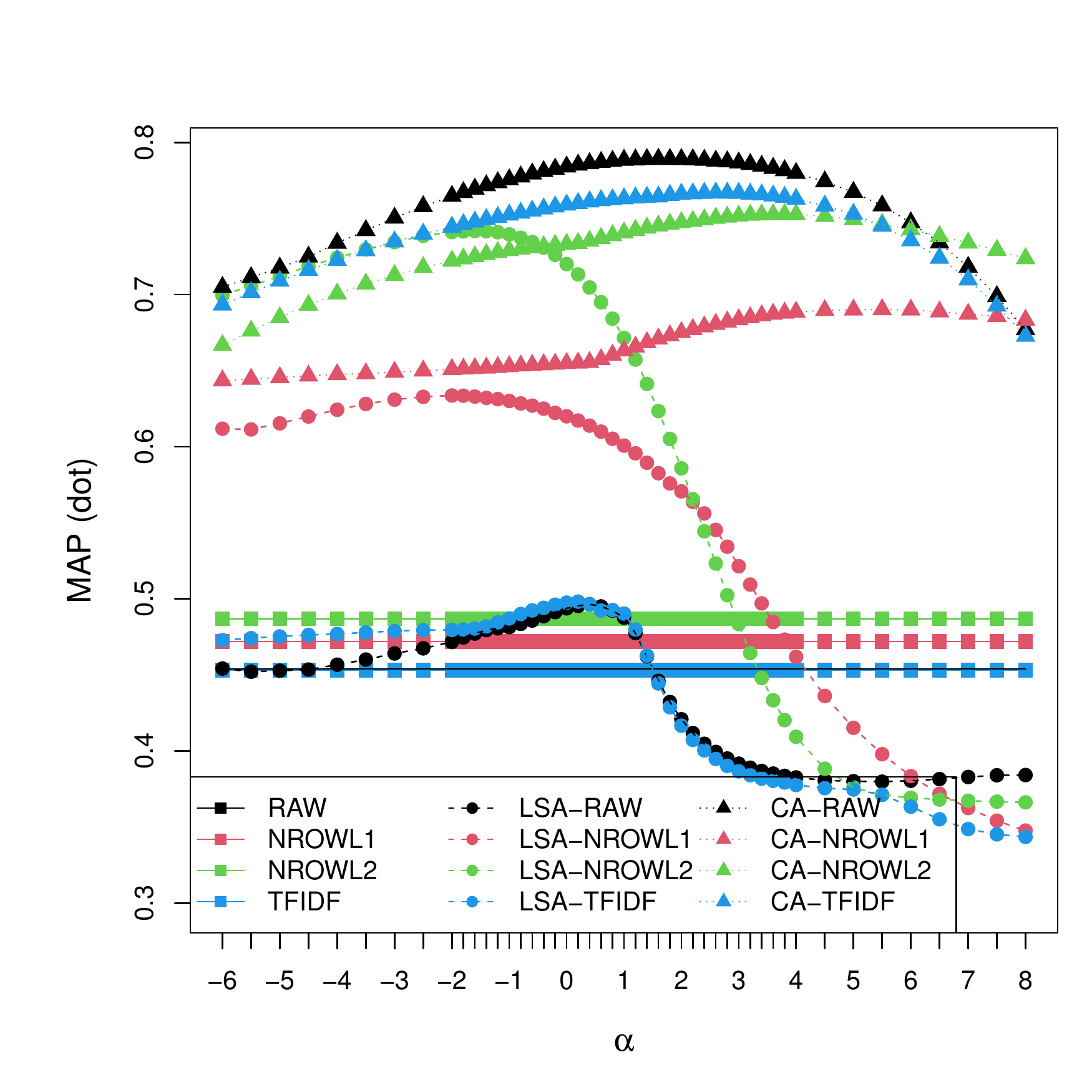}
         \caption{20 Newsgroups}\label{F: dotoptimalMAP20newsgroups}
         \end{subfigure}
        \begin{subfigure}[b]{0.45\linewidth}
         \centering
         \includegraphics[width=1\textwidth]{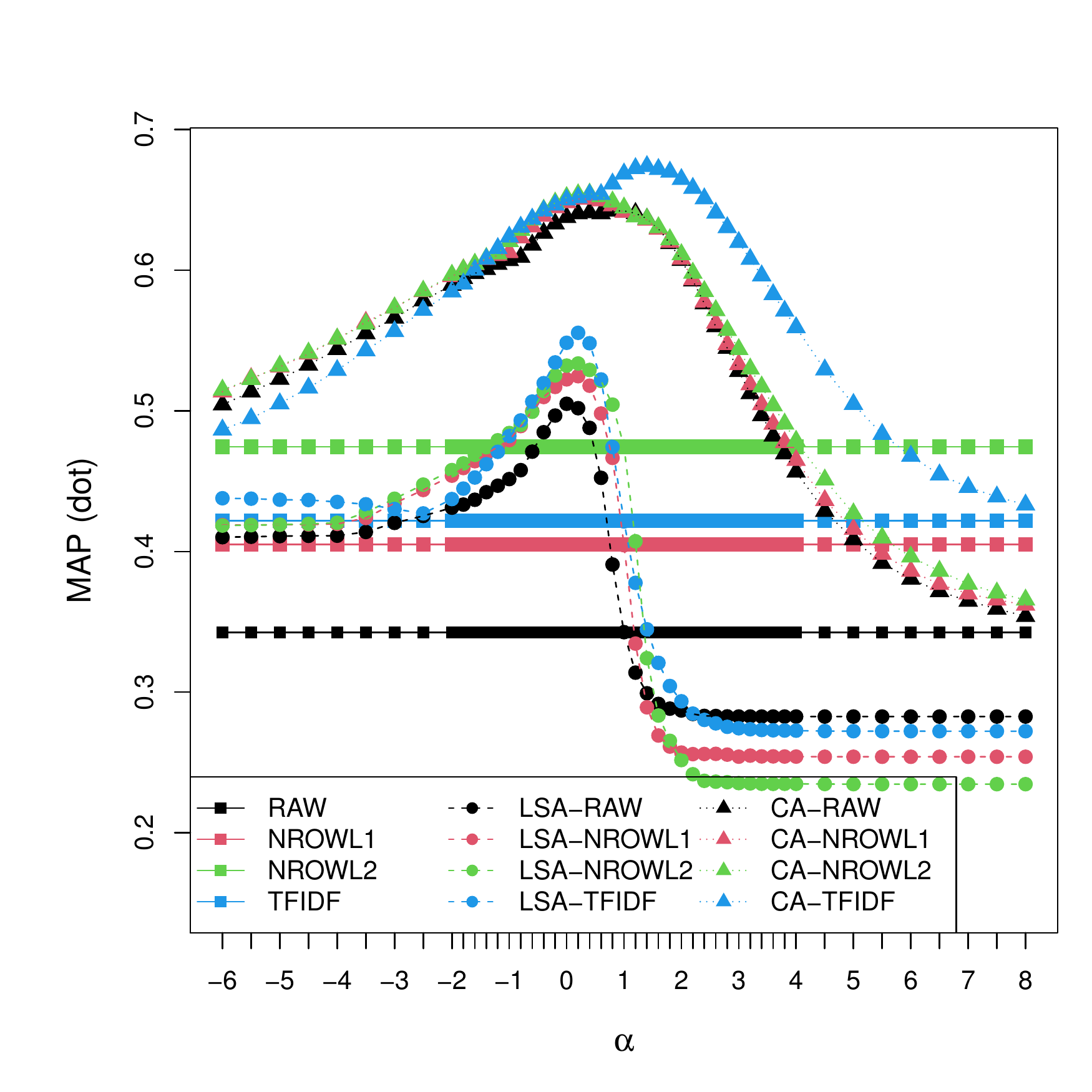}
         \caption{Wilhelmus}\label{F: dotoptimalMAPwilhelmus}
         \end{subfigure}
    \caption{MAP as a function of $\alpha$ under optimal dimension.}
    \label{F: dotoptimalMAP}
\end{figure}

\begin{table}[H]
\centering  
\caption{MAP under the optimal $\alpha$ and optimal dimension $k$ about dot similarity. Bold values are best within group; underlined values are best overall.} 
\label{TIRdot}
\begin{tabular}{ccccccccccccc}    
&\multicolumn{3}{c}{BBCNews} &\multicolumn{3}{c}{BBCSport}&\multicolumn{3}{c}{20 Newsgroups}&\multicolumn{3}{c}{Wilhelmus} \\
&$\alpha$&$k$&MAP&$\alpha$&$k$&MAP&$\alpha$&$k$&MAP&$\alpha$&$k$&MAP\\
\hline 
RAW & &&0.543&& &0.534& &&0.454&&&0.342\\
LSA-RAW &-0.8 &6 & 0.701& -1&6 &  0.687 &0.4 &17 & 0.496&0&13& 0.505\\
CA-RAW&0.4&4&\underline{\textbf{0.896}}&3.6&10&\textbf{0.885}&1.6&6&\underline{\textbf{0.789}}&1&12&\textbf{0.642}\\
\hdashline 
NROWL1& &&0.603 & &&0.600 & &&0.472 &&&0.405 \\
LSA-NROWL1 &-1.2&5&0.822&-0.6&5&0.756&-2&12&0.634&0.2&11& 0.525\\
CA-NROWL1&1.4&5&\textbf{0.893}&2.8&6&\textbf{0.889}&5.5&32&\textbf{0.690}&0.2&6&\textbf{0.651}\\
\hdashline 
NROWL2 & &&0.602  & &&0.615 & &&0.487 &&&0.474 \\
LSA-NROWL2 &-1.2&5&0.827&-0.6&5&0.792&-1.6&4&0.742& 0.2&10&0.534\\
CA-NROWL2&1.2&5&\underline{\textbf{0.896}}&3&6&\textbf{0.893}&3.8&10&\textbf{0.753}&0.4&6&\textbf{0.654}\\
\hdashline 
TFIDF & && 0.553 & && 0.609&  &&0.453 &&& 0.422 \\
LSA-TFIDF &-0.8&10&0.717&-0.4&9&0.724&0.2&24&0.498& 0.2&16&0.555\\
CA-TFIDF&2&6&\textbf{0.878}&4&10&\underline{\textbf{0.896}}&2.6&9&\textbf{ 0.767}& 1.4&16&\underline{\textbf{0.674}}\\
\hline 
\end{tabular}  
\end{table} 

\appendixqqsection{Cosine similarity}\label{SS: cosine similarity}

We performed identical experiments to the main paper, but using cosine similarity, instead of Euclidean distance, as similarity measurement method. The results follow the same trend of the main paper, leading similarity conclusions.

\subsection{Comparing LSA and CA for information retrieval}\label{Sub: standardcacos}

\subsubsection{MAP as a function of the number of dimensions for four versions of LSA with standard weighting exponent $\alpha = 1$ and CA}\label{Subsub: standardalphacos}

\begin{figure}[H]
    \centering
       \begin{subfigure}[b]{0.45\linewidth}
         \centering
         \includegraphics[width=1\textwidth]{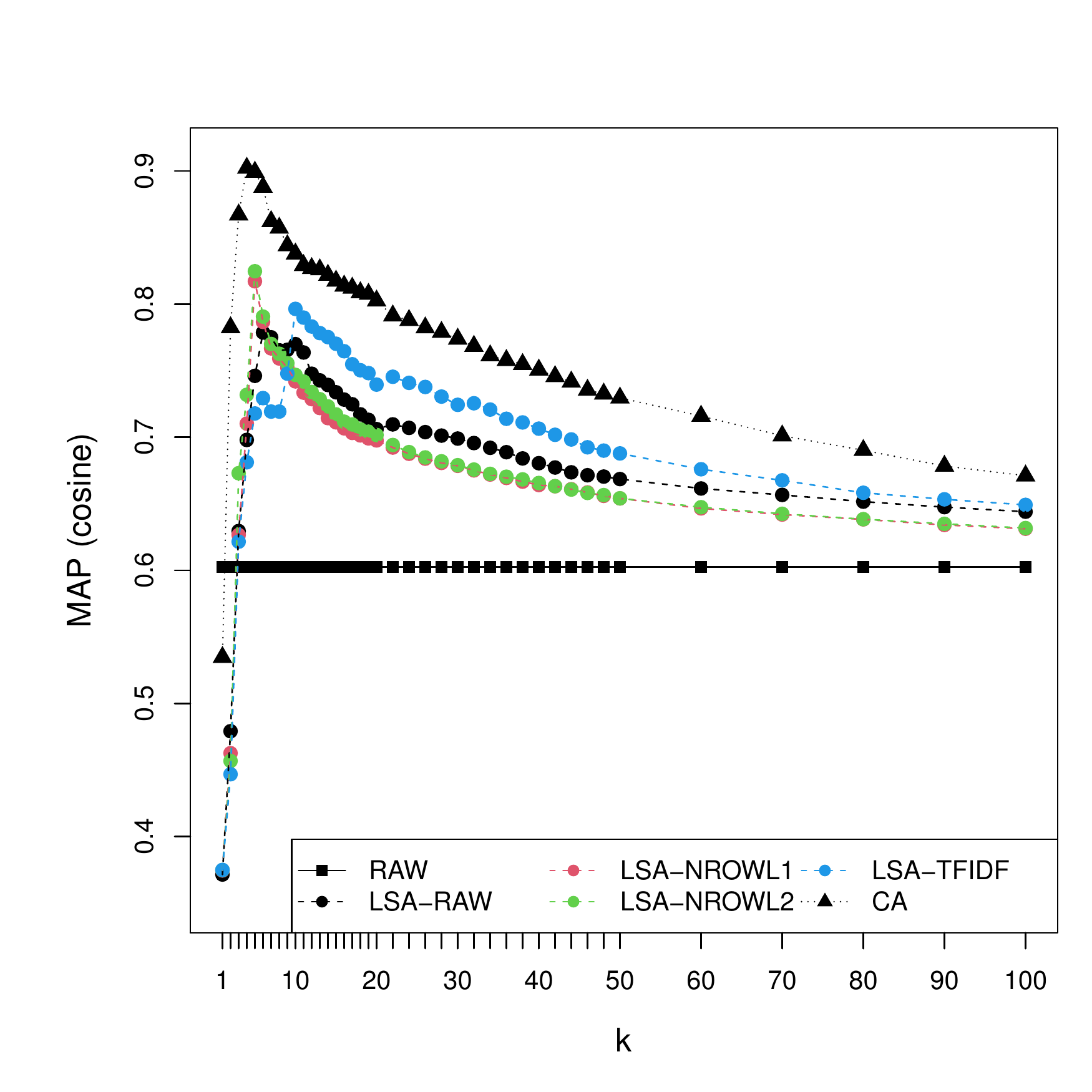}
        \caption{BBCNews}\label{F: cosstandardpBBCnews}
         \end{subfigure}
      \begin{subfigure}[b]{0.45\linewidth}
         \centering
         \includegraphics[width=1\textwidth]{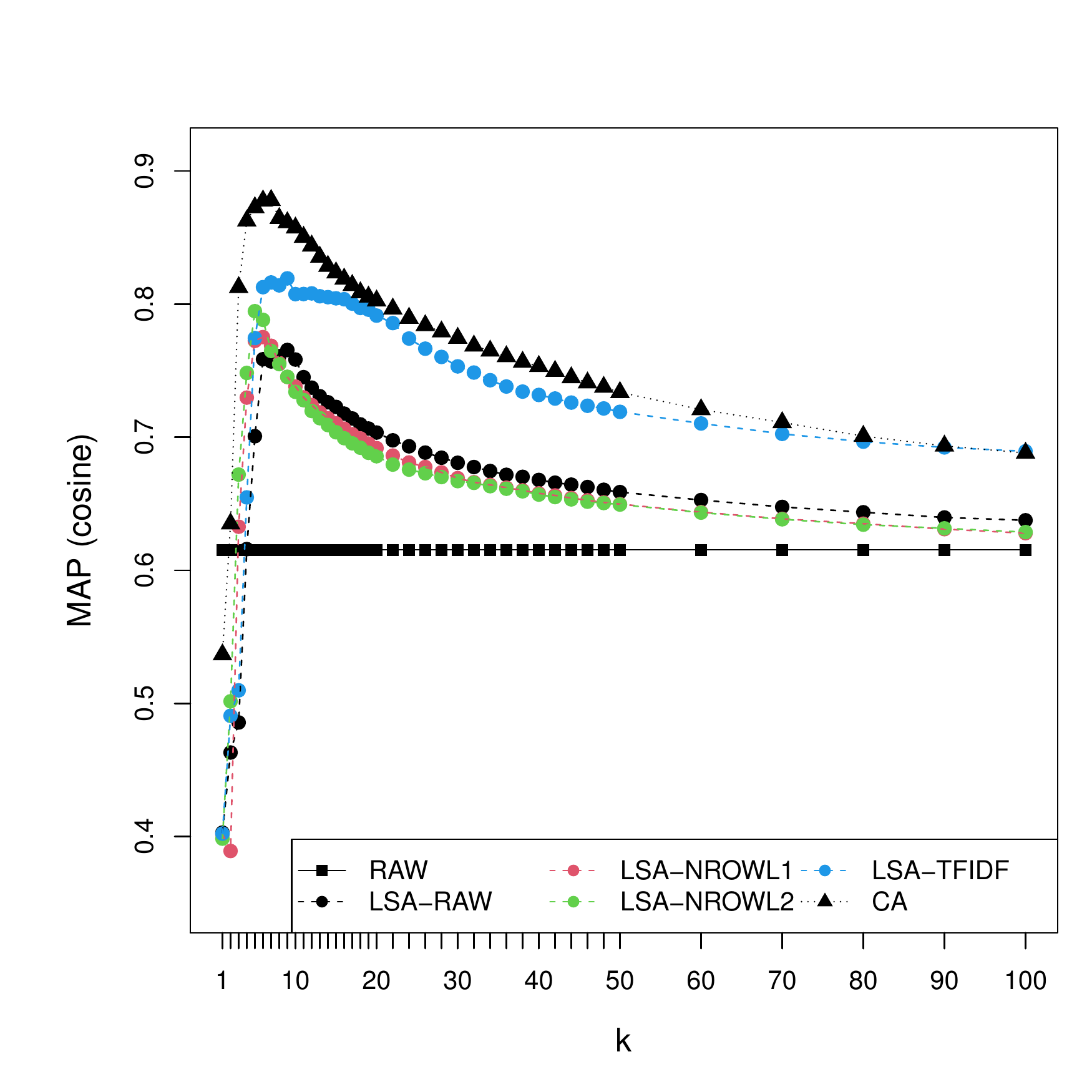}
         \caption{BBCSport}\label{F: cosstandardpBBCsport}
         \end{subfigure}
     \begin{subfigure}[b]{0.45\linewidth}
         \centering
         \includegraphics[width=1\textwidth]{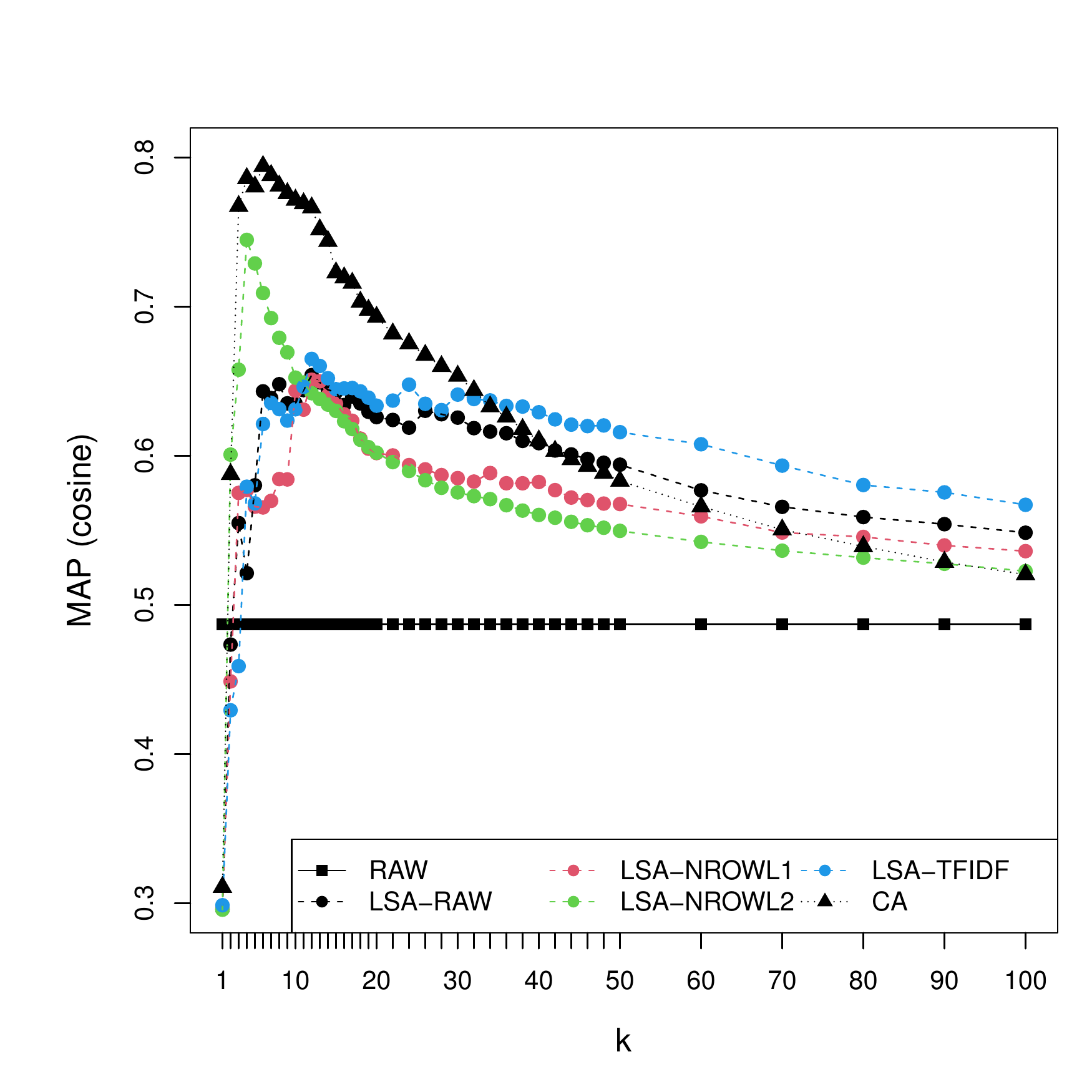}
        \caption{20 Newsgroups}\label{F: cosstandardp20newsgroups}
         \end{subfigure}
      \begin{subfigure}[b]{0.45\linewidth}
         \centering
         \includegraphics[width=1\textwidth]{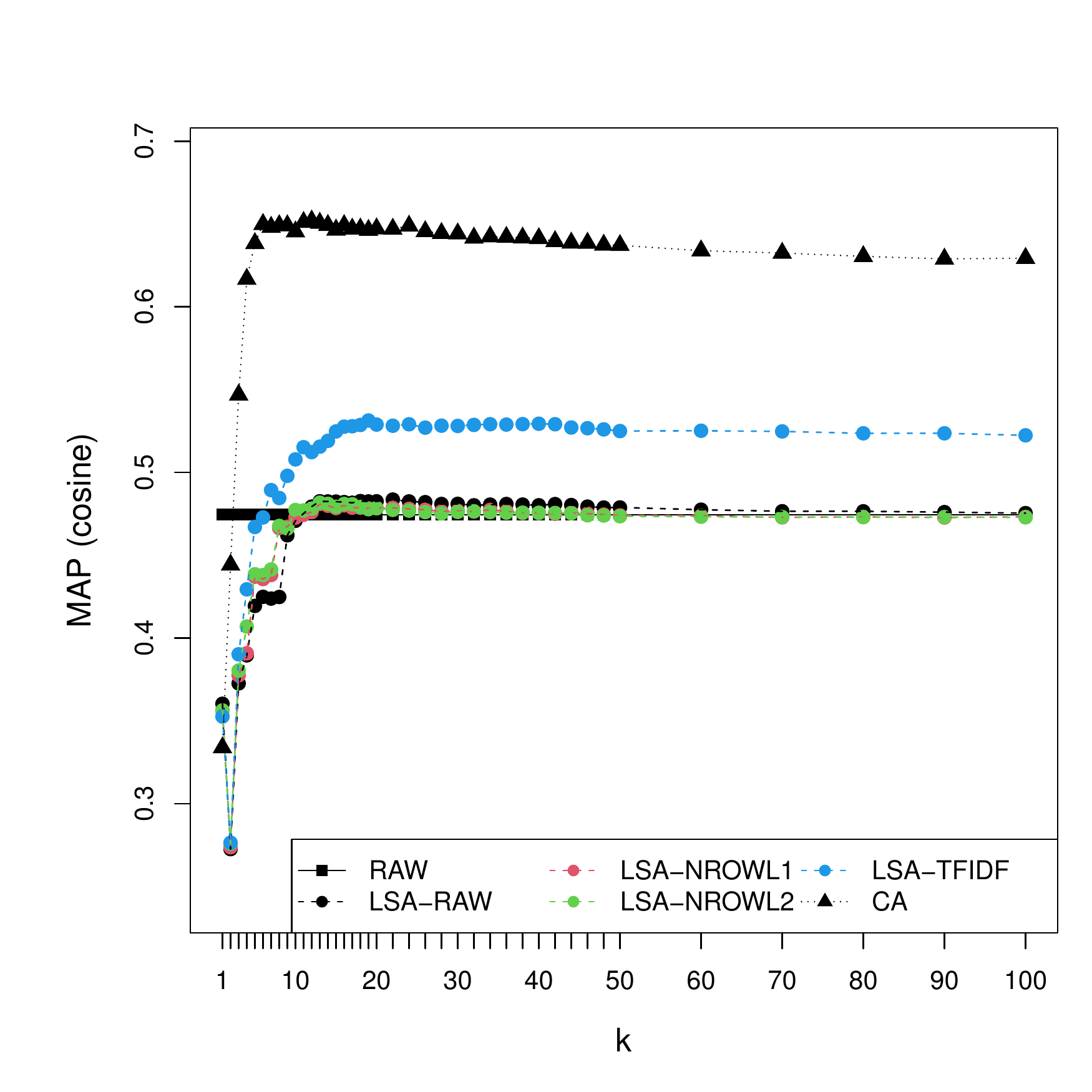}
         \caption{Wilhelmus}\label{F: cosstandardpwilhelmus}
         \end{subfigure}
    \caption{MAP as a function of the number of dimensions under standard coordinates.}
    \label{F: cosstandardp}
\end{figure}

\subsubsection{MAP as a function of the weighting exponent $\alpha$ about LSA and MAP about CA for various values of the number of dimensions}\label{Sub: constantkcos}

\begin{figure}[H]
    \centering
     \begin{subfigure}[b]{0.45\linewidth}
         \centering
         \includegraphics[width=1\textwidth]{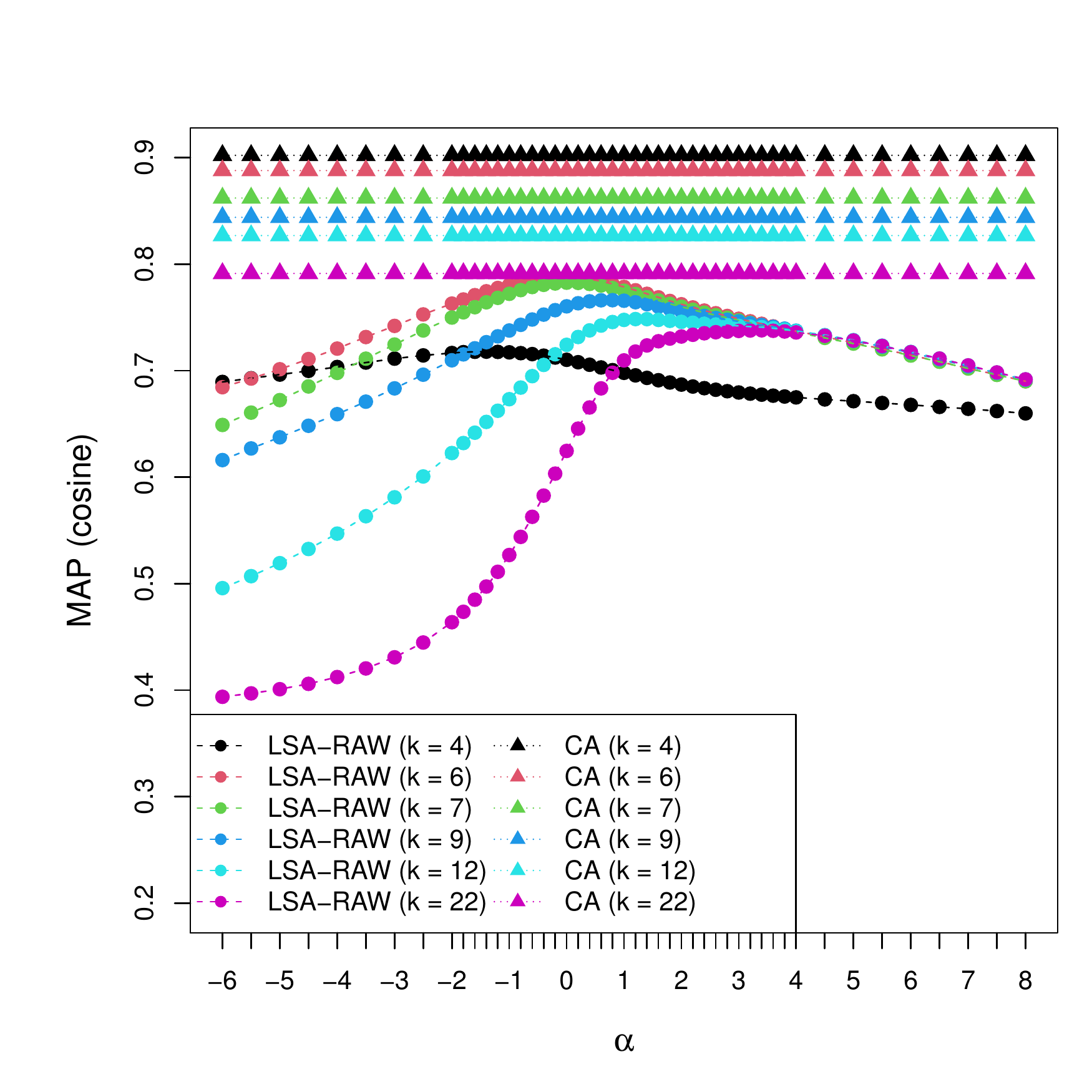}
         \caption{BBCNews}\label{F: cosstandarddBBCnews}
         \end{subfigure}
      \begin{subfigure}[b]{0.45\linewidth}
         \centering
         \includegraphics[width=1\textwidth]{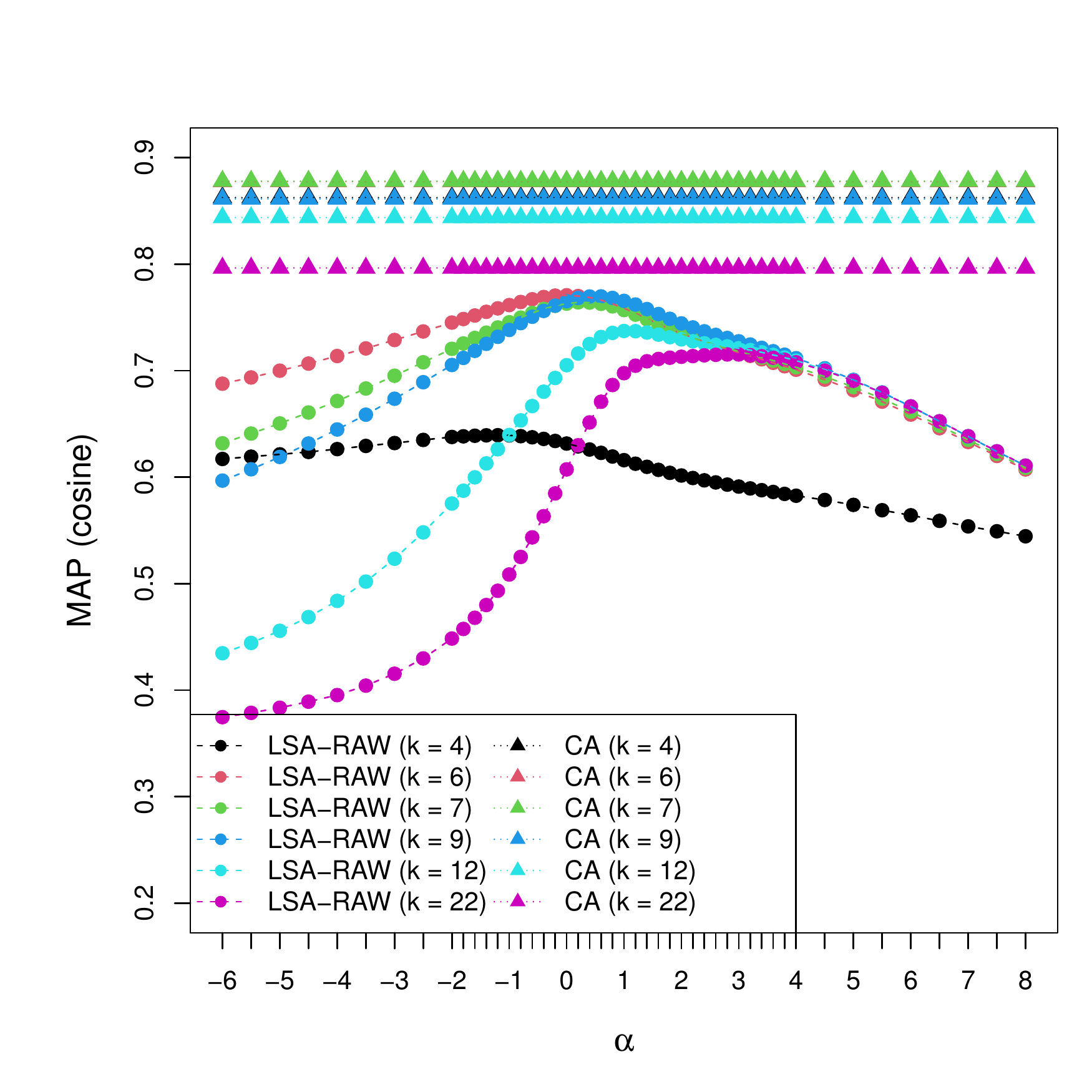}
         \caption{BBCSport}\label{F: cosstandarddBBCsport}
         \end{subfigure}
      \begin{subfigure}[b]{0.45\linewidth}
         \centering
         \includegraphics[width=1\textwidth]{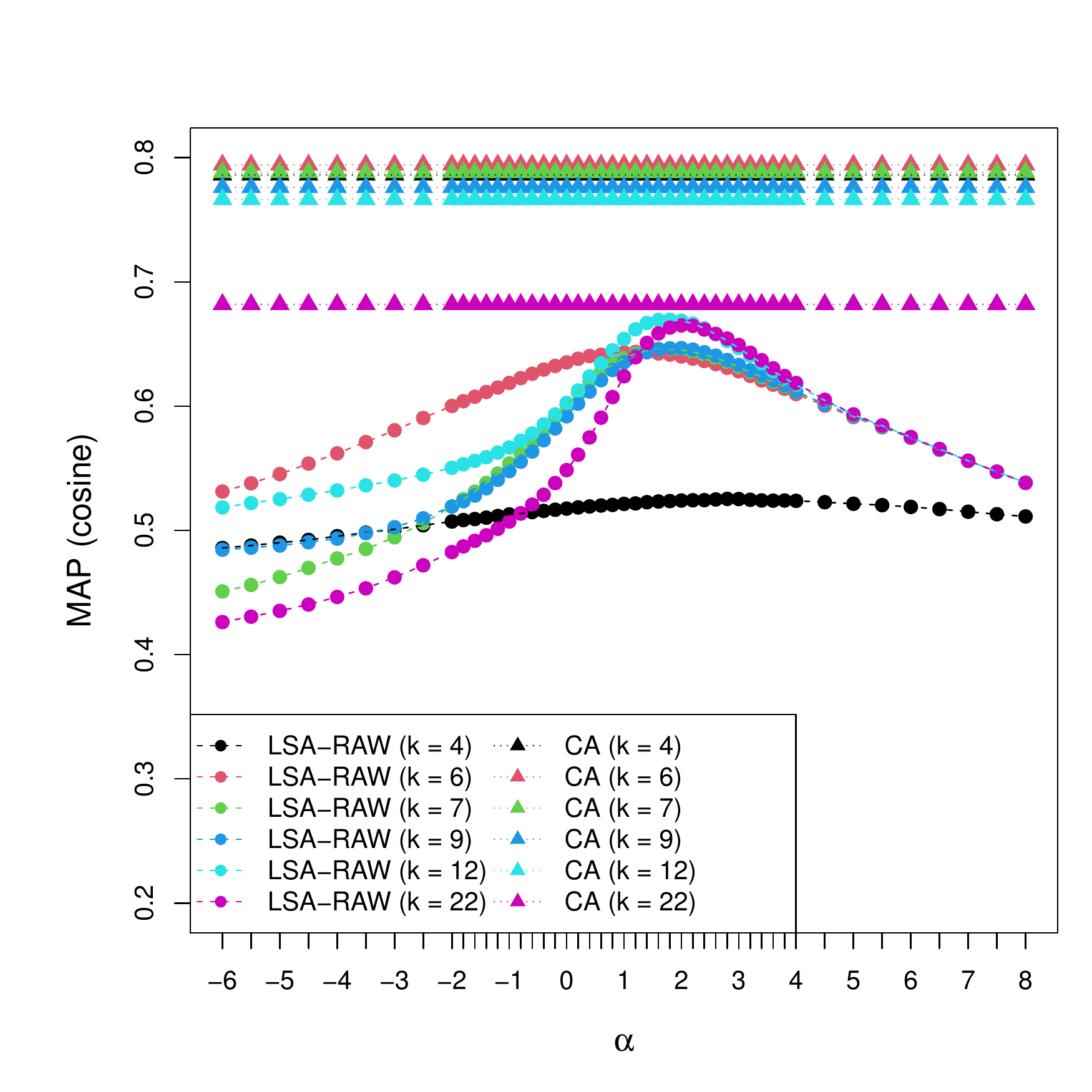}
         \caption{20 Newsgroups}\label{F: cosstandardd20newsgroups}
         \end{subfigure}
      \begin{subfigure}[b]{0.45\linewidth}
         \centering
         \includegraphics[width=1\textwidth]{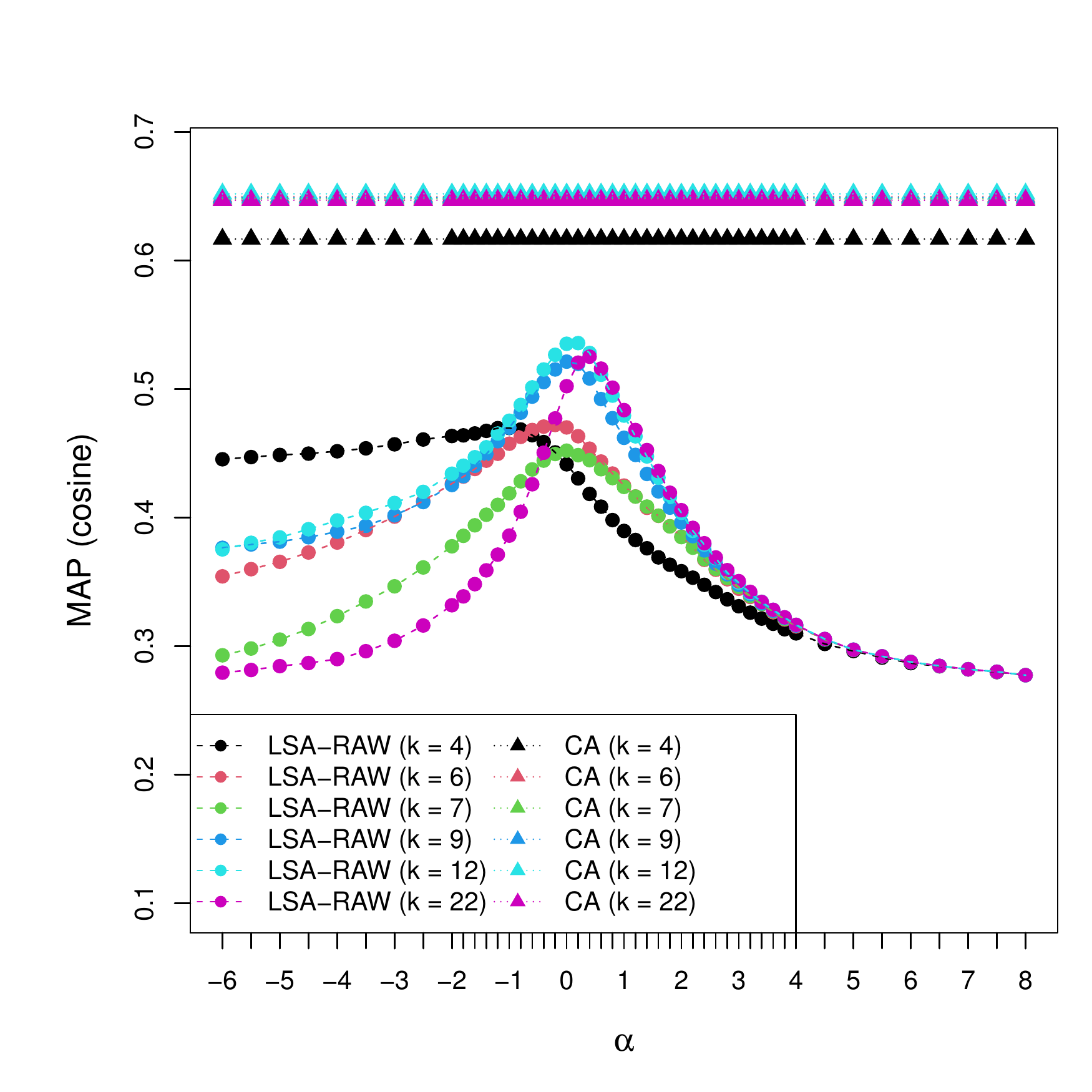}
         \caption{Wilhelmus}\label{F: cosstandarddwilhelmus}
         \end{subfigure}
    \caption{MAP as a function of $\alpha$ for LSA-RAW and MAP for CA under various values of $k$.}
    \label{F: cosstandardd}
\end{figure}

Figure~\ref{F: cosstandardd} shows MAP as a function of $\alpha$ about LSA-RAW and MAP about CA for the number of dimensions: $k = 4, 6, 7, 9, 12, \text{and }22$.

\begin{table}[H]
\centering  
\caption{MAP with the optimal weighting exponent $\alpha$ for LSA-RAW and MAP for CA under $k = 4, 6, 7, 9, 12, \text{and }22$ about cosine similarity. Bold values are best.} 
\label{TIRcosdim10}
\begin{tabular}{ccccccccc}    
&\multicolumn{2}{c}{BBCNews} &\multicolumn{2}{c}{BBCSport}&\multicolumn{2}{c}{20 Newsgroups}&\multicolumn{2}{c}{Wilhelmus} \\
&$\alpha$&MAP&$\alpha$&MAP&$\alpha$&MAP&$\alpha$&MAP\\
\hline 
LSA-RAW ($k = 4$) &-1.4& 0.718&-1.2&  0.639& 2.8 &0.525 & -1 &0.470 \\
LSA-RAW ($k = 6$) &0&0.788 &0& 0.771& 1.2&0.644 & -0.2 &0.472 \\
LSA-RAW ($k = 7$) &0& 0.783&0.2& 0.764&1.6&0.646 & 0&0.452 \\
LSA-RAW ($k = 9$) &0.8&0.766 &0.6 & 0.770& 2&0.647 & 0 &0.521\\
LSA-RAW ($k = 12$) &1.2&0.748 &1& 0.737& 1.8& 0.670 & 0.2 &0.536 \\
LSA-RAW ($k = 22$) &3.4&0.738&2.8 & 0.715& 2& 0.665 & 0.4 &0.525 \\
CA ($k = 4$) & &  \textbf{0.902}& &  0.863&  &0.786 &   &0.617 \\
CA ($k = 6$) & &  0.888& &  0.878&  &\textbf{0.794} &   &0.650 \\
CA ($k = 7$) & &  0.862& & \textbf{0.878}&  &0.788 &   &0.648 \\
CA ($k = 9$) & & 0.844& &  0.861&  &0.776 &   &0.649\\
CA ($k = 12$) & & 0.827& & 0.844&  &0.767 &  & \textbf{0.652}\\
CA ($k = 22$) & & 0.791& & 0.796&  &0.682 &  & 0.647\\
\hline 
\end{tabular}  
\end{table} 

\subsection{Improving performance of CA for information retrieval}\label{Sub: improvingcacos}

\subsubsection{Weighting scheme of raw document-term matrix for CA}\label{Subsub: caweightingschemecos}

\begin{figure}[H]
    \centering
       \begin{subfigure}[b]{0.45\linewidth}
         \centering
         \includegraphics[width=1\textwidth]{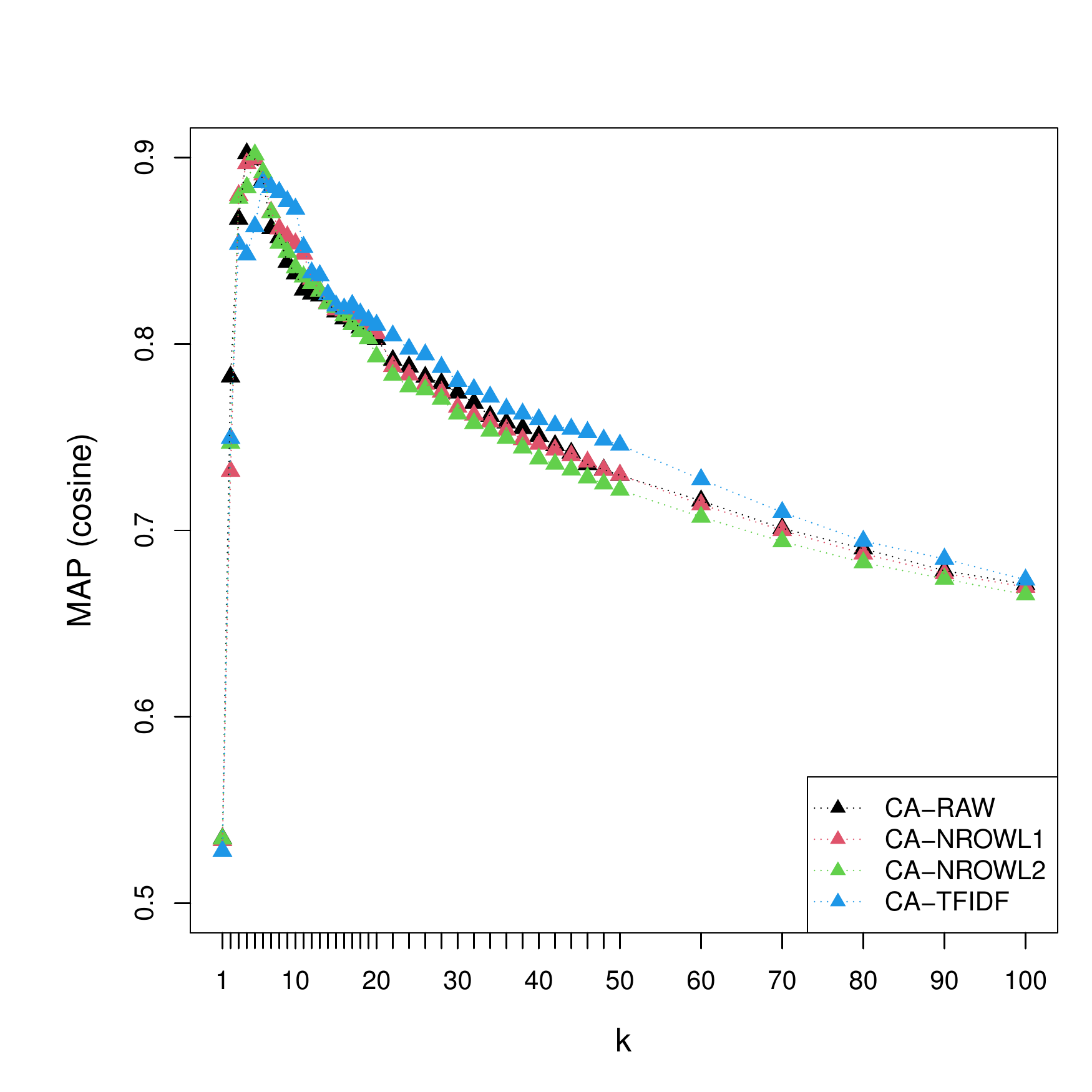}
         \caption{$\alpha = 1$ (BBCNews)}\label{F: cosstandardpBBCnewscaweisch}
         \end{subfigure}
      \begin{subfigure}[b]{0.45\linewidth}
         \centering
         \includegraphics[width=1\textwidth]{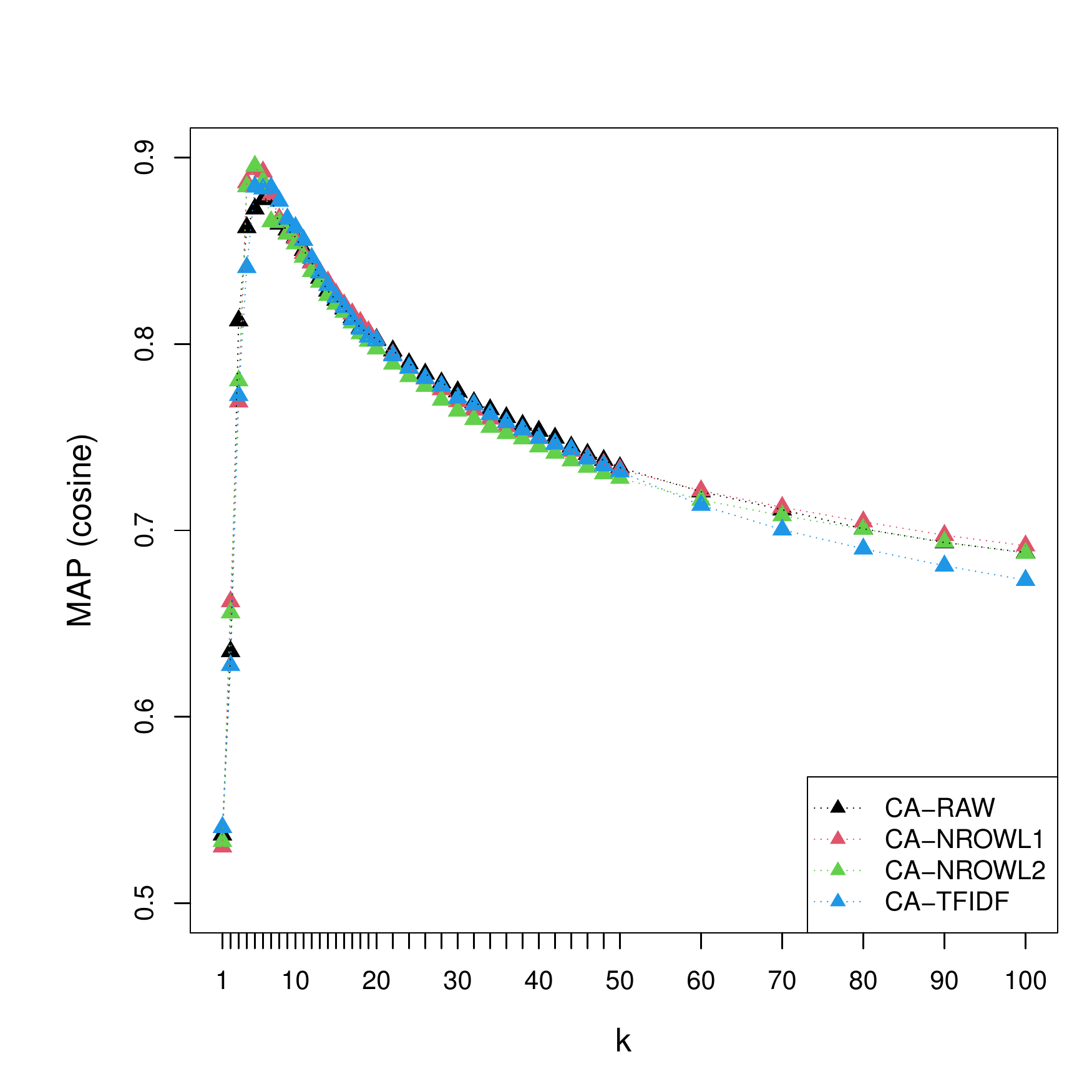}
         \caption{$\alpha = 1$ (BBCSport)}\label{F: cosstandardpBBCsportcaweisch}
         \end{subfigure}
     \begin{subfigure}[b]{0.45\linewidth}
         \centering
         \includegraphics[width=1\textwidth]{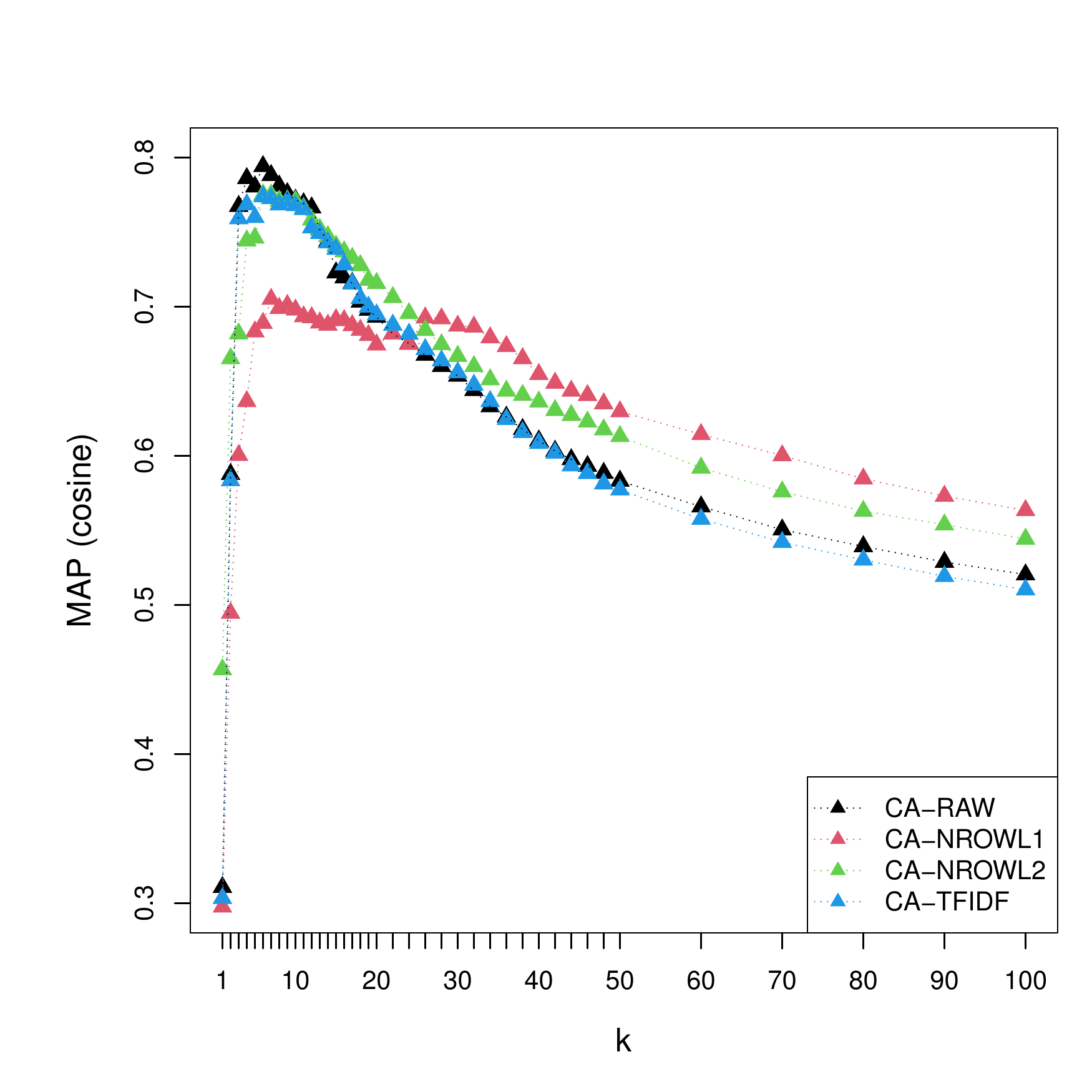}
         \caption{$\alpha = 1$ (20 Newsgroups)}\label{F: cosstandardp20newsgroupscaweisch}
         \end{subfigure}
      \begin{subfigure}[b]{0.45\linewidth}
         \centering
         \includegraphics[width=1\textwidth]{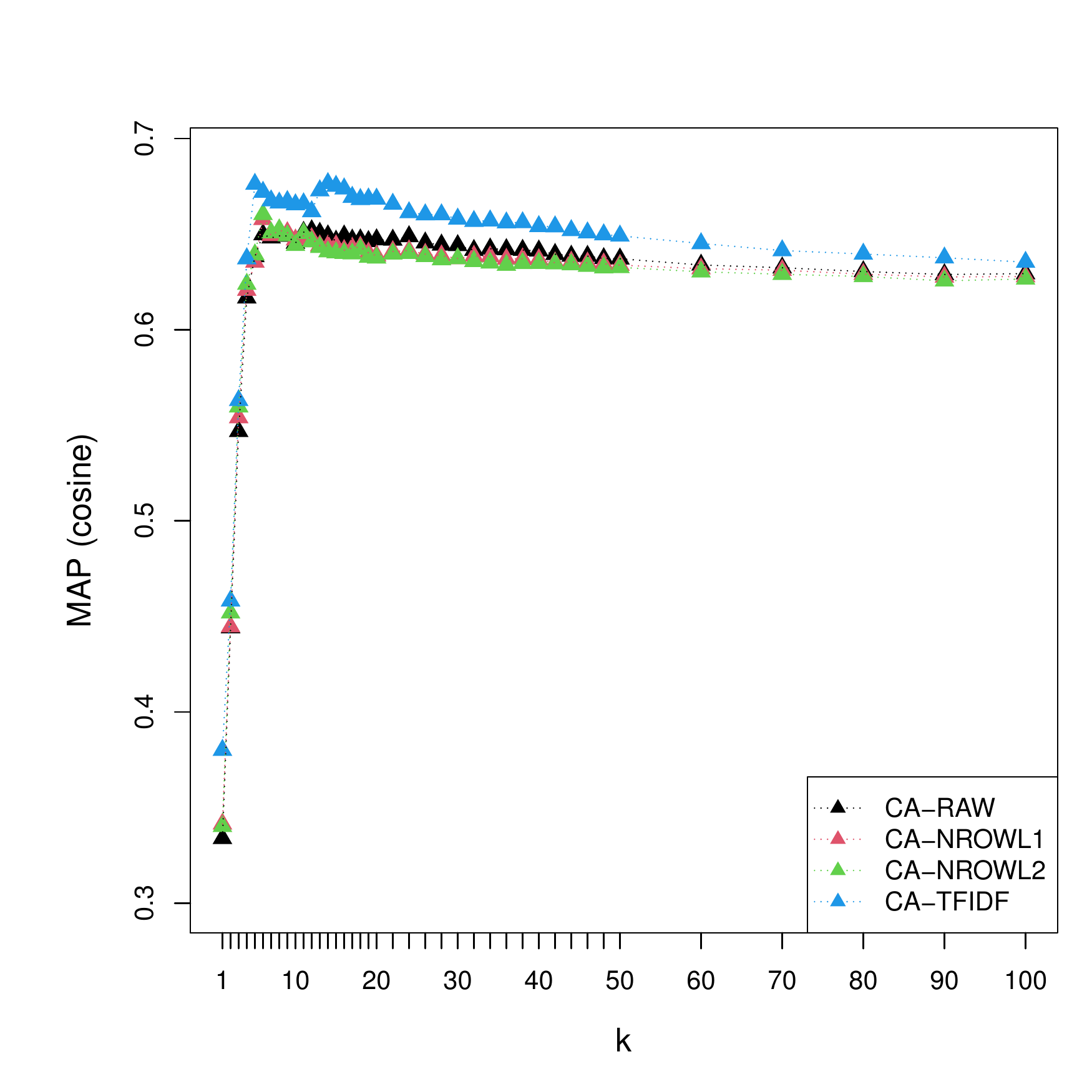}
         \caption{$\alpha = 1$ (Wilhelmus)}\label{F: cosstandardpwilhelmuscaweisch}
         \end{subfigure}
    \caption{MAP as a function of the number of dimensions $k$ for the four versions of CA under standard coordinates.}
    \label{F: cosstandardpcaweisch}
\end{figure}

\begin{table}[H]
\centering  
\caption{MAP with the optimal number of dimensions $k$ for the four versions of CA about cosine similarity. Bold values are best.} 
\label{TIRcosalpha1caweisch}
\begin{tabular}{lcccccccc}    
&\multicolumn{2}{c}{BBCNews} &\multicolumn{2}{c}{BBCSport}&\multicolumn{2}{c}{20 Newsgroups}&\multicolumn{2}{c}{Wilhelmus} \\
&$k$&MAP&$k$&MAP&$k$&MAP&$k$&MAP\\
\hline 
CA-RAW&4&\textbf{0.902}&7&0.878& 6&\textbf{0.794}&12&0.652\\
CA-NROWL1&5&0.900&5&0.894&7&0.705&6&0.658\\
CA-NROWL2&5&0.902&5&\textbf{0.896}&6&0.775&6&0.660\\
CA-TFIDF&6&0.887&5&0.884&6&0.774&14&\textbf{0.677}\\
\hline 
\end{tabular}  
\end{table} 

\subsubsection{Weighting exponent $\alpha$ in CA}\label{Subsub: caweightingexponentcos}

\begin{figure}[H]
    \centering
     \begin{subfigure}[b]{0.45\linewidth}
         \centering
         \includegraphics[width=1\textwidth]{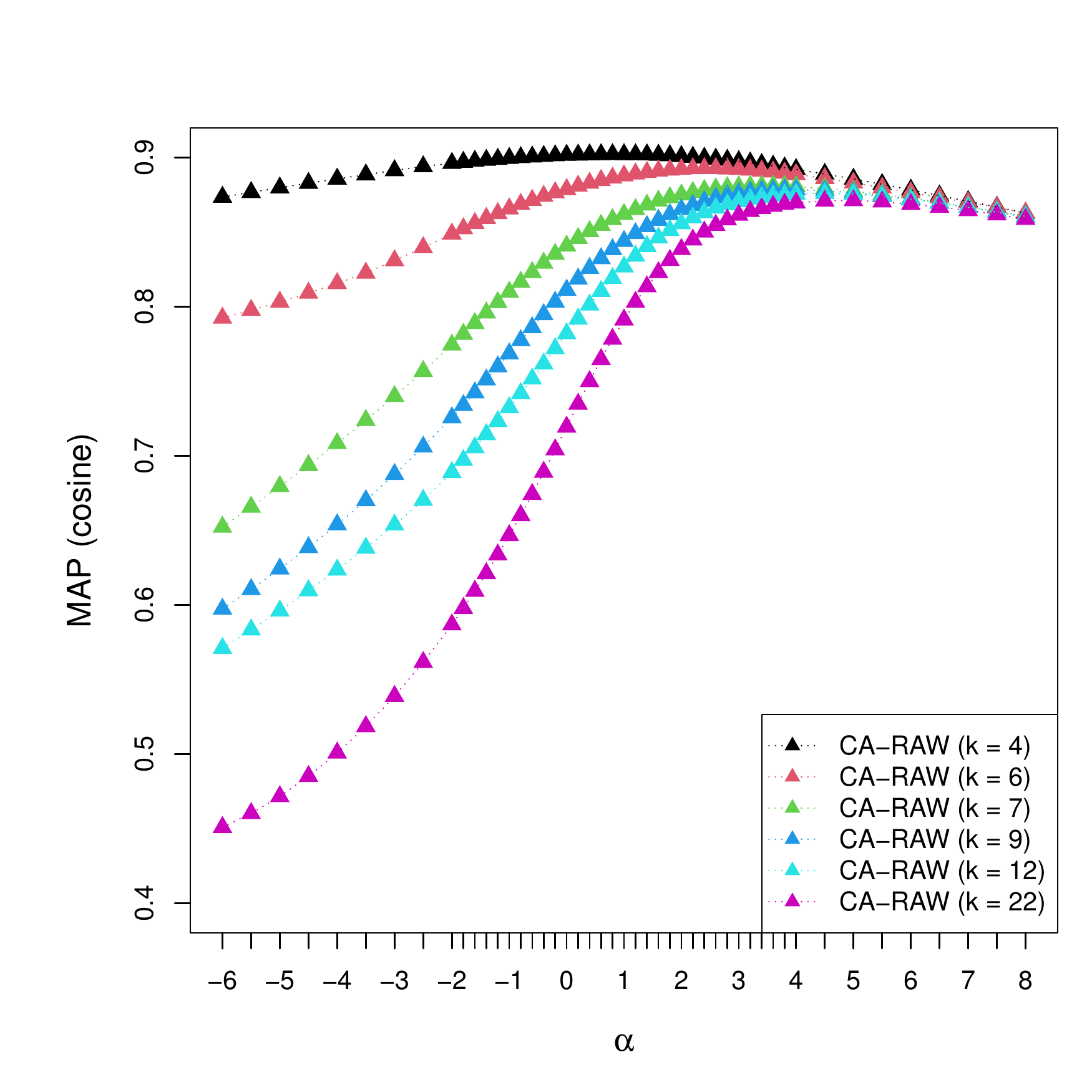}
         \caption{BBCNews}\label{F: cosstandarddBBCnewscaweiexp}
         \end{subfigure}
      \begin{subfigure}[b]{0.45\linewidth}
         \centering
         \includegraphics[width=1\textwidth]{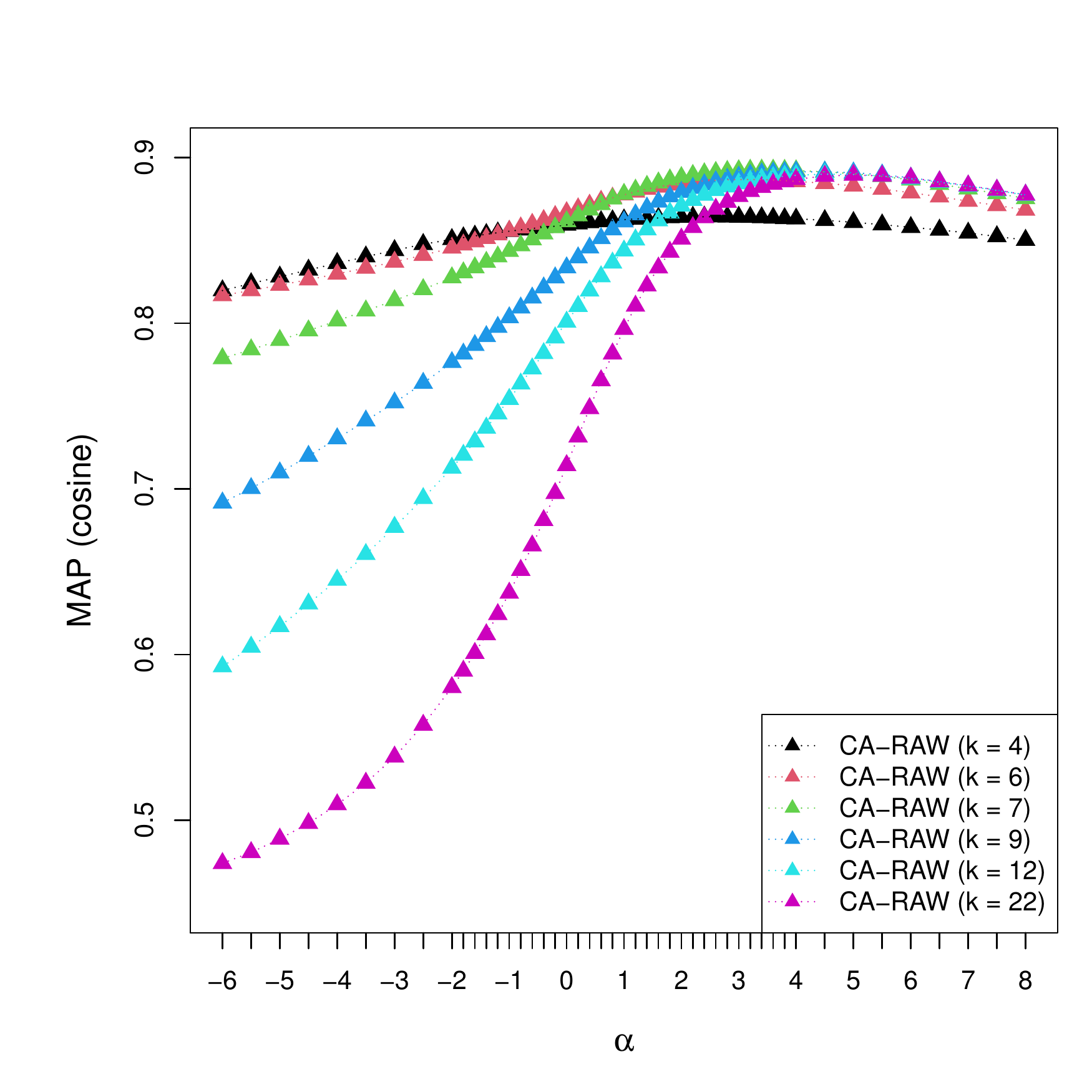}
         \caption{BBCSport}\label{F: cosstandarddBBCsportcaweiexp}
         \end{subfigure}
      \begin{subfigure}[b]{0.45\linewidth}
         \centering
         \includegraphics[width=1\textwidth]{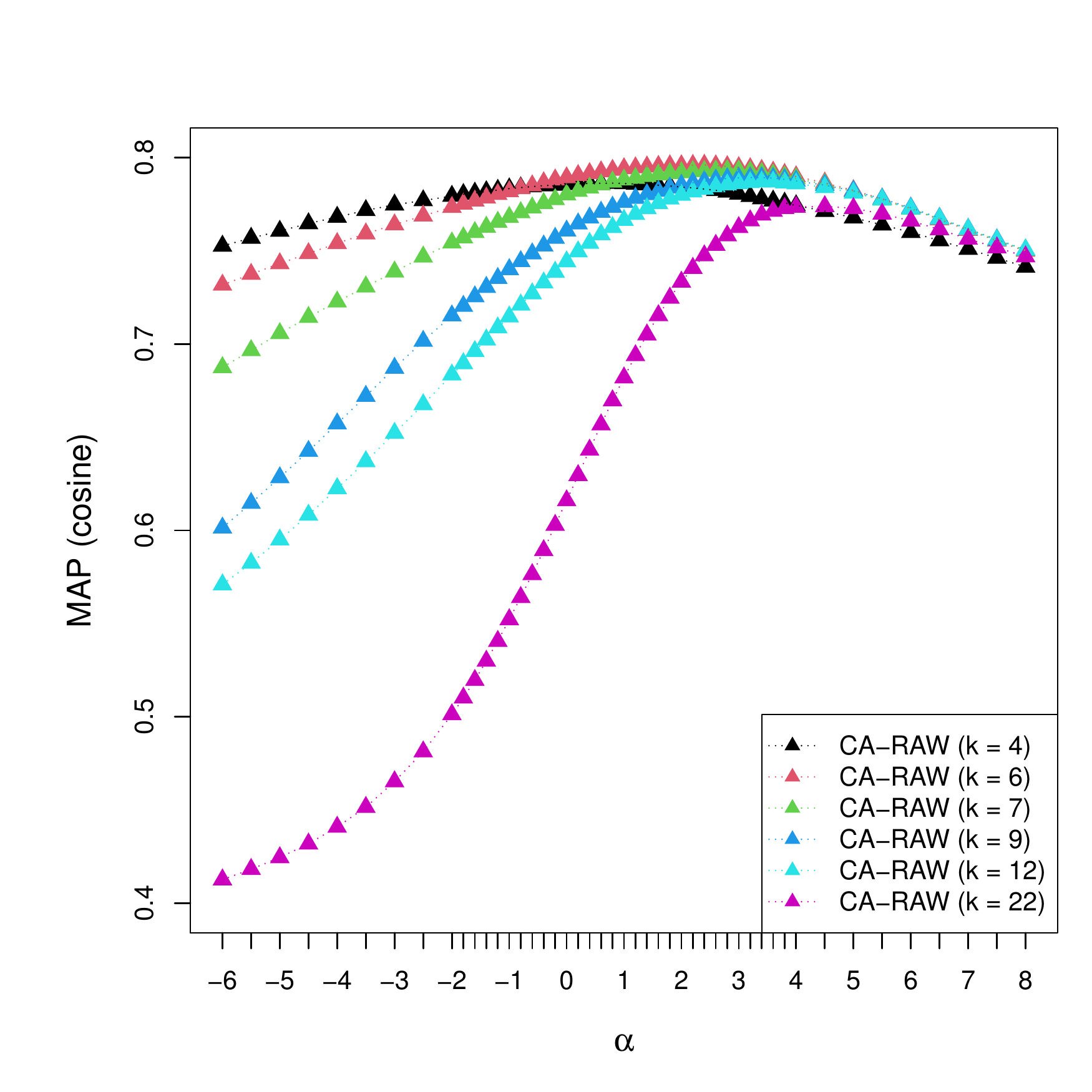}
         \caption{20 Newsgroups}\label{F: cosstandardd20newsgroupscaweiexp}
         \end{subfigure}
      \begin{subfigure}[b]{0.45\linewidth}
         \centering
         \includegraphics[width=1\textwidth]{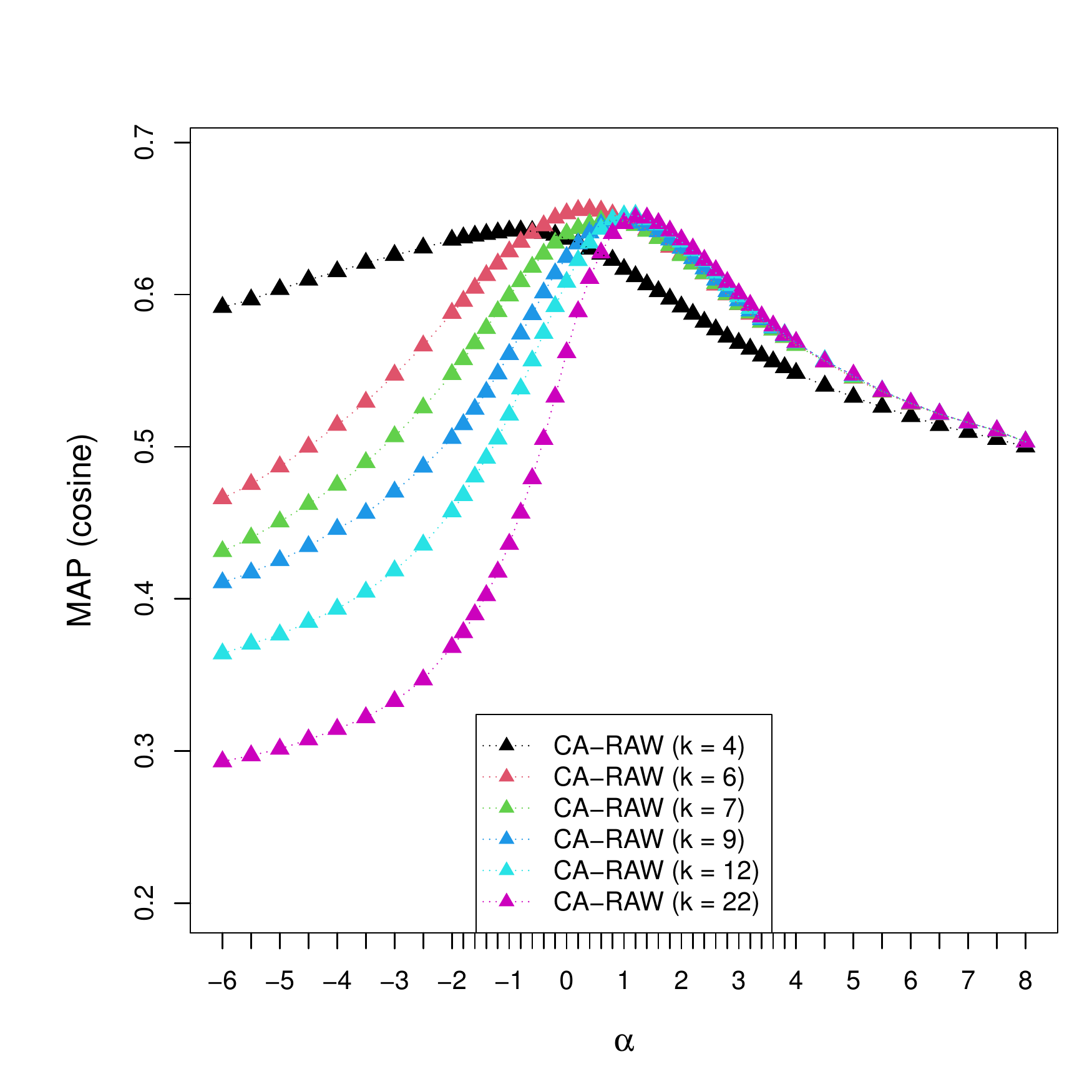}
         \caption{Wilhelmus}\label{F: cosstandarddwilhelmuscaweiexp}
         \end{subfigure}
    \caption{ MAP as a function of $\alpha$ for CA-RAW under various values of $k$.}
    \label{F: cosstandarddcaweiexp}
\end{figure}

\begin{table}[H]
\centering  
\caption{MAP with the optimal $\alpha$ for CA-RAW under $k = 4, 6, 7, 9, 12, \text{and }22$ about cosine similarity. Bold values are best.} 
\label{TIRcosdim10caweiexp}
\begin{tabular}{ccccccccc}    
&\multicolumn{2}{c}{BBCNews} &\multicolumn{2}{c}{BBCSport}&\multicolumn{2}{c}{20 Newsgroups}&\multicolumn{2}{c}{Wilhelmus} \\
&$\alpha$&MAP&$\alpha$&MAP&$\alpha$&MAP&$\alpha$&MAP\\
\hline 
CA ($k = 4$) &0.8 &  \textbf{0.902}&2.6 & 0.864& 0.8 &0.786 & -0.8  &0.642 \\
CA ($k = 6$) & 2.4& 0.893&3.2 & 0.887& 2.4 &\textbf{0.796} &  0.4 &\textbf{0.656} \\
CA ($k = 7$) & 3.6&  0.881& 3.4 & \textbf{0.893}&2.8  &0.793 &  0.8 &0.649 \\
CA ($k = 9$) & 3.8&0.879 &4 & 0.892& 3.4 & 0.789& 1  &0.649\\
CA ($k = 12$) & 4.5& 0.876&4.5 & 0.890&3.6  & 0.787& 1.2 & 0.652\\
CA ($k = 22$) &5 & 0.871&5 &0.890&4.5  &0.774 & 1.2 & 0.651\\
\hline 
\end{tabular}  
\end{table} 

\subsection{Exploring MAP as a function of $\alpha$ under the optimal number of dimensions for LSA and CA}\label{Sub: optimalalphacos}

\begin{figure}[H]
    \centering
       \begin{subfigure}[b]{0.45\linewidth}
         \centering
         \includegraphics[width=1\textwidth]{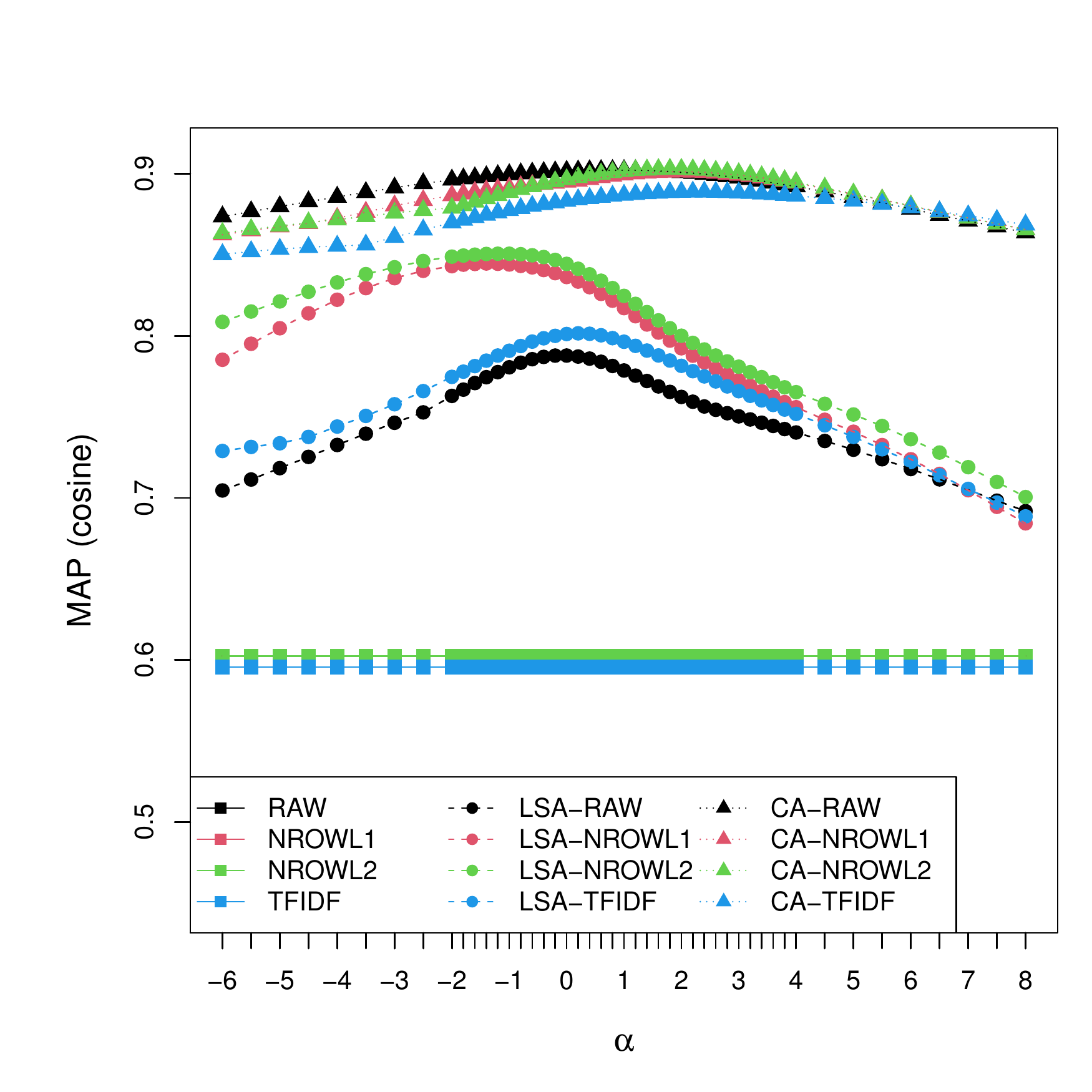}
         \caption{BBCNews}\label{F: cosoptimalMAPBBCnews}
         \end{subfigure}
      \begin{subfigure}[b]{0.45\linewidth}
         \centering
         \includegraphics[width=1\textwidth]{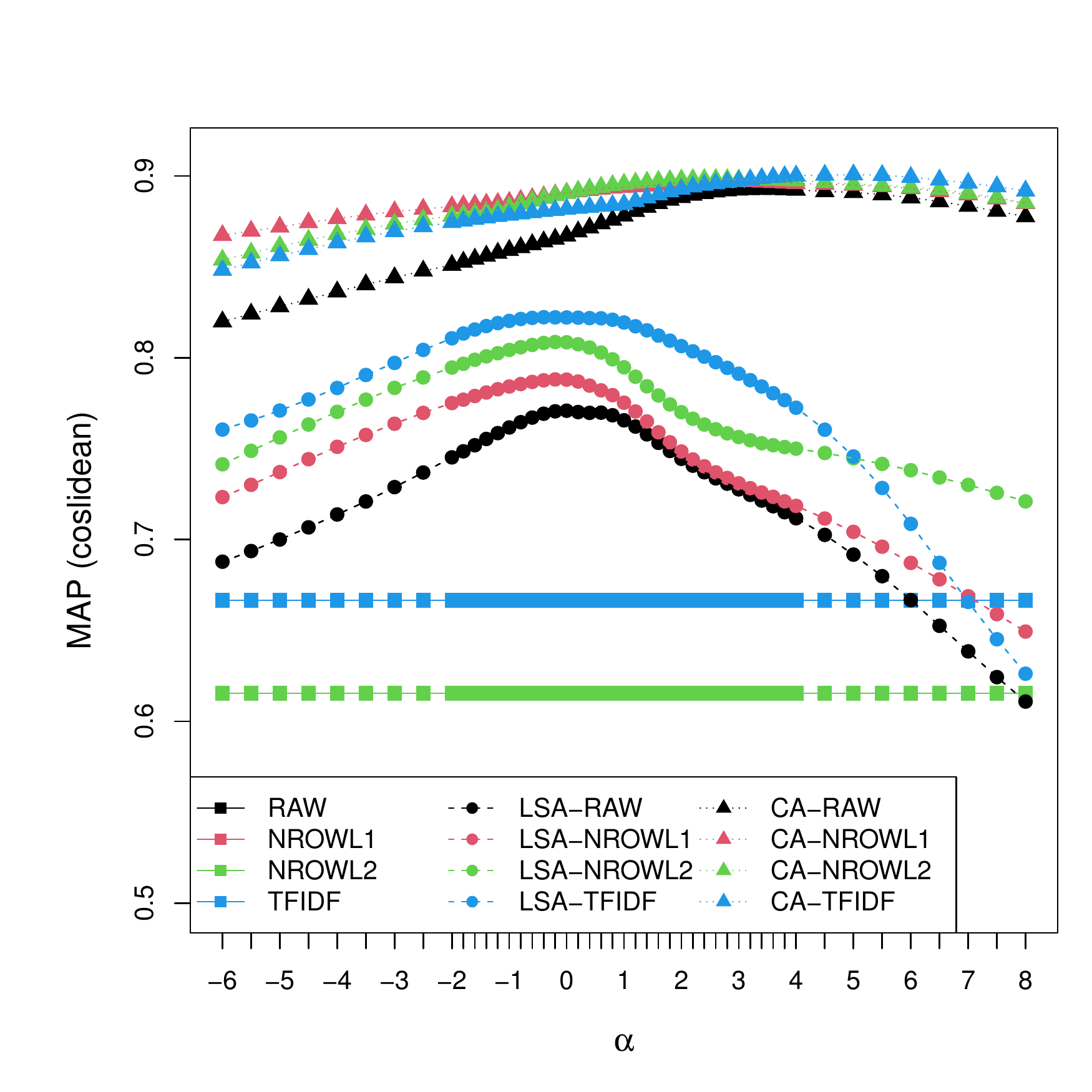}
         \caption{BBCSport}\label{F: cosoptimalMAPBBCsport}
         \end{subfigure}
        \begin{subfigure}[b]{0.45\linewidth}
         \centering
         \includegraphics[width=1\textwidth]{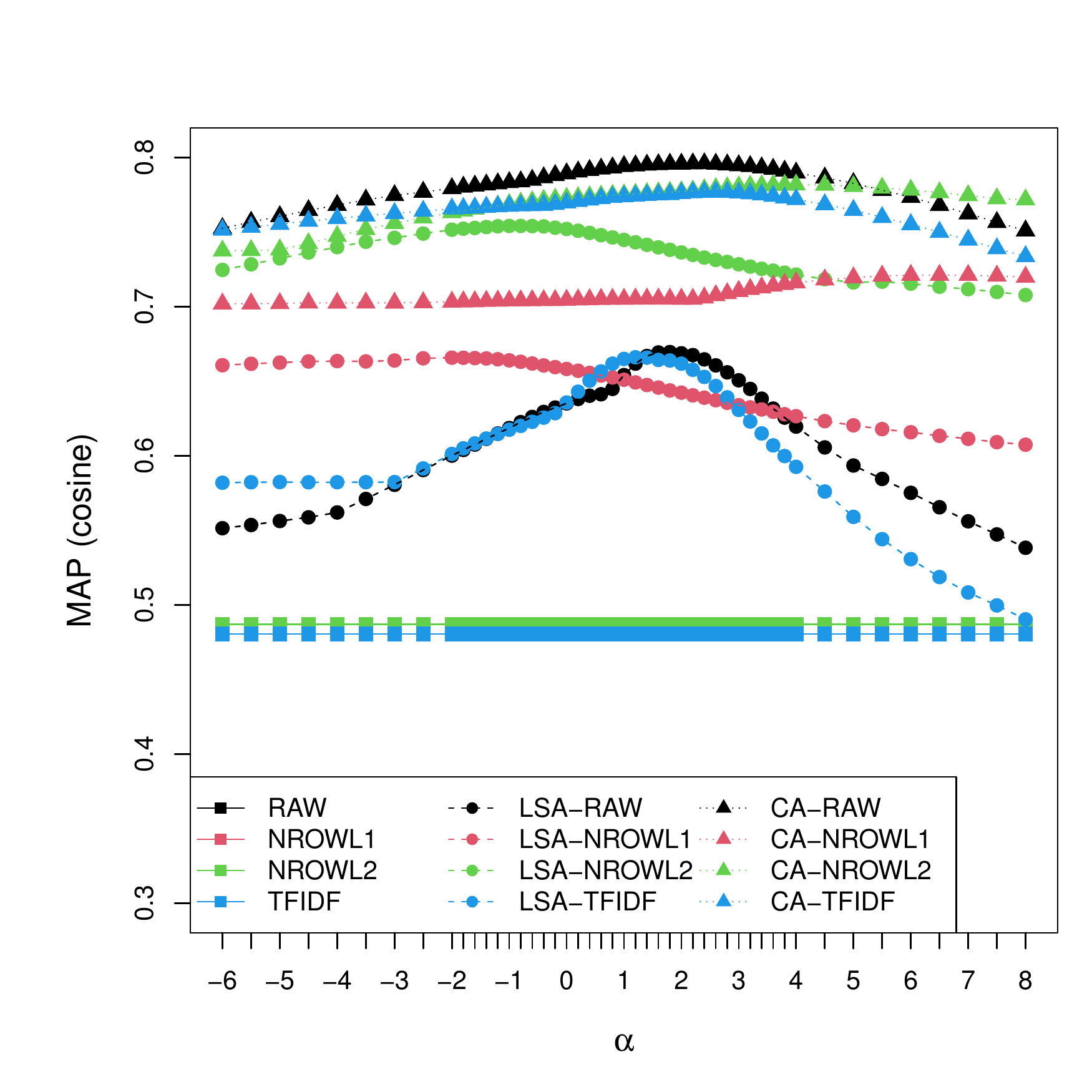}
         \caption{20 Newsgroups}\label{F: cosoptimalMAP20newsgroups}
         \end{subfigure}
        \begin{subfigure}[b]{0.45\linewidth}
         \centering
         \includegraphics[width=1\textwidth]{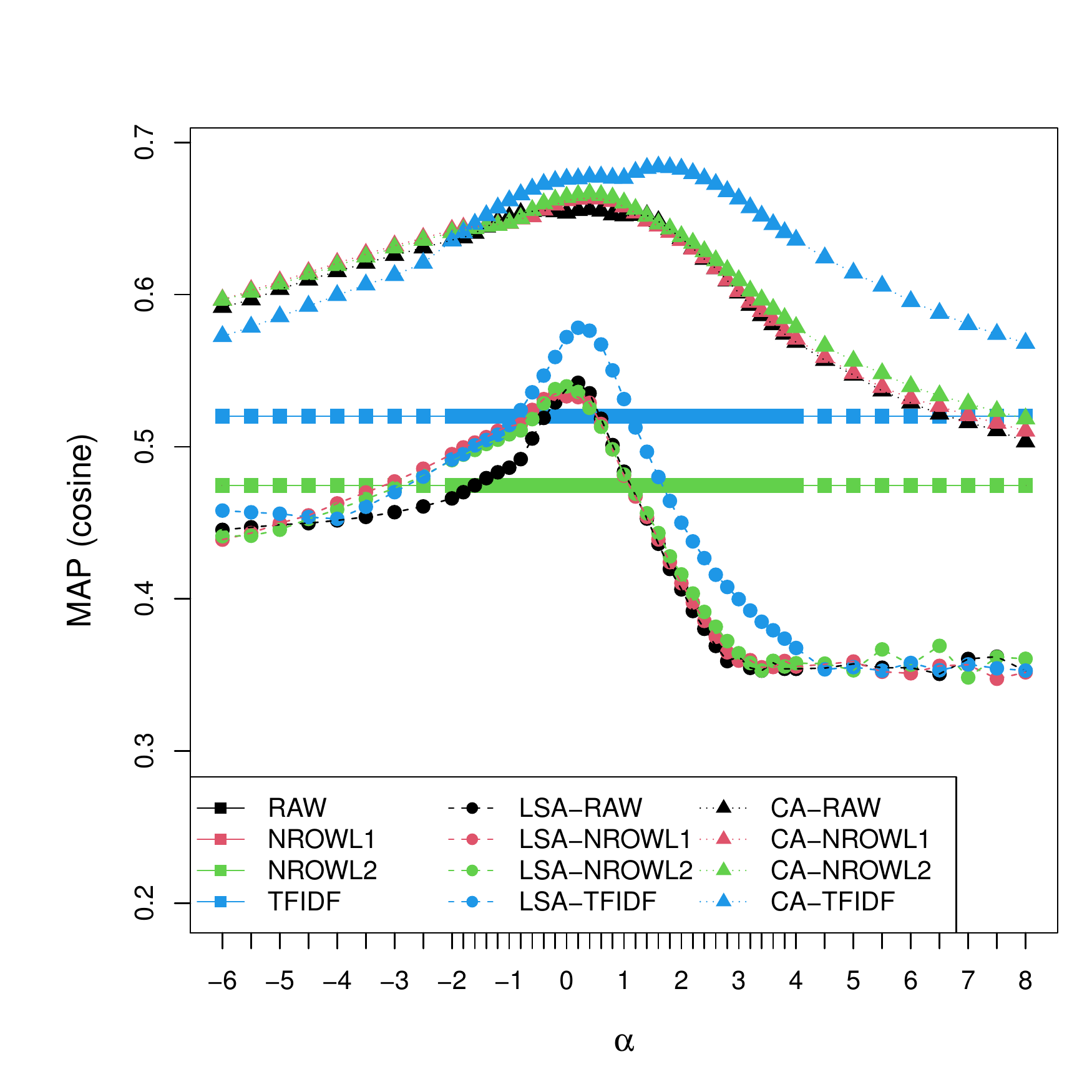}
         \caption{Wilhelmus}\label{F: cosoptimalMAPwilhelmus}
         \end{subfigure}
    \caption{MAP as a function of $\alpha$ under optimal dimension.}
    \label{F: cosoptimalMAP}
\end{figure}

\begin{table}[H]
\centering  
\caption{MAP under the optimal $\alpha$ and optimal dimension $k$ about cosine similarity. Bold values are best within group; underlined values are best overall.} 
\label{TIRcos}
\begin{tabular}{ccccccccccccc}    
&\multicolumn{3}{c}{BBCNews} &\multicolumn{3}{c}{BBCSport}&\multicolumn{3}{c}{20 Newsgroups}&\multicolumn{3}{c}{Wilhelmus} \\
&$\alpha$&$k$&MAP&$\alpha$&$k$&MAP&$\alpha$&$k$&MAP&$\alpha$&$k$&MAP\\
\hdashline 
RAW & &&0.602& &&0.615& &&0.487&&& 0.474\\
LSA-RAW &0 &6 & 0.788& 0&6 & 0.771 & 1.8 &12 & 0.670&0.2&13& 0.542\\
CA-RAW&0.8&4&\textbf{0.902}&3.4&7&\textbf{0.893}&2.4&6&\underline{\textbf{0.796}}&0.4&6&\textbf{0.656}\\
\hdashline 
NROWL1& &&0.602&  &&0.615 & &&0.487& &&0.474 \\
LSA-NROWL1 &-1.4&5&0.845&-0.2&5&0.788&-2&12&0.666&-0.2&8& 0.535\\
CA-NROWL1&2&5&\textbf{0.901}
&2.8&6&\textbf{0.897}& 6.5&32&\textbf{0.721}&0.4&6&\textbf{0.664}\\
\hdashline 
NROWL2 & &&0.602 & &&0.615 & &&0.487 &&&0.474 \\
LSA-NROWL2 &-1&5&0.851&-0.2&5& 0.809&-1&4&0.754& 0.0&10&0.540\\
CA-NROWL2&1.8&5&\underline{\textbf{0.903}}&2.6&5&\textbf{0.898}&4&10&\textbf{0.782}&0.4&6&\textbf{0.666}\\
\hdashline 
TFIDF & && 0.596& &&0.667 & &&0.481 &&& 0.520 \\
LSA-TFIDF &0.2&10&0.802&-0.4&6&0.822&1.2&12&0.666&0.2&16&0.578\\
CA-TFIDF&2.4&6&\textbf{0.889}&5&10&\underline{\textbf{0.901}}&2.6&9&\textbf{0.777}&1.6&14&\underline{\textbf{0.684}}\\
\hdashline 
\end{tabular}  
\end{table}

\end{document}